\documentclass{emulateapj}  
\usepackage{apjfonts}

\newcommand{\Lsol}{\mbox{$L_\odot$}}
\newcommand{\Msol}{\mbox{$M_\odot$}}

\newcommand{\kms}{\mbox{km s$^{-1}$}}
\newcommand{\kpc}{\mbox{kpc}}
\newcommand{\klambda}{\mbox{k$\lambda$}}
\newcommand{\squarepc}{\mbox{pc$^{2}$}}
\newcommand{\squarekpc}{\mbox{kpc$^{2}$}}

\newcommand{\perbeam}{\mbox{beam$^{-1}$}}
\newcommand{\perpc}{\mbox{pc$^{-1}$}}
\newcommand{\persquarepc}{\mbox{pc$^{-2}$}}
\newcommand{\persquarekpc}{\mbox{kpc$^{-2}$}}
\newcommand{\persquarearcsec}{\mbox{arcsec$^{-2}$}}
\newcommand{\persquarecm}{\mbox{cm$^{-2}$}}
\newcommand{\perkms}{\mbox{(km s$^{-1}$)$^{-1}$}}
\newcommand{\percubiccm}{\mbox{cm$^{-3}$}}
\newcommand{\peryr}{\mbox{yr$^{-1}$}}
\newcommand{\perhertz}{\mbox{Hz$^{-1}$}}
\newcommand{\permag}{\mbox{mag$^{-1}$}}
\newcommand{\unitofx}{\mbox{cm$^{-2}$ (K km s$^{-1}$)$^{-1}$}}

\newcommand{\HH}{\mbox{H$_2$}}
\newcommand{\minus}{\mbox{$-$}}
\newcommand{\xco}{\mbox{$x_{\rm CO}$}}
\newcommand{\xone}{\mbox{$x_{1}$}}
\newcommand{\xtwo}{\mbox{$x_{2}$}}

\newcommand{\twelveCO}{\mbox{$^{12}$CO}}
\newcommand{\thirteenCO}{\mbox{$^{13}$CO}}
\newcommand{\CeighteenO}{\mbox{C$^{18}$O}}

\renewcommand{\tilde}{\mbox{$\sim$}}
\newcommand{\hr}{\mbox{$^{\rm h}$}}
\newcommand{\mn}{\mbox{$^{\rm m}$}}

\newcommand{\twotoone}{\mbox{2--1}}

\newcommand{\uv}{\mbox{$u$--$v$}}
\newcommand{\tm}[1]{\tablenotemark{#1}}
\newcommand{\nd}{\nodata}
\newcommand{\sigmasfr}{\mbox{$\Sigma_{\rm SFR}$}}
\newcommand{\sigmahtwo}{\mbox{$\Sigma_{\rm H_2}$}}
\newcommand{\tdyn}{\mbox{$T_{\rm dyn}$}}

\shorttitle{Starburst Galaxy  NGC 1365} 
\shortauthors{Sakamoto et al. }
\slugcomment{Accepted for publication in ApJ}

\begin{document}
\title{Detection of CO Hotspots Associated with Young Clusters 
in the \\ Southern Starburst Galaxy NGC 1365}
\author{Kazushi Sakamoto\altaffilmark{1,2}, 
Paul T. P. Ho\altaffilmark{3,4}, 
Rui-Qing Mao\altaffilmark{5}, \\
Satoki Matsushita\altaffilmark{4},
and 
Alison B. Peck\altaffilmark{1}}
\altaffiltext{1}{Harvard-Smithsonian Center for Astrophysics,
Submillimeter Array, 645, N. A'ohoku Place, Hilo, HI 96720}
\altaffiltext{2}{National Astronomical Observatory of Japan,
Mitaka, Tokyo 181-8588, Japan. email: sakamoto.kazushi@nao.ac.jp}
\altaffiltext{3}{Harvard-Smithsonian Center for Astrophysics,
60 Garden Street, Cambridge, MA 02138}
\altaffiltext{4}{Academia Sinica, Institute of Astronomy and Astrophysics, 
P.O. Box 23-141, Taipei 106, Taiwan}
\altaffiltext{5}{Purple Mountain Observatory, Chinese Academy of Sciences, 
Nanjing, 210 008, China}

\begin{abstract}
We have used the Submillimeter Array for the first interferometric CO imaging toward the
starburst-Seyfert nucleus of the southern barred spiral galaxy NGC 1365, which
is one of the four galaxies within 30 Mpc that have $L_{8-1000 \; \mu {\rm m}} \geq 10^{11} \Lsol$.
Our  mosaic maps of \twelveCO, \thirteenCO, and \CeighteenO (J=2--1) emission 
at up to 
2\arcsec\ (200 pc) resolutions have revealed 
a circumnuclear gas ring and several CO clumps in the central 3 kpc.
The molecular ring 
shows morphological and kinematical signs of bar-driven gas dynamics,
and the region as a whole is found to follow the star formation laws of Kennicutt.
We have found that some of the gas clumps and peaks in CO brightness temperature, which
we collectively call CO hotspots, coincide with  
the radio and mid-infrared sources previously identified as dust-enshrouded super star clusters.
This hotspot-cluster association suggests that 
either the formation of the most massive clusters took place
in large molecular gas concentrations 
(of $\Sigma_{\rm mol} \sim 10^{3} $ \Msol\ \persquarepc\ in 200 pc scales) 
or  the clusters have heated their ambient gas to cause or enhance the CO hotspots.
The active nucleus is in the region of weak CO emission and is not associated with distinctive
molecular gas properties.
\end{abstract}


\section{Introduction \label{s.intro} }
Central regions of spiral galaxies sometimes host vigorous star formation called (circum)nuclear starbursts.
This phenomenon forms stars at a higher rate and more preferentially in compact massive clusters
than a non-starburst environment \citep[e.g.,][]{Telesco93,Barth95}.
Since stars form from molecular gas, properties of the molecular gas 
are expected to control the starbursts.
The study of such properties in nearby starbursts tells us about
physical processes involved in the phenomenon, and the information
serves as a basis for modeling starbursts throughout galaxy evolution.

The typical extent of a (circum)nuclear starburst in our vicinity is $\lesssim$1 kpc, which is only 20\arcsec\ for
a distance of 10 Mpc.  High angular resolution is therefore essential for starburst studies.
In particular, high resolution imaging of molecular gas with millimeter interferometers has been 
a powerful tool in the field \citep[e.g.,][]{Jogee05}.
Until very recently, however, such arrays had been operated only at the mid-latitudes in the northern
hemisphere, leaving many southern galaxies unattainable.

We have now observed \object{NGC 1365}, a starburst-Seyfert composite galaxy in the southern sky,
using the Submillimeter Array (SMA)\footnote{
The Submillimeter Array is a joint
project between the Smithsonian Astrophysical Observatory and the
Academia Sinica Institute of Astronomy and Astrophysics, and is
funded by the Smithsonian Institution and the Academia Sinica.
}
.
This is part of our SMA survey of southern starbursts, aiming to explore molecular gas and dust in
some of the most important but undeservedly less observed starburst galaxies in the southern sky. 
NGC 1365 is one of our targets 
because it has the circumnuclear starburst typical of barred spirals, 
and because the infrared luminosity of the starburst is among the largest in non-merging galaxies, 
bridging more luminous starbursts in distant mergers and less luminous ones in undisturbed spirals.
Other galaxies reported so far from our survey are \object{M83}, \object{NGC 253}, and \object{NGC 3256} 
\citep{Sakamoto04,Sakamoto06a,Sakamoto06b}.

We have observed CO emission to trace the molecular gas. 
Our resolution of 2\arcsec\ is a significant improvement over the best previous ones of 5\arcsec--10\arcsec.
In addition, the three isotopic lines that were simultaneously imaged, 
i.e., the J=2--1 transitions of \twelveCO, \thirteenCO, and \CeighteenO,
provide information about the physical properties of the molecular gas. 
Among other things, we have discovered bright CO spots, or CO hotspots, that are  
associated with young massive star clusters in the starburst.
We discuss their properties and relation to the starburst.

The goals of this paper are 
(1) to report the detailed structure, kinematics, and properties of molecular gas
in the central region of NGC 1365 on the basis of the new high-resolution data, 
(2) to search for the relation between circumnuclear starburst and the gas properties, 
and
(3) to see if there is anything special in the molecular gas in the vicinity of the Seyfert nucleus.
The next section further introduces NGC 1365, and
\S \ref{s.observations} describes our SMA observations and data reduction.
The observational results are presented in \S \ref{s.results}, and
are discussed in \S \ref{s.bardynamics} --  \S \ref{s.starformation}.
A summary is given in \S \ref{s.summary}.

\section{NGC 1365} \label{s.n1365}
NGC 1365 is an archetype of barred spiral galaxies and hosts a circumnuclear starburst and a Seyfert nucleus.
A comprehensive review of the galaxy is given in \citet{Lindblad99}.
A short summary of the properties of this galaxy is provided below and in Table \ref{t.galparm}.
We adopt a Cepheid-based distance of 17.95 Mpc \citep{Freedman01}.  
At this distance, 1\arcsec\ corresponds to 86 pc and 1 kpc to 11\farcs5.
The galaxy has a long stellar bar, spanning
about 3\arcmin\ on the sky \citep{Regan97} or 18 kpc on the galaxy plane.
The bar has a pair of dust lanes in its front edge, assuming that the spiral arms are trailing in the galaxy.
The assumption makes the northwest side of the galaxy near to us.
The dust lanes curve inward near the galactic center to suggest
an oval ring around the nucleus (see, e.g., the optical images in NED).

Active star formation in the circumnuclear region was first noticed 
as optical hotspots by  \citet{Morgan58} and \citet{Sersic67}.
Later optical imaging using the Hubble Space Telescope (HST) revealed dozens of
bright compact clusters around the dust ring \citep{Kristen97}. 
The brightest of them, after extinction correction, is 300 times brighter than the most luminous 
globular cluster in our Galaxy. 
Mid-IR (10--13 \micron) imaging  by \citet{Galliano05} also revealed several compact
sources, presumably compact clusters, in the region. 
These clusters have been identified as the so-called super star clusters, or
young clusters as massive as $\tilde 10^6$ \Msol\ containing thousands of OB stars.
The infrared luminosity of NGC 1365 is $L_{\rm 8-1000\; \mu m}=10^{11.00}$ \Lsol\ \citep{Sanders03},
which makes it one of only four objects having $\geq 10^{11}$ \Lsol\ 
within 30 Mpc\footnote{
The other three are NGC 1068, a merger NGC 2146, and NGC 7552
at $\delta=-42\arcdeg$.}. 
At 15 \micron, 71\% of the total flux of the galaxy originates from within the central 43\arcsec\ \citep{Roussel01}. 
Thus most of the infrared luminosity probably comes from inside the starburst region of less than
2 kpc radius.

In the radio, where dust obscuration is minimal,  a circumnuclear ring of about $20'' \times 9''$ 
containing several compact ($< 1$\arcsec) peaks have been found at 2--20 cm
\citep{Sandqvist82, Sandqvist95, Saikia94, Forbes98, Morganti99, Thean00}.
Some of the radio and the mid-IR sources coincide.
The brightest radio hotspots have a power of $P_{\rm 6 cm} \tilde 10^{20}$ W \perhertz\ for each.
In other words, they are more than 100 times as luminous as Cas A, are comparable to the 
brightest radio sources in \object{M82} and NGC 253 \citep{Kronberg85,Ulvestad97}, 
and are also comparable in radio power with radio supernovae \citep{Weiler02}.
The center of the galaxy is also bright in molecular lines \citep[e.g.,][]{Sandqvist95,Sandqvist99}.
Previous observations found bar-driven twin peaks of molecular gas 
in the region \citep{Sandqvist99,Ott05}, 
and higher resolution observations fully resolving the gas structure have been needed.

The Seyfert 1.5 nucleus of the galaxy is the driver, or at least the ionizing source, 
of a biconical plasma outflow \citep{Phillips83,Kristen97,Veilleux03},
but the AGN is of low-luminosity and a small contributor to the luminosity of 
the central region \citep{Iyomoto97,Komossa98,Stevens99}.
The  mid- and far-IR spectral indexes of the galaxy,
$\alpha(60,25)=-2.2$ and $\alpha(100,60)=-1.1$ for $S_{\lambda} \propto \lambda^{-\alpha}$
and the IRAS fluxes in \citet{Sanders03},
are not those of `warm' AGN-dominated sources
but those of \ion{H}{2}-type galaxies \citep{deGrijp85,Miley85}. 
The position of the nucleus in Table \ref{t.galparm} is 
the average of optical, near-IR, and radio positions. 
We use it as a coordinate reference for observations without accurate astrometry.

\section{SMA Observations and Data Reduction} \label{s.observations}
NGC 1365 was observed using the Submillimeter Array on the 4100 m summit of
Mauna Kea, Hawaii. 
The array consists of eight 6-meter antennas,  cryogenically cooled 
SIS receivers, and digital correlators.
More information on the array is presented in \citet{Ho04}.
Parameters of our observations are summarized in Table \ref{t.obsparm}.

Two nights of observations were made in the 2003-2004 winter
with the array in its compact and extended configurations.
The SIS receivers were tuned to simultaneously detect three CO lines in the
1 mm band, namely \twelveCO(2--1) in the upper sideband (USB) and 
\thirteenCO(2--1) and \CeighteenO(2--1) in the lower sideband (LSB).
The digital correlators have a 2 GHz bandwidth in each sideband and were
configured for a spectral resolution of 0.8125 MHz.
The southern galaxy was observed between the elevations of 15\arcdeg\ and 34\arcdeg\
under excellent weather with zenith opacity of 0.05--0.08.
The median double-sideband system temperature toward the galaxy
was 210 K and 160 K for each night.
After flagging bad scans, the galaxy data have projected baselines
in the range of 6.8 -- 175.0 m and the total integration time of 6.5 hr.

We made mosaics with a half-power field of view of about $100'' \times 50''$ by observing three positions, 
namely the galactic center and two offset positions 25\arcsec\  from the center 
along the position angle of 70\arcdeg.
They were cycled every 12 minutes to give similar \uv\ coverage for each position.
We also observed quasars J0132\minus169 and J0457\minus234 for gain calibration,
Jupiter, Saturn, Uranus, Callisto, Ganymede, 3C279, and J2258\minus279 for passband calibration, 
and used Mars and Uranus for flux calibration\footnote{
       Mars was not used in the USB because of the significant \twelveCO\ absorption in the
       sideband.}.
A small correction, 7\% in amplitude, is made to our data to compensate for the lack of automatic
focus tracking at the time of the observations \citep{Matsushita06}.

The SMA data were reduced in three steps.
First, the standard passband and gain calibrations were made 
using MIR, which is an IDL version of MMA \citep{Scoville93}.
Then mosaic images were made from the calibrated visibilities using  
MIRIAD \citep{Sault95}. 
Finally, the NRAO AIPS package \citep{Bridle94} was used for data analysis.
In the imaging process, the spectral-line data were binned to 10 \kms\ 
or 20 \kms\  resolutions
and various types of \uv\ weighting were employed.
For one set of images, each line was given a different \uv\ weighting (i.e., {\sf robust} value) in order
to maximize spatial resolution while retaining reasonable signal-to-noise ratio.
For another, we used exactly the same range of \uv\ radii and the
same \uv\ weighting for different lines in order to accurately measure line intensity ratios.
Continuum data were made by combining channels without line emission, and
the total bandwidth of the data was 0.7 GHz in each sideband.
Weak continuum emission was detected at about 5$\sigma$.
The continuum was subtracted from the line data in the \uv\ domain.

\begin{figure}[h]
\epsscale{0.3}
\plotone{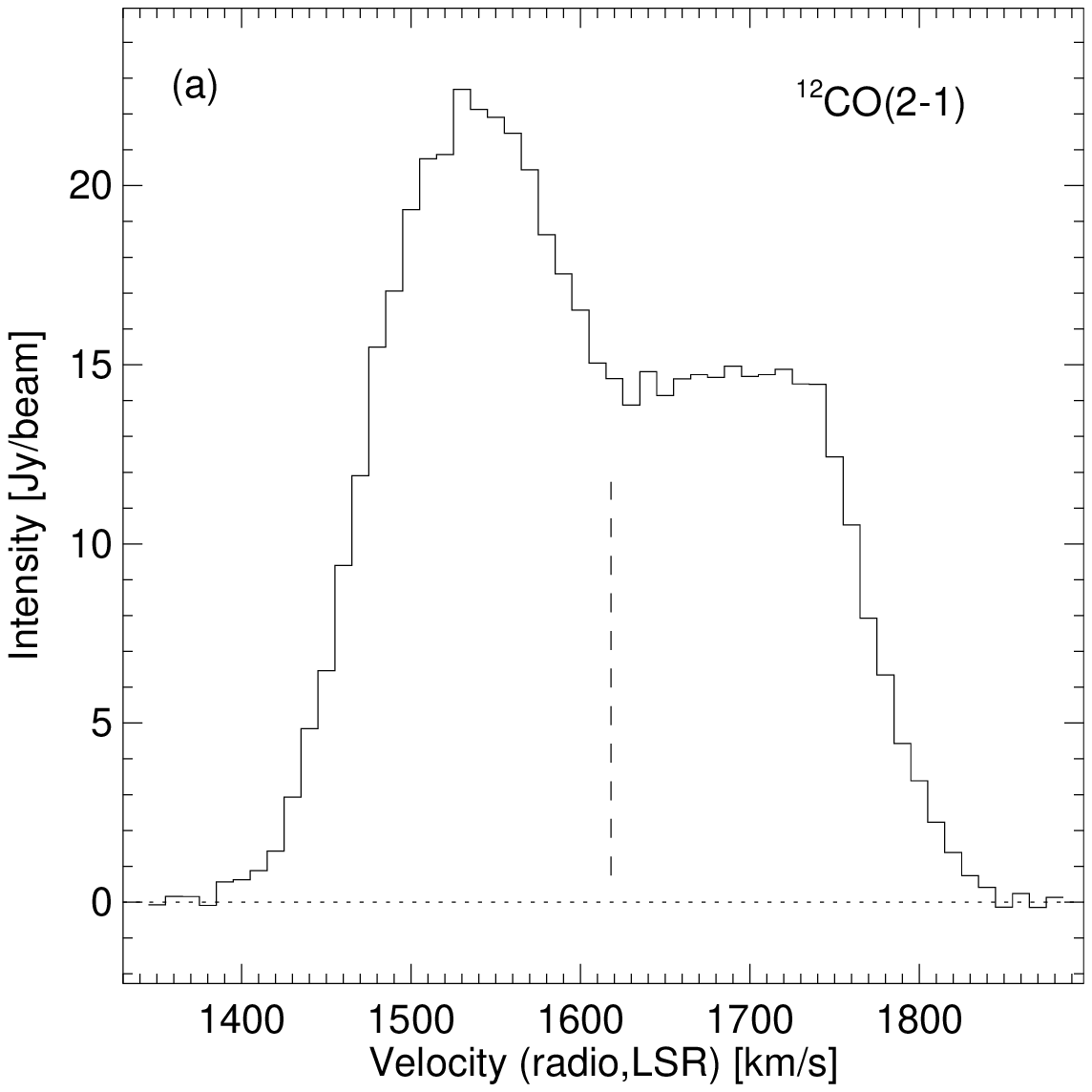} 
\plotone{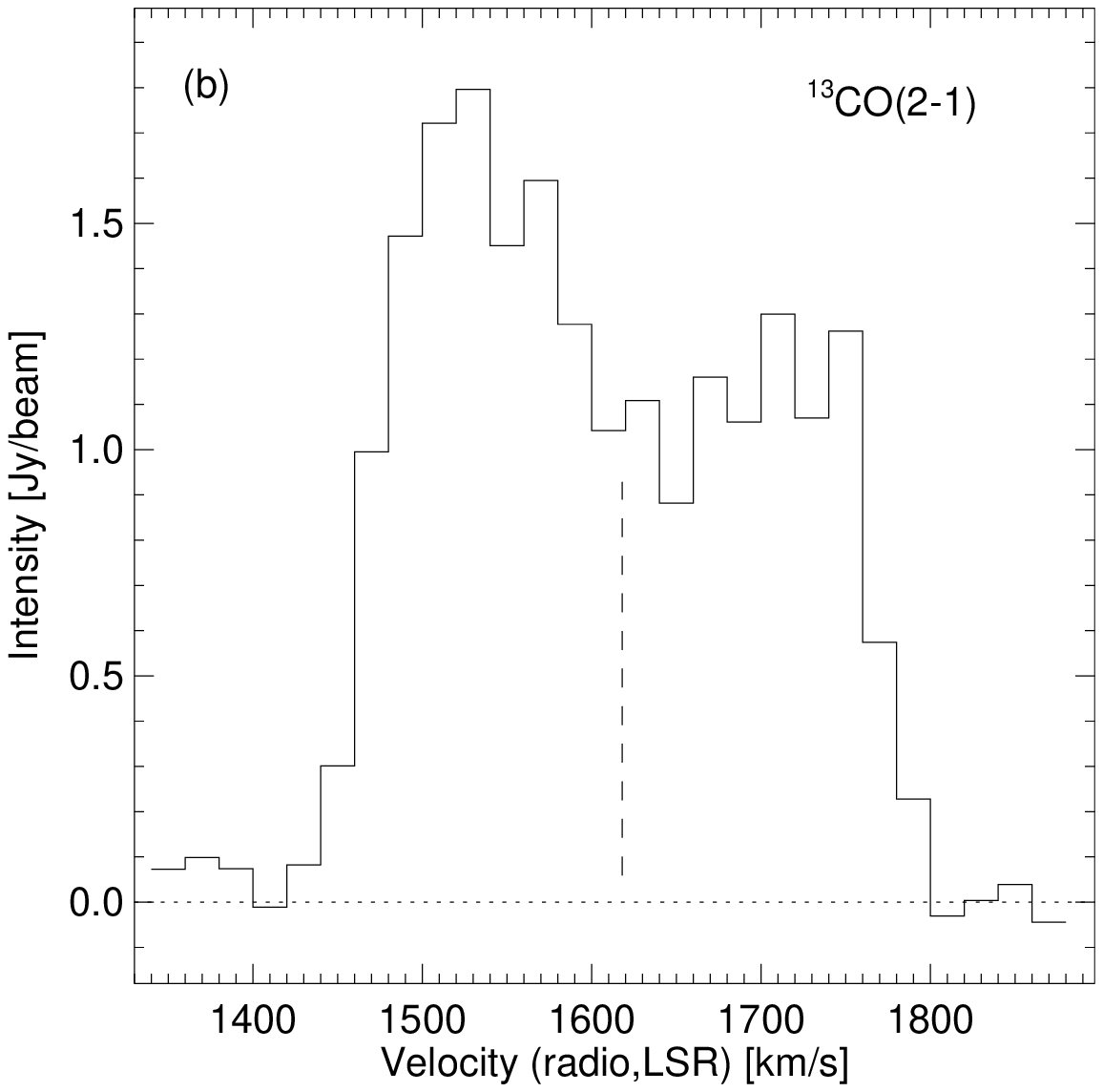}  
\plotone{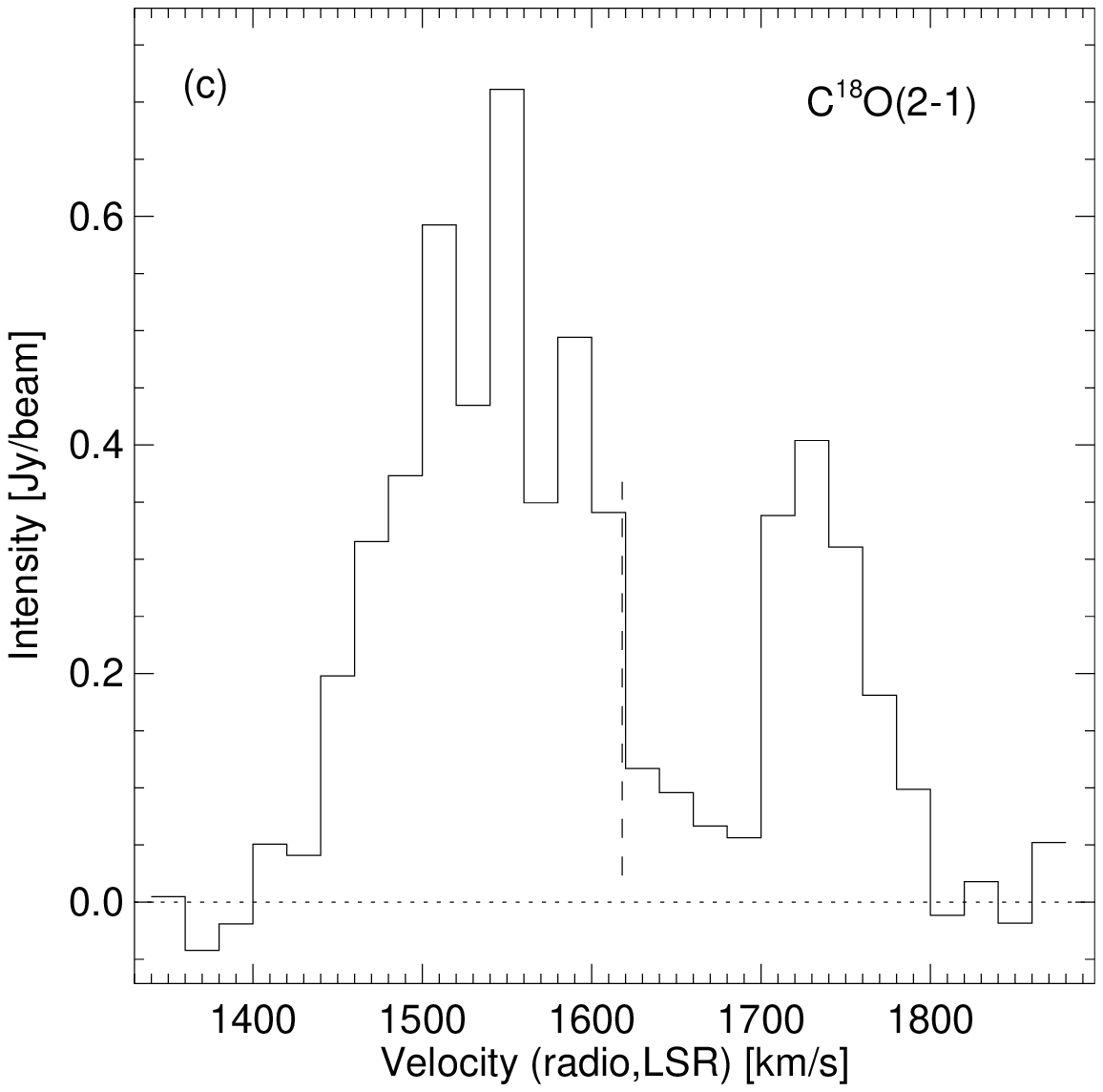} 
\epsscale{1.0}
\caption{Spectra of \twelveCO, \thirteenCO, and \CeighteenO\ (J=2--1) emission in the central 25\arcsec\ of NGC 1365.
For these spectra, the SMA data cubes also used for Fig. \ref{fig.comaps} are first corrected for primary beam attenuation, then convolved to a resolution of 25\arcsec\ (FWHM), and sampled at the galactic center
position of \citet{Sandqvist82}, which was used in many single-dish observations.
The dashed line shows the systemic velocity of the galaxy, 
$V_{\rm sys} {\rm (radio, LSR)} = 1618$ \kms.
 \label{fig.cospectra} }
\end{figure}

We compared our data with previous single-dish observations to check consistency.
Fig. \ref{fig.cospectra} shows the CO spectra to be compared 
with 25\arcsec\ resolution observations at the galactic center. 
They are made from the data cubes used for Fig. \ref{fig.comaps}.
The asymmetry of the line profile seen in all the lines, with stronger blue-shifted emission, is
due to an intrinsic asymmetry of gas distribution.
The total flux in each line is 5.5$\times 10^3$ Jy \kms, 4.1$\times 10^2$ Jy \kms, and
1.1$\times 10^2$ Jy \kms\ for  \twelveCO(2--1),  \thirteenCO(2--1), and  \CeighteenO(2--1), respectively.
\citet{Sandqvist95} observed the same position in \twelveCO(2--1) with a 25\arcsec\ beam
at the Swedish-ESO-Submillimeter Telescope (SEST) and obtained a total flux of 4.9 $\times 10^3$ Jy \kms.
Thus the SMA and SEST observations agree with each other considering that each
probably has an uncertainty of $\sim$10\%.
The SEST line profile has a symmetric double peak unlike SMA's, and 
the small pointing offset implied by this may be a part of the reasons for the smaller SEST flux.
We conclude from the comparison 
that  the SMA observations detected most of the CO flux in the galactic center.
We also infer that  our \thirteenCO(2--1) and \CeighteenO(2--1) data similarly detected most of the flux
in the observed area because they have almost the same \uv\ coverage as the \twelveCO(2--1) data 
and because the \uv\ weighting used for them are closer to the natural weighting.

In this paper, velocities are with respect to the Local Standard of Rest (LSR) 
and are defined in the radio convention. 
Maps are presented without correcting for the attenuation by the mosaicked primary beam,
unless otherwise noted.
However, all the flux and intensity measurements are made after the correction.
Line spectra are also made from data cubes after the correction.

\section{Results } \label{s.results}
\subsection{Gas Distribution \label{s.gas_distribution} }
The CO maps in Figure \ref{fig.comaps} show that the emission is mostly in
the central $4\times2$ kpc of our $9\times 5$ kpc field of view 
and that brighter emission is on an elliptical ring.
There are gas lanes on the leading sides of the bar which
curve inward near the nucleus to form the incomplete circumnuclear ring.
This gas morphology agrees with that of dust lanes seen in the optical (Fig. \ref{fig.cohst}),
except in its western part (i.e., the near side) where only the dark lane is conspicuous.
The oval ring has a major axis of about 2 kpc (\tilde20\arcsec), 
is made of several distinct gas clumps,
and is surrounded by weaker emission.
Particularly in the rare CO isotopes, a pair of condensations stands out 
near the ring's apocenters, where the ring and leading-edge gas lanes connect.
This feature,  resembling the gas `twin-peaks'
observed in the centers of many barred galaxies \citep{Kenney92},
is the one previously observed in CO(3--2) and NH$_3$ \citep{Sandqvist99,Ott05}.
The CO(2--1) emission does not peak at the Seyfert nucleus.
The 1.3 mm continuum is detected at about 5$\sigma$ near the brighter one of the twin peaks.

Radial distribution of the \twelveCO\ integrated intensity, sampled
in concentric circular rings in the galaxy plane and shown in Figure \ref{fig.coradial}, 
confirms the molecular ring and the compact size of the central gas concentration.
The peak at the radius of about 5\arcsec\ is due to the ring smeared
by the circular sampling.
The radial profile rapidly declines outwards, falling to 1/10 of the peak at a radius
of 2 kpc (23\arcsec).
The extent of this molecular gas concentration is comparable to that of
dust emission at 7 and 15 \micron\ and that of ionized gas seen at 3 and 6 cm \citep{Roussel01, Forbes98}.
The warm dust and hot gas  reflect the active star formation in the region.

The mass of molecular gas in the galactic center is estimated using a
conversion factor from \twelveCO\ integrated intensity to \HH\ column density,
$X_{\rm CO}=0.5\times 10^{20} $ \persquarecm\ (K \kms)$^{-1}$.
The adopted coefficient is a middle value of the empirical estimates for the
center of our Galaxy \citep{Sodroski95, Oka98, Strong04}, 
and is about a factor of 3--6 smaller than the ones estimated for the disk of our
Galaxy \citep[e.g.][]{Sanders84, Hunter97, Dame01}.
The tendency for a galactic center, especially a starburst nucleus, to have a lower
conversion factor than the Galactic disk has been suggested by 
a number of authors \citep[e.g.][]{Maloney88, Bryant96, Meier01}.
We also assume that the scaling factor is constant over the region we observed, and
that the \twelveCO(1--0) and \twelveCO(2--1) lines have the same brightness temperature.
Single-dish observations of the 2--1 to 1--0 ratio range from 0.55 to 1.1 for the galactic center
\citep{Sandqvist88,Sandqvist95,Papadopoulos98,Curran01}.
The latter approximation makes $X_{\rm CO(2-1)}=X_{\rm CO(1-0)}$, and hence we
omit the transition index.
For brevity, $X_{\rm CO}$ in units of $10^{20} $ \persquarecm\ (K \kms)$^{-1}$
is written as $x_{\rm CO}$ in the following; we adopted  $x_{\rm CO}=0.5$.
The  \twelveCO(\twotoone) flux within a 1 kpc radius
of the galactic center in the galaxy plane is 4.3$\times 10^3$ Jy \kms.
The mass of molecular gas in the central 2 kpc is thus estimated to be 
$M_{\rm mol}(r_g < 1\; \kpc) = 9 \times 10^{8}$ \Msol\ 
from the above assumptions,  the \twelveCO(\twotoone) flux,
and a factor of 1.36 for the contribution of helium.
Molecular mass as a function of radius is in Figure \ref{fig.coradial}.

An alternative way to estimate the mass of molecular gas uses the 
CO isotope lines that are probably optically thin.
The molecular mass in the central 2 kpc is estimated to be $4\times 10^{8}$ \Msol\ 
from the \CeighteenO(\twotoone) flux of $1.1\times 10^2$ Jy \kms\ in the region.
For this, we used the
\CeighteenO\ abundance of $[\HH/\CeighteenO]=10^{6.3}$, 
i.e., [\HH/\twelveCO] $=10^{4}$ 
and [\twelveCO/\CeighteenO] $= 200$ \citep{Henkel93, Wilson94},
and
$N(\CeighteenO)/I_{\rm C^{18}O(2-1)} \approx 10^{14.6} $ \persquarecm\ (K \kms)$^{-1}$
suitable for warm and moderately dense  gas 
\citep[$T_{\rm kin} > 40$ K, $N_{\rm H_2}=10^3$--$10^{5.5}$ \percubiccm, and 
$\tau(\CeighteenO\, \twotoone) < 0.1$ :][]{Stutzki90, Wild92}. 
This estimate uses the fact that CO molecules can be excited only to a limited number of 
levels for the density and, in that circumstance, the level population becomes independent of
temperature when the gas is warm enough.
If one instead assumes the LTE, 
then the mass is the same for 20 K gas and larger for warmer gas.
The gas mass from \CeighteenO\ is about a factor of 2 smaller than the one from \twelveCO.
This may mean that the adopted $X_{\rm CO}$ is too large. 
However, it is also possible that there is lower density gas between or in the outskirts of 
the molecular clouds, and that \CeighteenO\ can not reliably trace such gas because
of difficulty in excitation or lower abundance of the molecule in the thin gas.
For the Galactic center, \citet{Dahmen98} estimated the abundance of this type of thin, invisible, molecular gas to be
as much as the \CeighteenO-visible gas.
We therefore estimate gas mass from \twelveCO\ and the abovementioned $X_{\rm CO}$ through the paper.
For completeness, the \thirteenCO\ flux of $3.8\times 10^2$ Jy \kms\ in the region
gives a $M_{\rm mol}$ of $3\times 10^{8}$ \Msol\ for 
[\twelveCO/\thirteenCO] = 40.
The line is more prone to saturate than \CeighteenO, and hence the mass is more likely
an underestimate.
The mass-related and other key parameters measured or estimated from our data are summarized in Table \ref{t.parameters}.

The uncertainty in the mass estimates is dominated by that in the CO
flux-to-mass conversion and may be as large as a factor of a few.
It is much larger than the \tilde10\% error in the CO flux.
Though we took the center of our Galaxy as a template for our mass estimate, it is
quite possible that the scaling relation is offset in the starburst nucleus of NGC 1365.
The rare CO isotopes give us an independent check, and the estimate of the error.
Note that the scaling factor can plausibly have local scatter within the galactic center, 
because gas properties are not expected to be uniform throughout the region.
It is only the lack of observational information sufficient to locally determine the gas mass, and the
scaling relation, that led us to use the constant conversion factor.

Bearing the abovementioned caveats in mind, we can calculate the mass and surface density
of molecular gas in small regions.
Figure \ref{fig.coihisto} shows the distribution of CO integrated intensity and
molecular surface density in the circumnuclear region.
The maximum integrated intensity of 42 Jy \kms\ \persquarearcsec\ translates to
the mass surface density of molecular gas 
$\Sigma_{\rm mol} = 9\times 10^2$ \Msol\ \persquarepc\ in the galaxy plane.
The surface density is averaged in our beam having
$5.1\times 10^{4}$ \squarepc\ area in the plane.
The peak is at a compact gas clump 7\arcsec\ north of the nucleus.
The total mass of molecular gas in the clump is \tilde$5\times 10^7$\Msol. 
We see later that the gas clump coincides with a prominent radio and mid-IR source called D/M4.

\subsection{Kinematics  } \label{s.kinematics}
The velocity field shown in Fig. \ref{fig.comaps} indicates the presence of noncircular motion,
most easily seen as the curved ({\sf S}-shaped) velocity contours around the systemic velocity.
The noncircular motion is, however, rather systematic. The contours show a $m=2$ symmetry,
i.e., they can be rotated by 180\arcdeg\ without breaking the overall pattern of the velocity field.

The CO velocity field was fitted with the AIPS task {\sf GAL} to derive kinematical parameters;
integrated intensity was used for data weighting.
The large noncircular motion and the limited spatial coverage of the CO data preclude us 
from determining the inclination and the position angle of the major axis.
Hence we adopt those parameters from \ion{H}{1} observations of the entire galaxy by \citet{Jorsater95}.
We expect, however, that the systemic velocity and the dynamical center of the galaxy are little biased by the
symmetric noncircular motion.
The systemic velocity of the galaxy is estimated as $V_{\rm sys} {\rm (radio, LSR)} = 1618$ \kms\ from the fit,
and the dynamical center of the galaxy is estimated to be
$\alpha =03^{\rm h}33^{\rm m}36\fs35$, 
$\delta  =-36\arcdeg08\arcmin25\farcs8$ (J2000).
The position coincides with the Seyfert nucleus within 0\farcs4, well within the measurement errors of the
dynamical center and the active nucleus.

The major-axis position-velocity (PV) diagram of \twelveCO(2--1) emission in
Figure \ref{fig.majpv} shows that the molecular ring is around the turnover of the rotation velocity
and that no anomalous velocity is observed at the active nucleus.
The rotation velocity at the galactocentric radius of 1 kpc (11\farcs5) is
approximately 150 \kms, according to the PV diagram.
For the adopted inclination of 40\arcdeg, the orbital period at the radius
and the dynamical mass within the radius are 
$T_{\rm dyn}(R_g=1 \mbox{ kpc}) \sim 3\times 10^{7} $ yr 
and
$M_{\rm dyn}(R_g\leq 1 \mbox{ kpc}) \sim 1 \times 10^{10} $ \Msol,
respectively.
Formal errors in these values can not be properly determined without modeling noncircular motion
of gas in the nonaxisymmetric potential. For a $\pm 20$\% uncertainty in the rotation velocity, which
we assume to be reasonable for our data, the fractional uncertainties in $T_{\rm dyn}$ and $M_{\rm dyn}$
are $\pm 20$\% and $\pm 40$\%, respectively.
The gas-to-dynamical mass ratio in the central 2 kpc is $\sim$0.1.

\subsection{Line Ratios}  \label{s.line_ratio} 
The ratio of brightness temperatures is measured 
between  \twelveCO(2--1) and \thirteenCO(2--1) lines
in data cubes of matching spatial and velocity resolutions.
The line ratio should reflect physical parameters of the molecular gas such as
density and column density \citep[e.g.,][]{Meier04}, and the SMA 
enables us to accurately measure it at high spatial resolutions through simultaneous observations
of the CO lines.
Care was taken to make the reduction procedures for the two lines as close as possible.
We used visibilities in the range of \uv\ radii 5--128 \klambda\
where both lines have data.
The same weighting parameter, i.e., a {\sf robust} value of 0.5, and the same velocity binning, 20 \kms,
were used.
The resulting data cubes, which had slightly different resolutions, were convolved to the 
same resolution of 3\farcs3 $\times$ 2\farcs0.
The line brightness temperatures were sampled  in
each velocity channel in a grid of 1\farcs2 spacing,
which is close to the Nyquist frequency for the spatial resolution of the data.
The line ratio is calculated only at the $(\alpha,\delta,v)$ pixels where both lines are detected above $3\sigma$.

Figure \ref{fig.12vs13} shows the scatter diagram of the line intensities 
and the histogram of the \twelveCO(2--1) to \thirteenCO(2--1) intensity ratio, 
$R_{12/13}^{(2-1)} \equiv I(\twelveCO\ \twotoone) /  I(\thirteenCO\ \twotoone)$,
where $I(\mbox{line}) \equiv \int \! T_{\rm b}(\mbox{line})dv$.
The ratio $R_{12/13}^{(2-1)}$ is mostly in the range of 4--14; 92\% of the data are in this range. 
The mean and the standard deviation of $R_{12/13}^{(2-1)}$ are 8.7 and 2.8, respectively.
The former is consistent with the single-dish measurement of $R_{12/13}^{(2-1)}=10\pm2$
in the central 21\arcsec\ of the galaxy \citep{Papadopoulos98}.
The latter is an upper limit of the intrinsic variation of the ratio because of the noise in the data.
Small number of pixels having extreme ratios are likely due to noise.
In particular, pixels having $R_{12/13}^{(2-1)} \leq 2 $, which are less than 2\% of the data, 
are almost certainly noise judging from their mostly isolated locations in the channel maps.

\begin{figure}[t]
\epsscale{1.1}
\plottwo{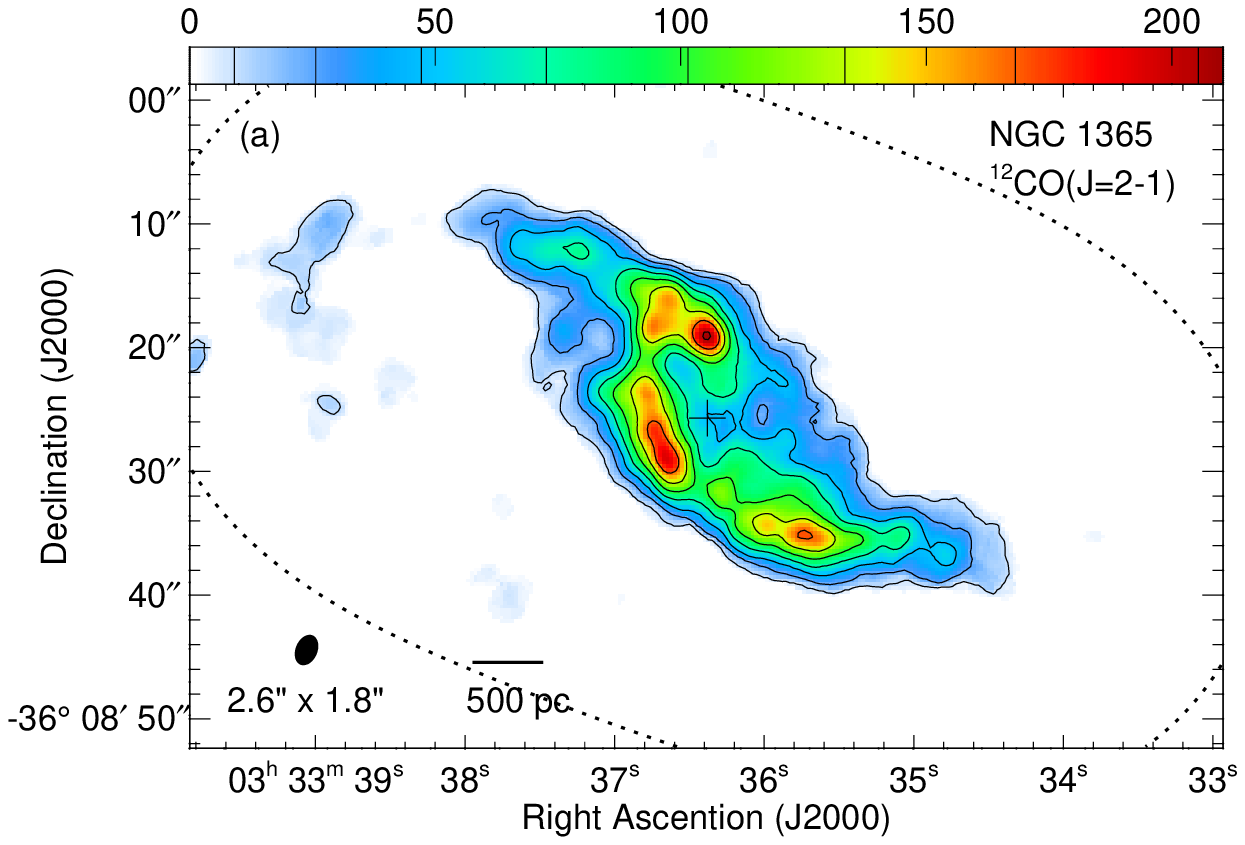}{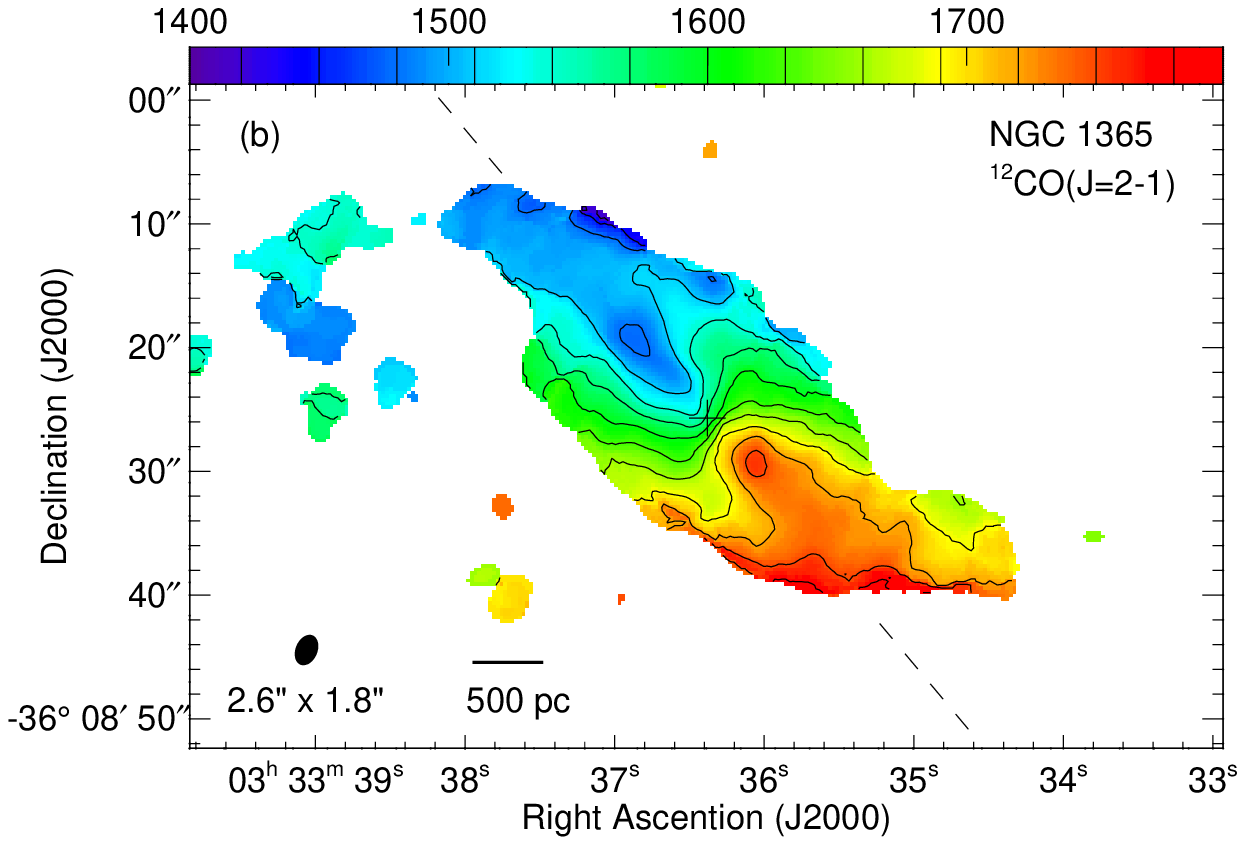}
\plottwo{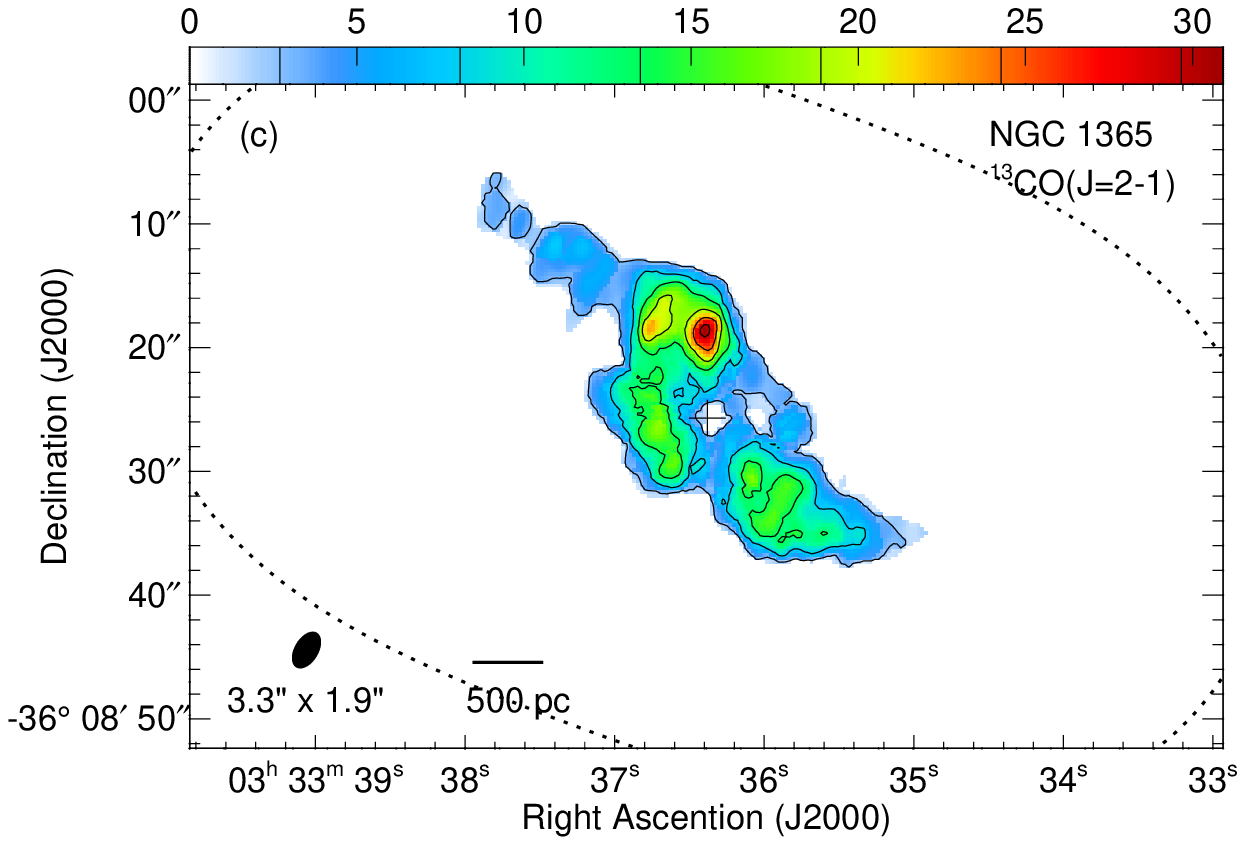}{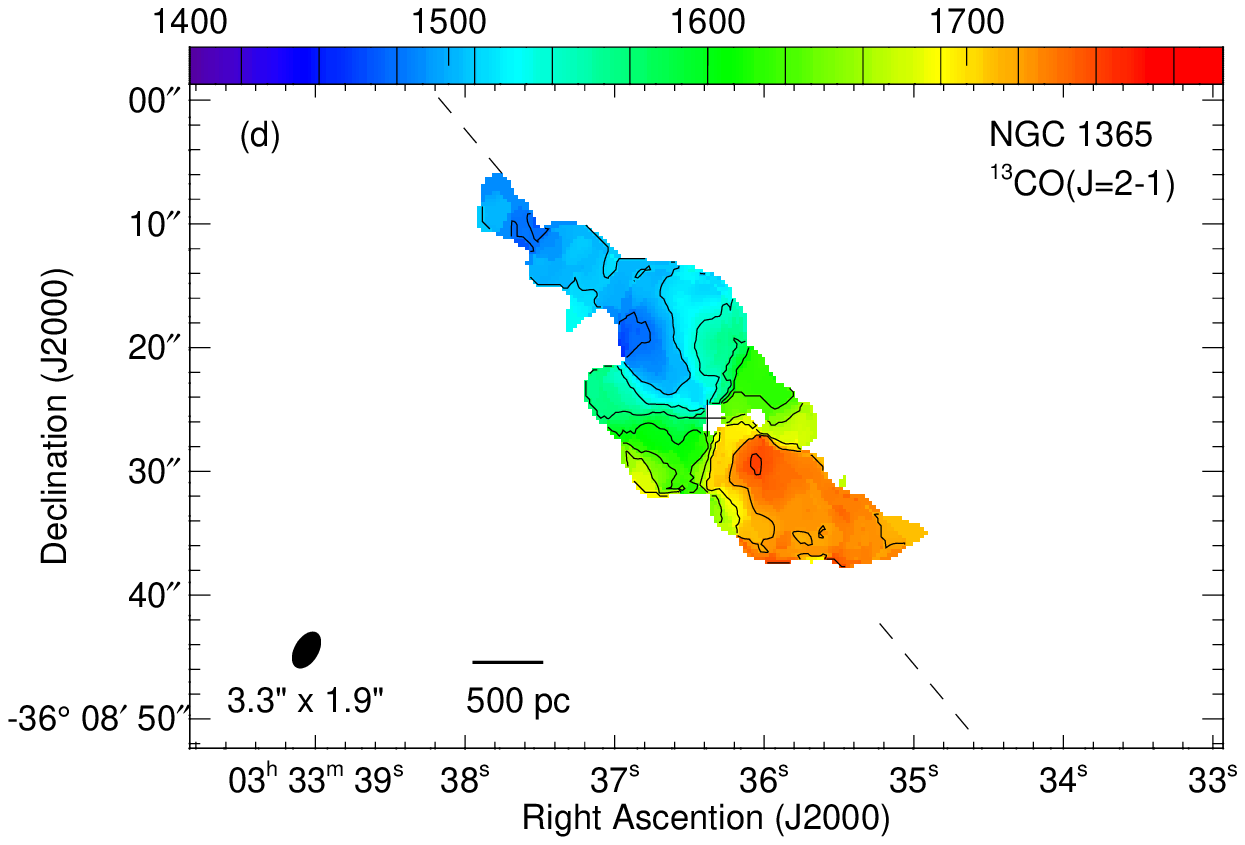} 
\plottwo{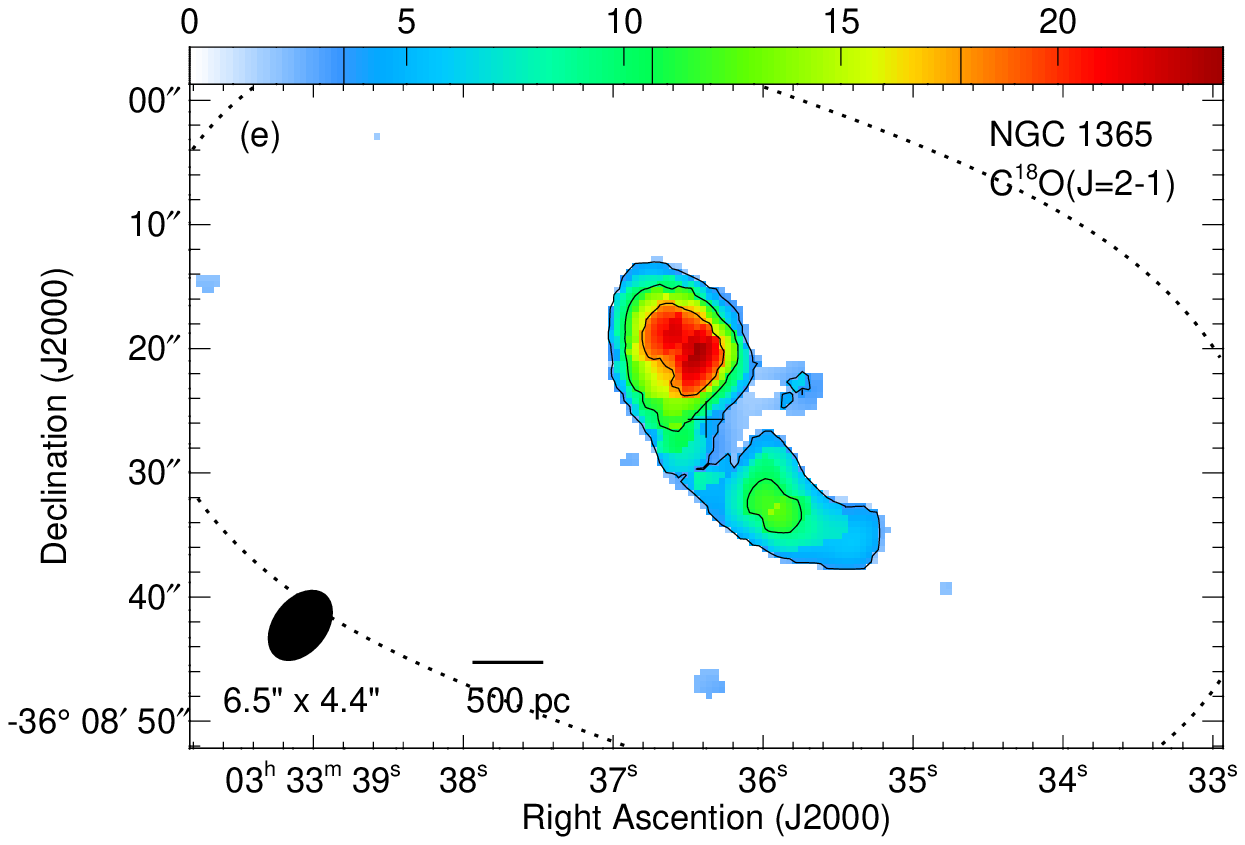}{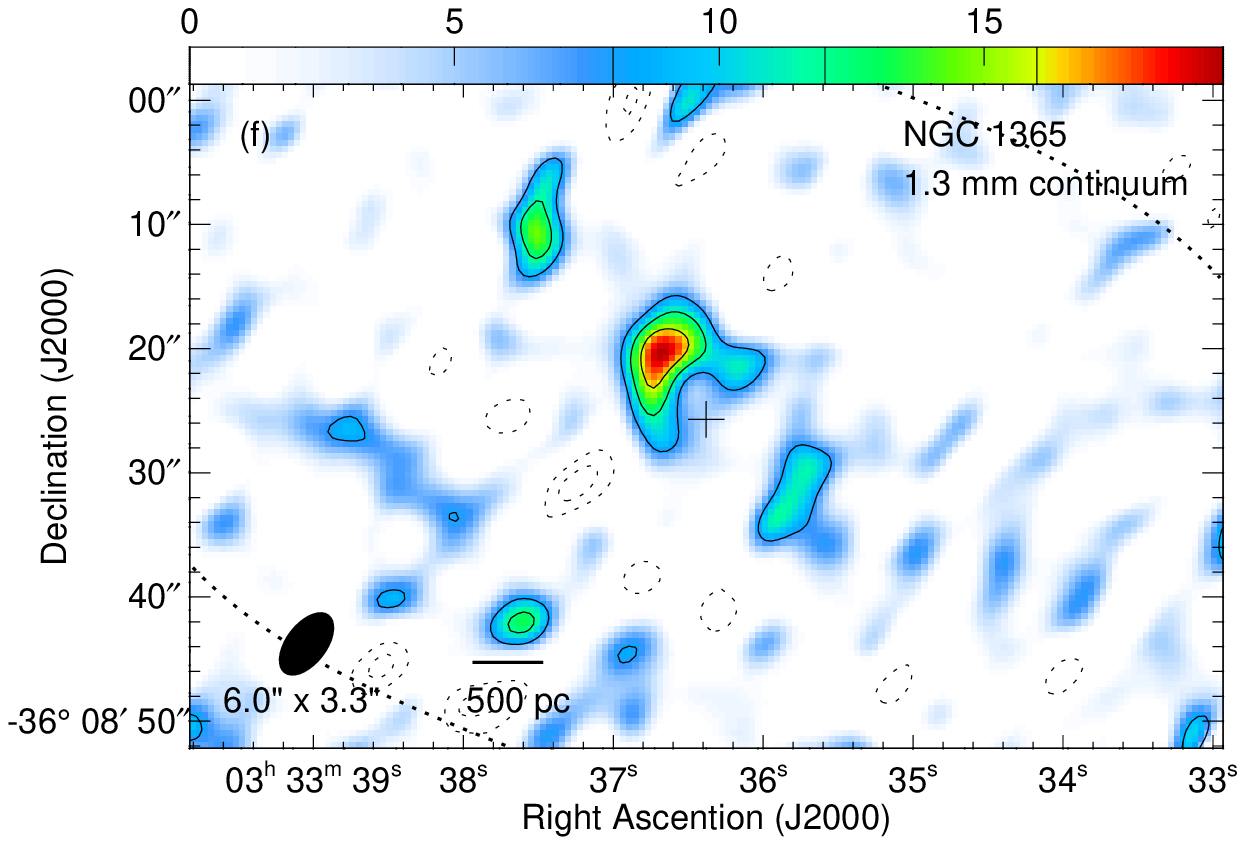} 
\caption{\small
CO line and continuum images of the center of NGC 1365.
(a)  \twelveCO(J=2--1) integrated intensity map. The $n$-th contour is at 9.1$\times n^{1.5}$ Jy \perbeam\ \kms.
(b) Mean velocity map of the \twelveCO(J=2--1) emission. Contours are in 30 \kms\ steps. 
(c) \thirteenCO(J=2--1) integrated intensity map. Contours are at 2.7$\times [1,3,5,7,9,11]$ Jy \perbeam\ \kms.
(d) Mean velocity map of the \thirteenCO(J=2--1) emission.
(e) \CeighteenO(J=2--1) integrated intensity map. Contours are at 3.6$\times [1,3,5]$ Jy \perbeam\ \kms.
(f) Continuum map at the effective frequency of 224 GHz, made by averaging the LSB and USB maps.
Contours are at $[-3,-2,2,3,4] \times 4.0$ mJy \perbeam\ $(=1\sigma)$.
Negative contours are dashed.
The cross at the center of each image marks the active nucleus.
The FWHM of the synthesized beam and a 500 pc scale are at the bottom left corner of each image.
The dotted oval shows the mosaicked primary beam at its 50\% level. 
The dashed line in the velocity maps is the line of nodes (P.A. $= 220$\degr).
\label{fig.comaps} }
\end{figure}

\begin{figure}[h]
\epsscale{1.0}
\plotone{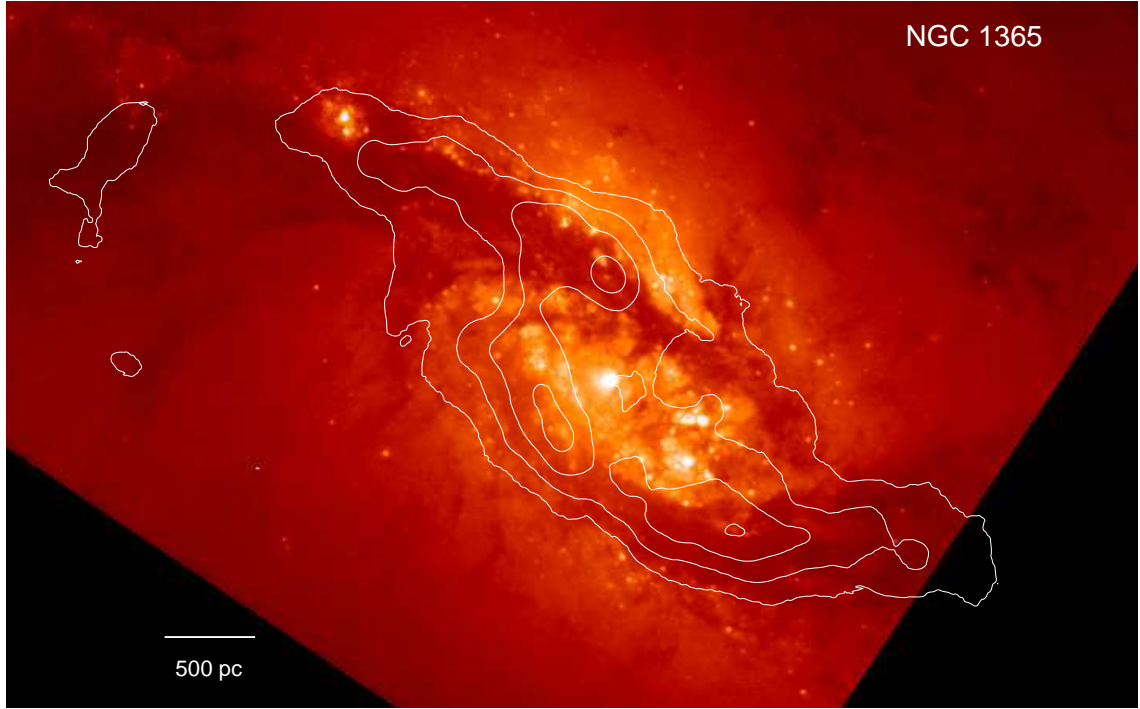}  
\epsscale{1.0}
\caption{\twelveCO\ contours on an archival HST image taken with the F606W filter.
Every 2nd contour in Fig. \ref{fig.comaps} (a) is plotted on
the logarithmically-scaled optical image.
North is up, east is to the left.
The HST image is shifted east by $1\farcs0$ from its header position to place the nucleus on its radio
position.
 \label{fig.cohst} }
\end{figure}

\begin{figure}[h]
\epsscale{1.0}
\plottwo{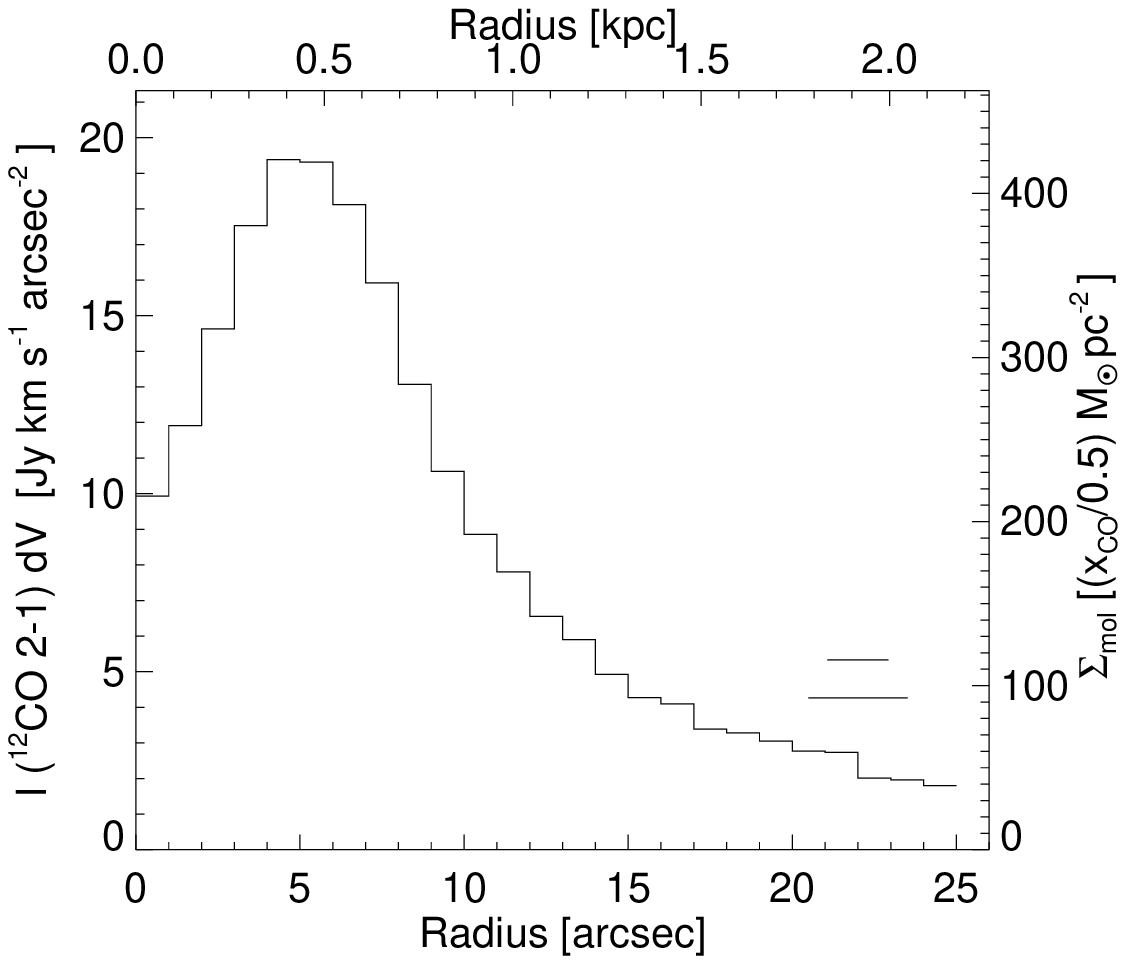}{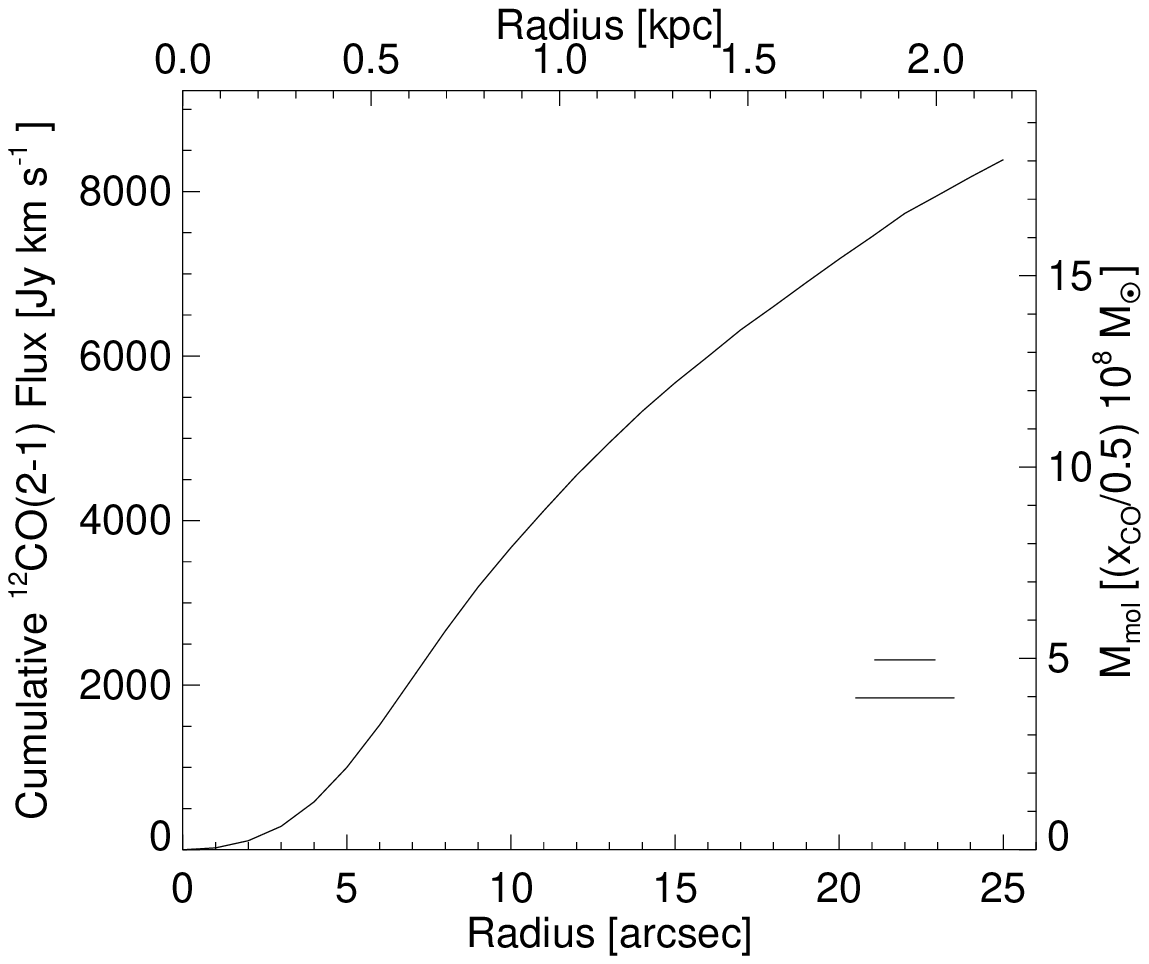}
\epsscale{1.0}
\caption{Radial distribution of \twelveCO(2--1) emission around the galactic center
measured in concentric circular annuli or circles in the galaxy plane.
(left) 
The mean integrated intensities on the sky plane.
The right axis is molecular gas surface density corrected for the inclination of the galaxy.
(right) 
The cumulative flux of \twelveCO(2--1) emission as a function of galactocentric radius.
The right axis is the mass of molecular gas in units of $10^{8} \Msol$.
Both plots are from the dataset of $2\farcs6 \times 1\farcs8$ resolution.
See \S \ref{s.gas_distribution} for the conversion of \twelveCO\ data to gas surface density and gas mass.
The two horizontal bars in each plot show FWHM of the elliptical beam along the galaxy's major axis (upper bar)
and the minor axis (lower bar). The lengths of both bars are calculated in the galaxy plane.
 \label{fig.coradial} }
\end{figure}

\begin{figure}[h]
\epsscale{0.5}
\plotone{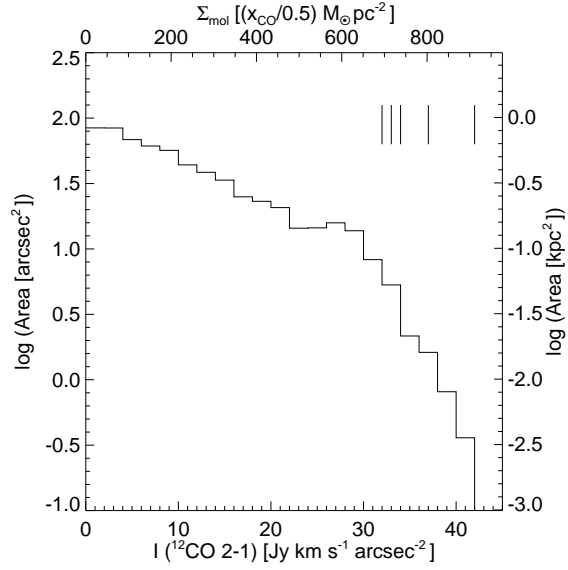} 
\epsscale{1.0}
\caption{Histogram of \twelveCO(2--1) integrated intensity within 2.5 kpc of the nucleus in the galaxy plane.
The vertical lines in the upper right mark the data for the CO hotspots in Table \ref{t.clumps}.
The moment map with $2\farcs6 \times 1\farcs8$ resolution was used for this plot after
corrected for the mosaicked primary beam. 
The area covered by the synthesized beam is $5.1 \mbox{ arcsec}^2$
on the sky and $5.1 \times 10^{-2}$ \squarekpc\ on the galaxy plane.
The top axis is the inclination-corrected surface density of molecular gas ($\Sigma_{\rm mol}$) in units of \Msol \persquarepc.
See \S \ref{s.gas_distribution} for the mass estimate from CO flux.
The left axis is the surface area in the galaxy plane.
Multiply the bottom axis by 1.3 to obtain the total visual extinction in our line of sight in mag 
for $N(\HH)/A_{V} = 1 \times 10^{21}$ \persquarecm\ \permag.
 \label{fig.coihisto} }
\end{figure}

\begin{figure}[h]
\epsscale{0.6}
\plotone{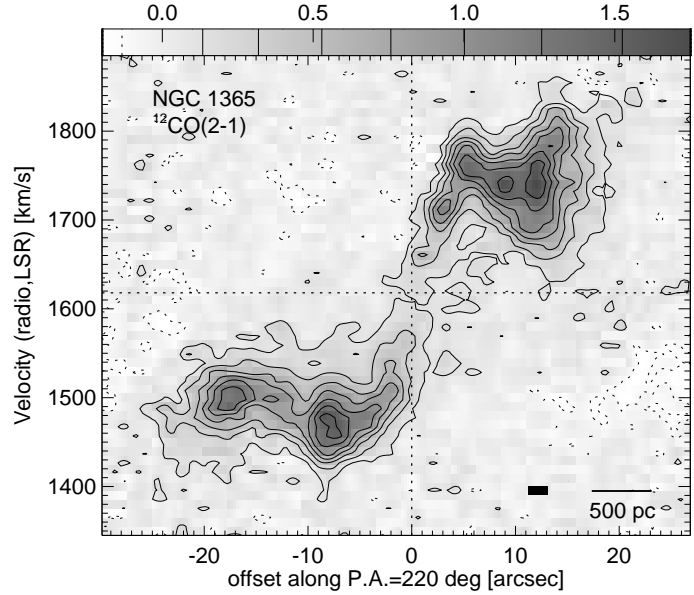} 
\epsscale{1.0}
\caption{Position-velocity diagram of \twelveCO(2--1) emission along the line of nodes.
The $n$-th solid contour is at $n^{1.25} \times 134$ mJy \perbeam; the lowest contour is at 2$\sigma$.
Negative contours are dashed and have the same absolute values as positive ones.
Conversion to brightness temperature in K can be made by multiplying by 5.2.
The diagram is corrected for the mosaicked primary beam.
The abscissa is measured from the active nucleus.
The horizontal dotted line indicates the systemic velocity of 1618 \kms.
The black rectangle at the bottom right corner represents a resolution element.
\label{fig.majpv}
}
\end{figure}

\begin{figure}[h]
\epsscale{1.0}
\plottwo{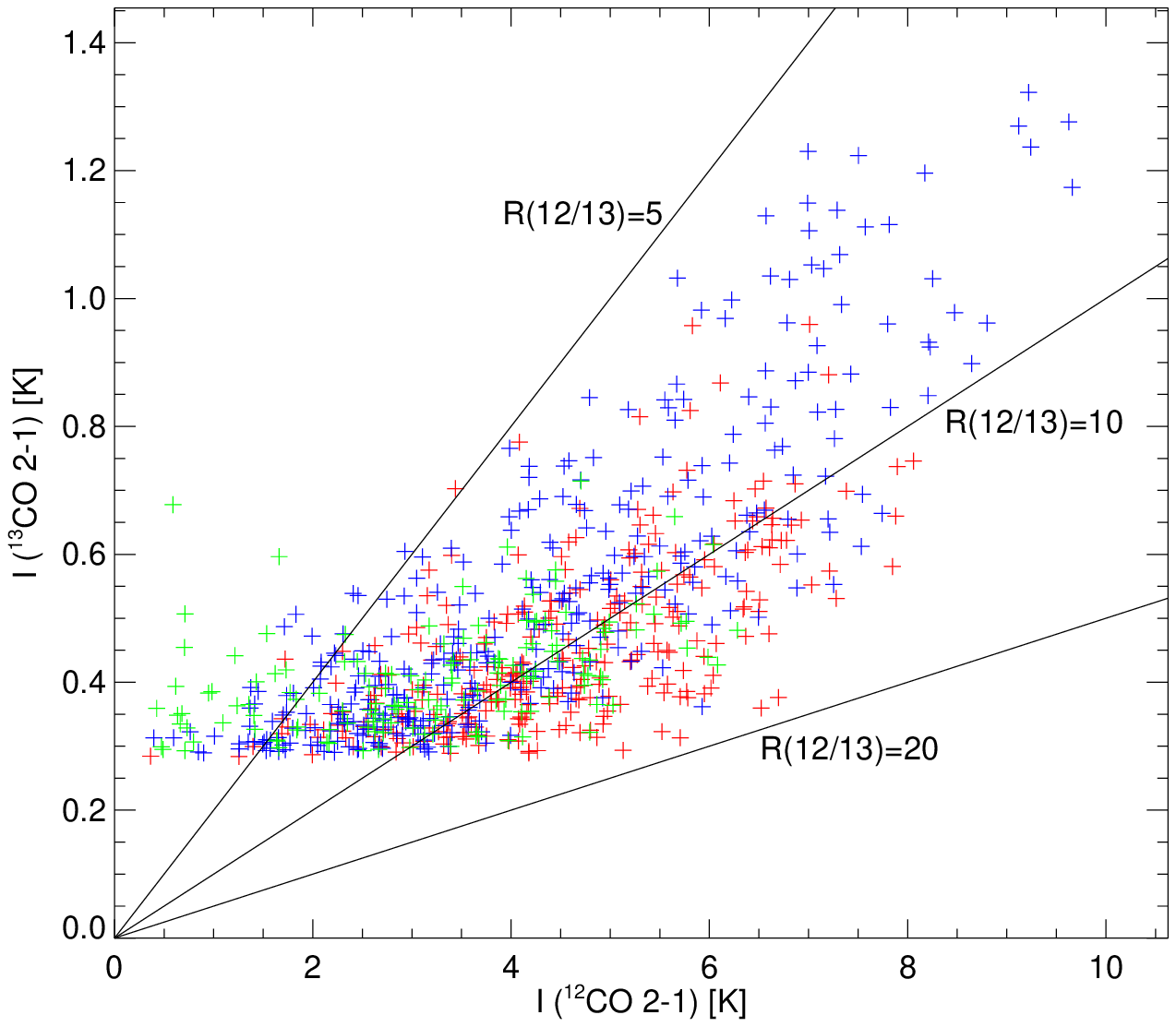}{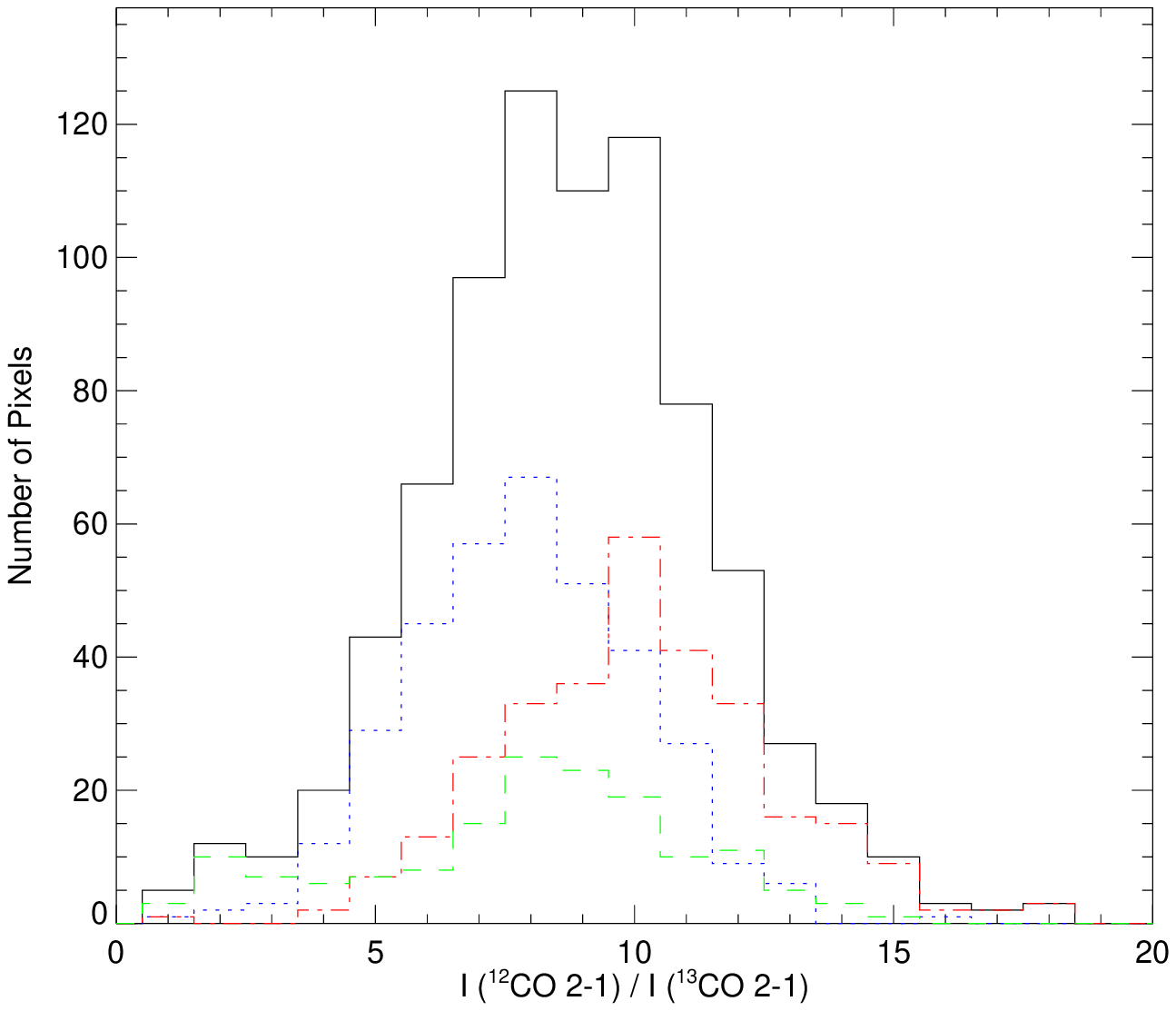}
\epsscale{1.0}
\caption{(left) Comparison of \twelveCO(2--1) and \thirteenCO(2--1) line intensities  in the center of
NGC 1365. 
The line intensities are sampled from data cubes of matching spatial and velocity resolutions, 
3\farcs3 $\times$ 2\farcs0 and 20 \kms.
The sampling spacing is 1\farcs2. 
Only those points where both lines are detected above 3$\sigma$ are plotted.
The $1\sigma$ noise is 0.12 K and 0.09 K for  \twelveCO(2--1) and \thirteenCO(2--1), respectively, 
at the center of the mosaics. 
The noise varies across the maps corrected for primary beam, and so does the
cutoff threshold in kelvin.
(right) Histogram of  \twelveCO(2--1) to \thirteenCO(2--1) intensity ratio.
The one in black is from the entire data. 
The blue and red ones are from the near and far side, respectively, and within 
the galactocentric radius of 1 kpc (= 12\arcsec).
Data from outside the circumnuclear disk are plotted in green.
\label{fig.12vs13}
}
\end{figure}

\begin{figure}[h]
\epsscale{0.7}
\plotone{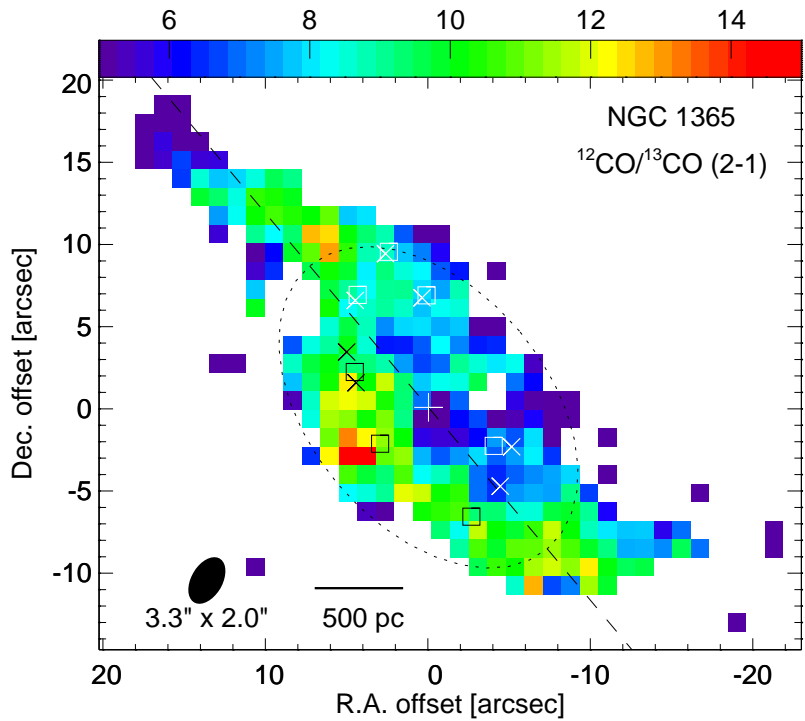} 
\epsscale{1.0}
\caption{Spatial distribution of  \twelveCO(2--1) to \thirteenCO(2--1) brightness temperature ratio. 
Ratios of different velocities are averaged at each position.
The $+$ sign marks the active nucleus and the dashed line is the line of nodes; the near side is to the northwest.
The dotted ellipse is at the galactocentric radius of 1 kpc (= 12\arcsec), and
the boxes and crosses mark the radio and mid-IR sources, respectively, in Fig. \ref{fig.guidemap}.
\label{fig.ratiomap}
}
\end{figure}

The spatial distribution of the ratio $R_{12/13}^{(2-1)}$ is shown in Figure \ref{fig.ratiomap}.
For this map, line ratios are averaged at each position if there are
multiple velocity channels with a valid ratio.
The overall trend in the map is that the ratio is relatively high at the ridges
of molecular gas. 
Specifically, higher ratios are seen 
on the southern molecular ridge at the southern dust lane in the bar,
on the southeastern half of the molecular ring,
and
on the northern molecular ridge in the bar.
Lower ratios are mainly observed in the northwestern half of the circumnuclear ring and
inside the ring.
The \twelveCO(2--1) to \CeighteenO(2--1) ratio has the same trend, as seen
from the comparison of Figures \ref{fig.comaps} (a) and  \ref{fig.comaps} (e).
Another way to describe the spatial distribution of the line ratio is that 
there is a trend of the ratio being higher
on the far (SE) side than the near (NW) side in the circumnuclear disk of 1 kpc radius
(see Figs. \ref{fig.12vs13} and \ref{fig.ratiomap}).
The mean and standard deviation of  $R_{12/13}^{(2-1)}$  are
(mean $\pm\; \sigma) = 7.9 \pm 2.2$ for the near side and $10.0 \pm 2.6$ for the far side.
A similar trend of higher $R_{12/13}$  in the far side has been noticed in the circumnuclear 
regions of IC 342 and NGC 253  \citep{Meier01, Sakamoto06a}, but the illumination 
model proposed for them may not apply to NGC 1365.
This is because the starburst ring coincident with the molecular ring can not
preferentially illuminate the side of clouds facing us, 
nor is there a radial gradient of $R_{12/13}$ expected from AGN illumination.
We further discuss the ratio in \S\ref{s.gas_properties}.

\subsection{Comparison of CO with Radio and mid-IR sources} \label{s.comparison.hotspot_ssc}
\begin{figure}[h]
\epsscale{0.4}
\plotone{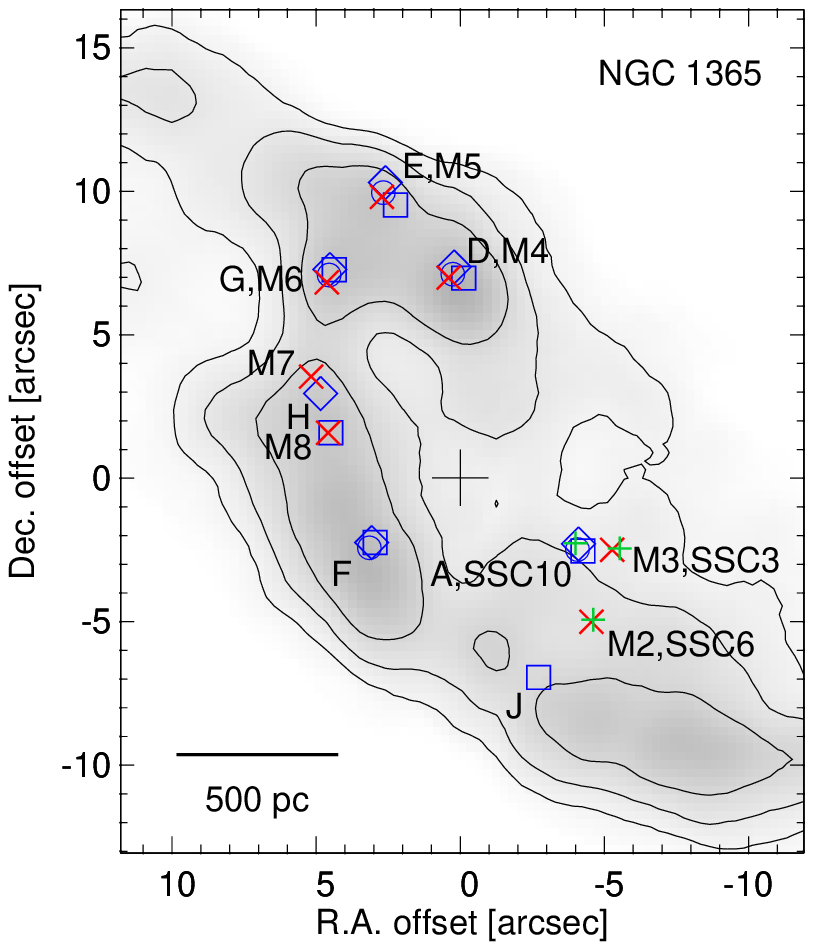}  
\epsscale{1.0}
\caption{A guide map of radio, mid-IR, and optical sources in the center of NGC 1365.
Sources A, D, E, F, G, H, and J are radio hotspots.
Their positions are shown as 
diamonds \citep[$\lambda_{\rm obs}=2, 6$ cm]{Sandqvist95}, 
squares \citep[$\lambda_{\rm obs}=3.5, 6$ cm]{Stevens99}, 
and circles \citep[$\lambda_{\rm obs}=3.5$ cm]{Thean00}.
Sources M2 through M8 shown as {\sf X} symbols are mid-IR sources of \citet{Galliano05}.
Super star clusters (SSCs) 3, 6, and 10,  shown with plus signs, are optically identified by \citet{Kristen97}.
SSC 3 and SSC 6 are the two brightest ($M_{\rm B}= -16.6$ mag) clusters in the region.
Positions are measured with respect to the Seyfert nucleus; see Table \ref{t.galparm}  for its absolute coordinates.
The background is our map of \twelveCO(2--1) integrated intensity to guide the eyes.
 \label{fig.guidemap} }
\end{figure}

We show in Figure \ref{fig.guidemap} the locations and conventional names of 
the compact radio sources, luminous mid-IR sources, and three of the optically identified super clusters
in the circumnuclear region.
The radio sources have alphabet names, and the
mid-IR sources are called  M\#, where \# is a number. 
As seen in the figure, some of the radio and mid-IR sources coincide with each other, 
and are thought to be the same objects, 
likely young massive clusters embedded in dust \citep{Sandqvist95,Galliano05}.
In contrast, the mid-IR sources M2 and M3 do not have a radio counterpart but coincide with 
optically identified clusters called SSC 3 and SSC 6, respectively, which
\citet{Kristen97} found to be the brightest  optical sources  in the region except the nucleus.
Their mid-IR colors suggest less dust emission than toward other sources \citep{Galliano05},
consistent with their optical detection.

\begin{figure}[t]
\epsscale{1.1}
\plottwo{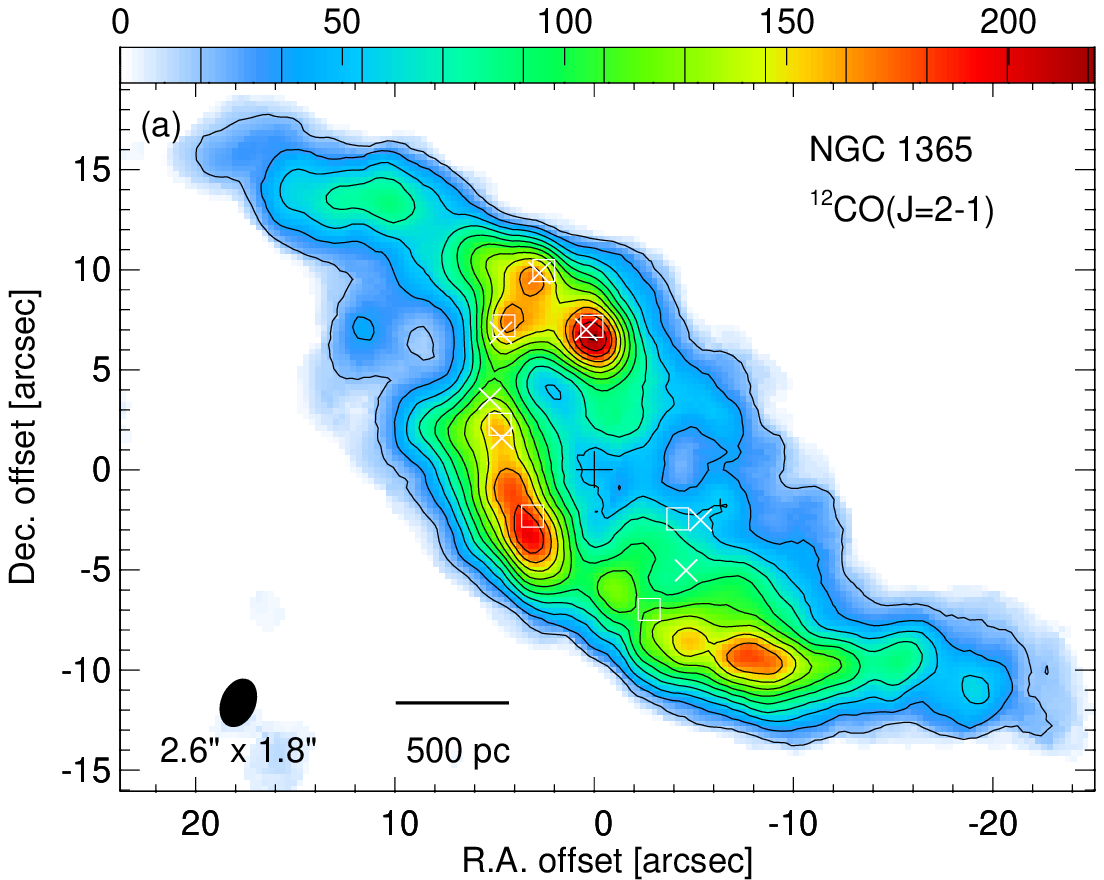}{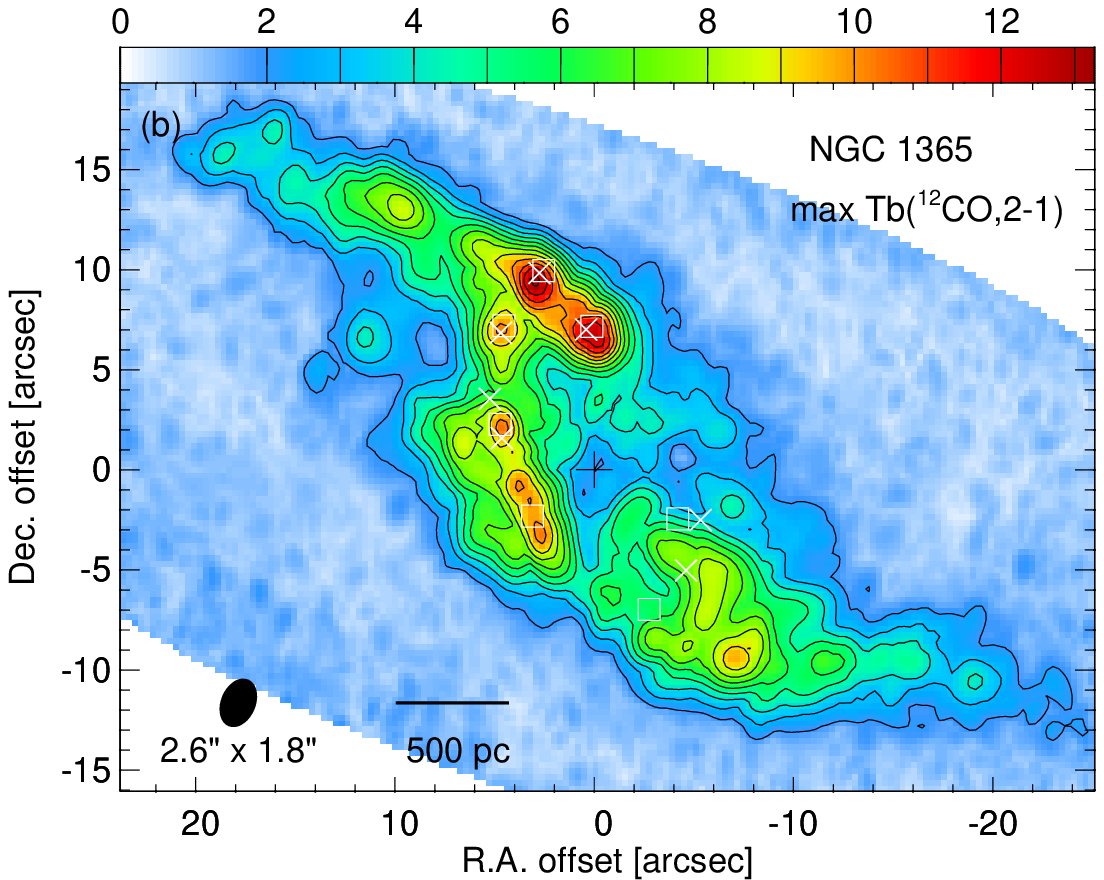} \\
\plottwo{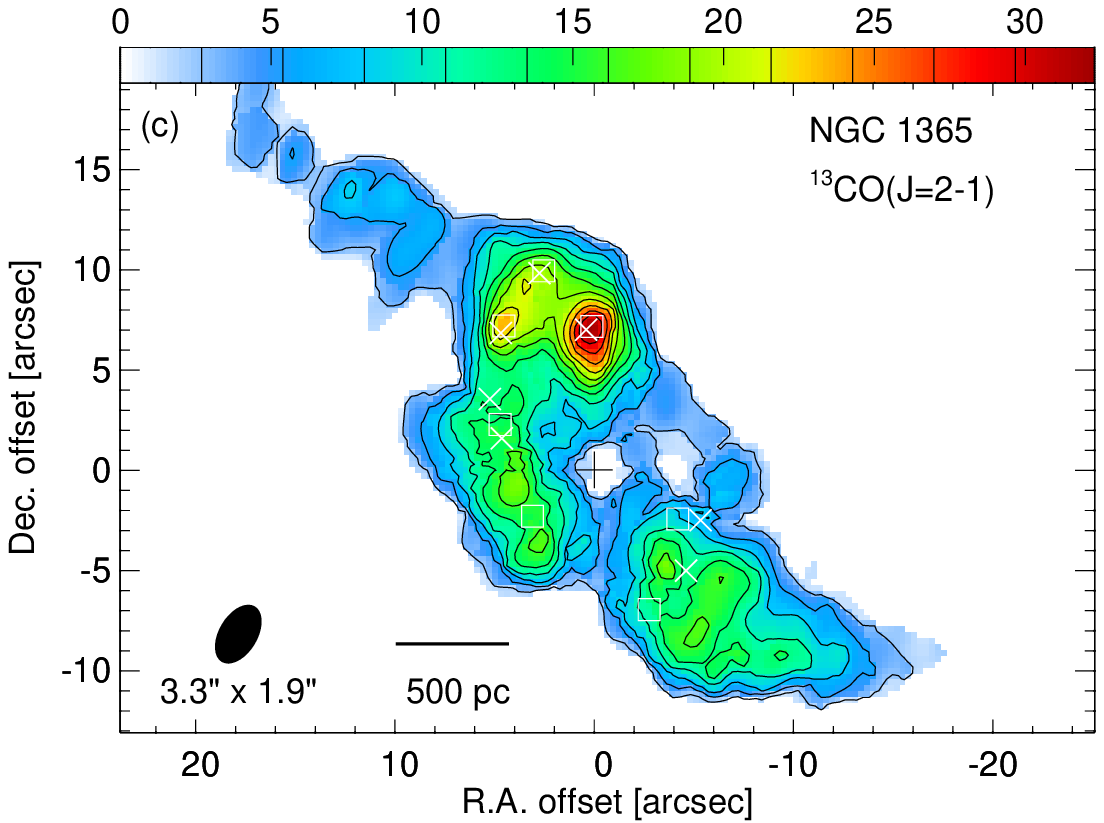}{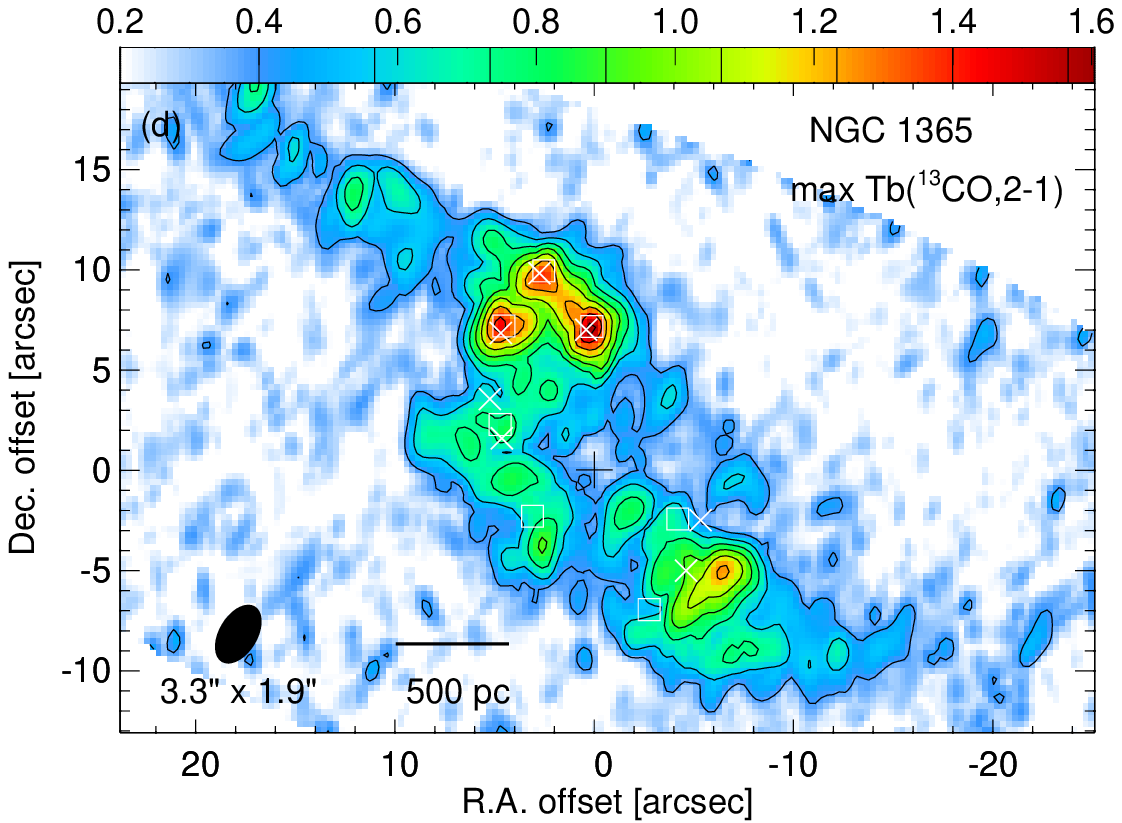} \\
\plottwo{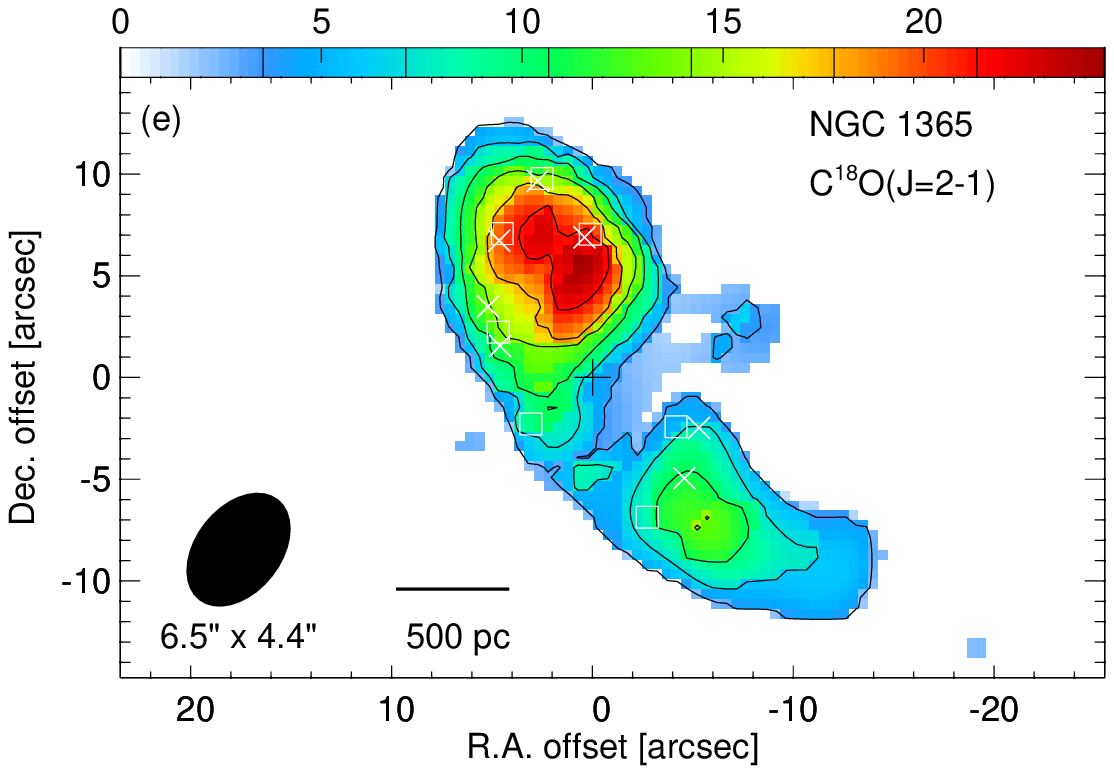}{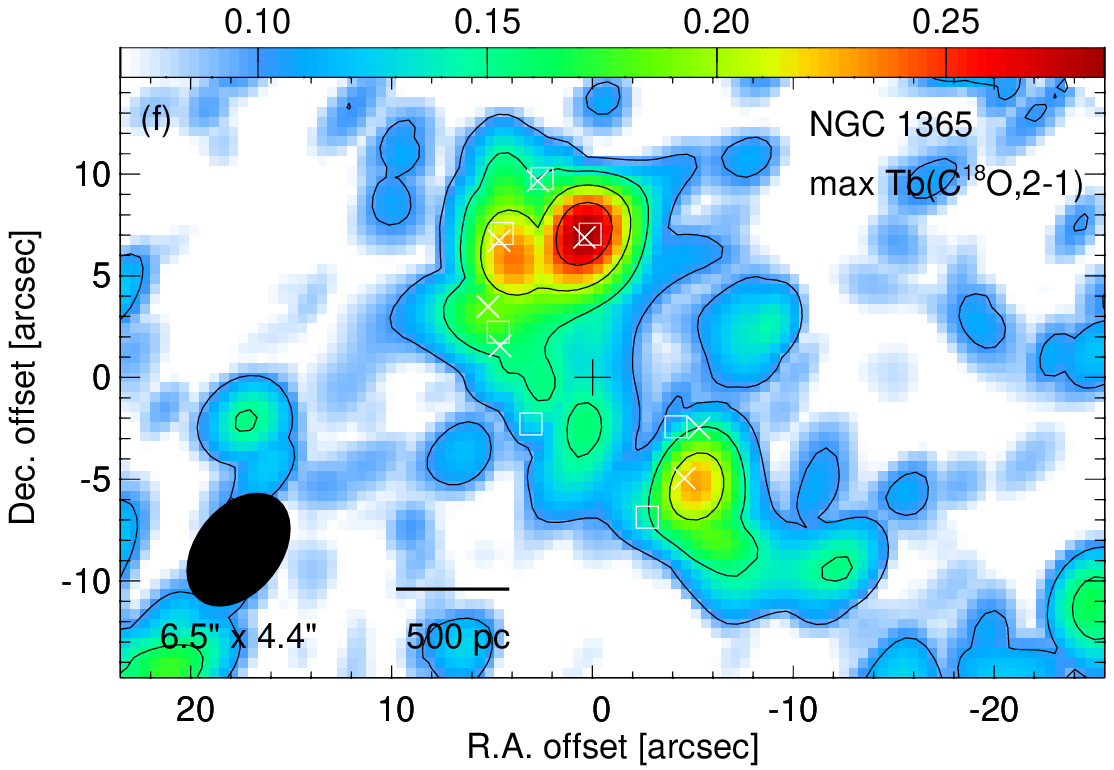} 
\caption{Radio and mid-IR sources plotted on the maps of CO integrated intensity (left)
and peak brightness temperature (right). 
Squares are seven radio peaks observed in centimeter wavelengths \citep{Stevens99, Sandqvist95, Thean00}
and crosses are mid-IR sources observed by \citet{Galliano05}.
The background images are integrated intensities in Jy \perbeam\ \kms\ for the left column and 
brightness temperatures in K for the right.
Each map is corrected for the primary beam attenuation. 
The $+$ sign at the center of each image marks the galaxy nucleus.
The synthesized beam and a 500 pc scale are at the bottom left corner of each panel. 
\label{fig.co_ssc} }
\end{figure}

Our CO data are compared with the radio and the mid-IR sources in Figure \ref{fig.co_ssc}.
We have produced for the comparison peak brightness temperature maps 
in addition to the CO integrated intensity maps.
Since the peaks in these maps often coincide, we collectively call the peaks 
CO hotspots.
In the maps, radio sources are plotted as squares at average positions of previous observations\footnote{
	The radio source H was given positions 1\farcs4 apart by \citet{Sandqvist95} and \citet{Stevens99}.
	We use the average position and associate the source with M8, which is 0\farcs7 from the radio position.
	There remains a possibility, however, 
	that there are two radio sources and M7 and M8 are associated with each of them.
},
while mid-IR sources are marked as crosses.

The comparison shows that some of the radio peaks and mid-IR sources coincide
with CO hotspots, i.e., peaks of CO integrated intensity or CO brightness temperature or both. 
Specifically, the sources D/M4, E/M5, G/M6, and H have a local maximum of \twelveCO\ integrated intensity
within 1\arcsec. 
A local peak in \thirteenCO\ integrated intensity is also seen within 1\arcsec\ of D, E, and G.
Each of the four sources also coincides with a local maximum 
of peak brightness temperature in \twelveCO\ and \thirteenCO,
and the source D is at the maximum of the \CeighteenO\ brightness temperature map.
Though not coincident with an isolated peak,
the radio source F is within 1\arcsec\ of a ridge of \twelveCO\ integrated intensity, and is on a sharp
ridge of \twelveCO\  peak brightness temperature.
In contrast, the sources M2/SSC6 and M3/SSC3 that are thought to be less embedded in dust are
indeed in the areas of less molecular gas.
The radio sources A and J do not have a nearby peak in the CO maps.
Table \ref{t.co_others} summarizes the associations of peaks in CO maps 
to the sources in other wavelengths. 
In general, the correlation to the sources appears better in peak CO temperature than in 
the line integrated intensity.
The weak 1.3 mm continuum does not coincide with any of the radio and mid-IR sources.

\begin{figure}[h]
\epsscale{0.18}
\plotone{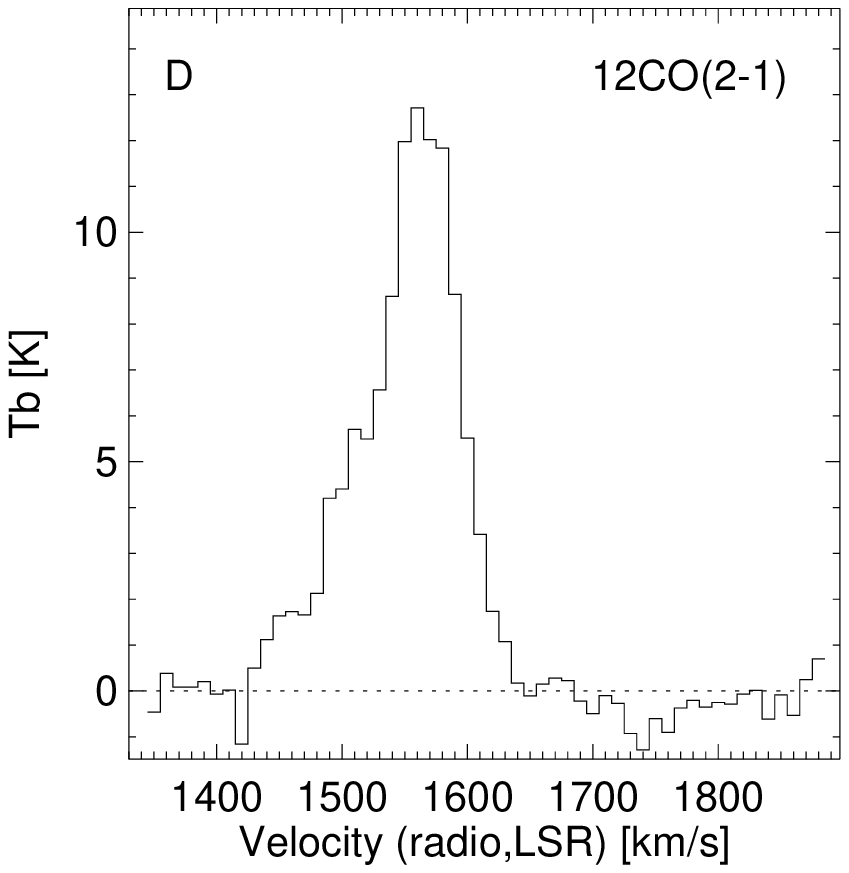} 
\plotone{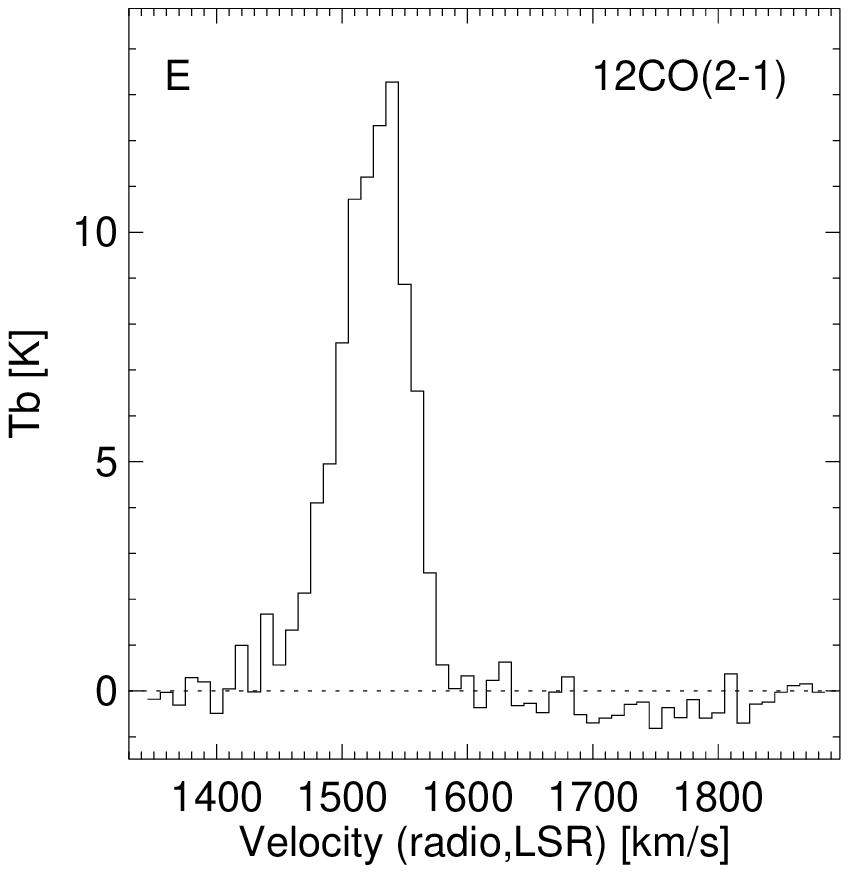} 
\plotone{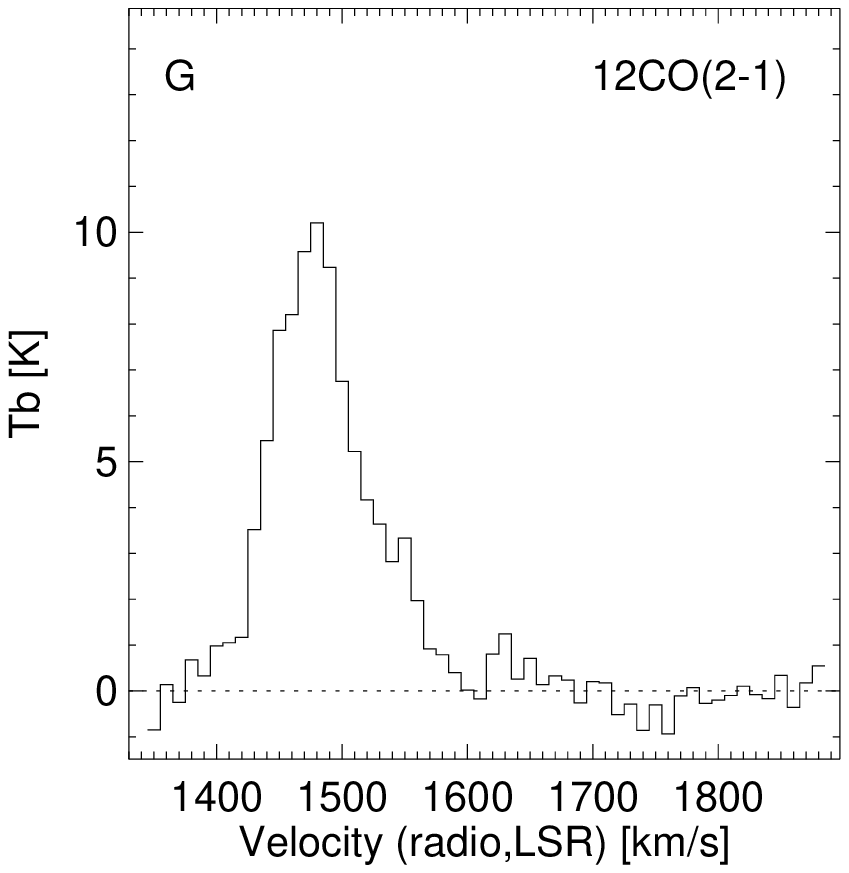} 
\plotone{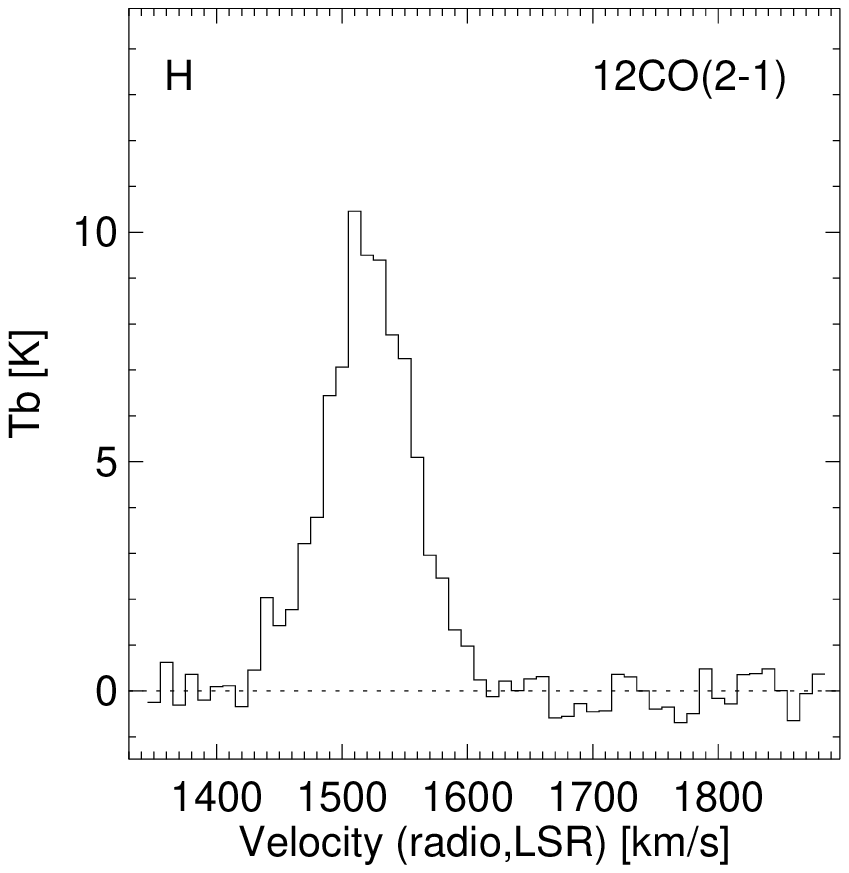} 
\plotone{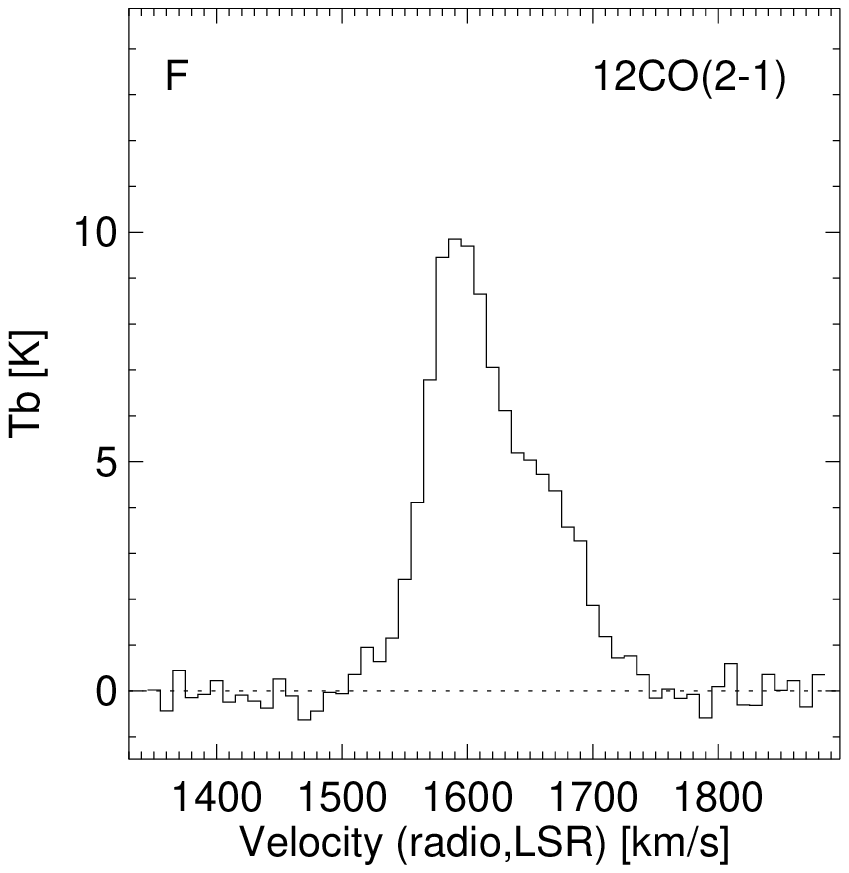} 
\\
\plotone{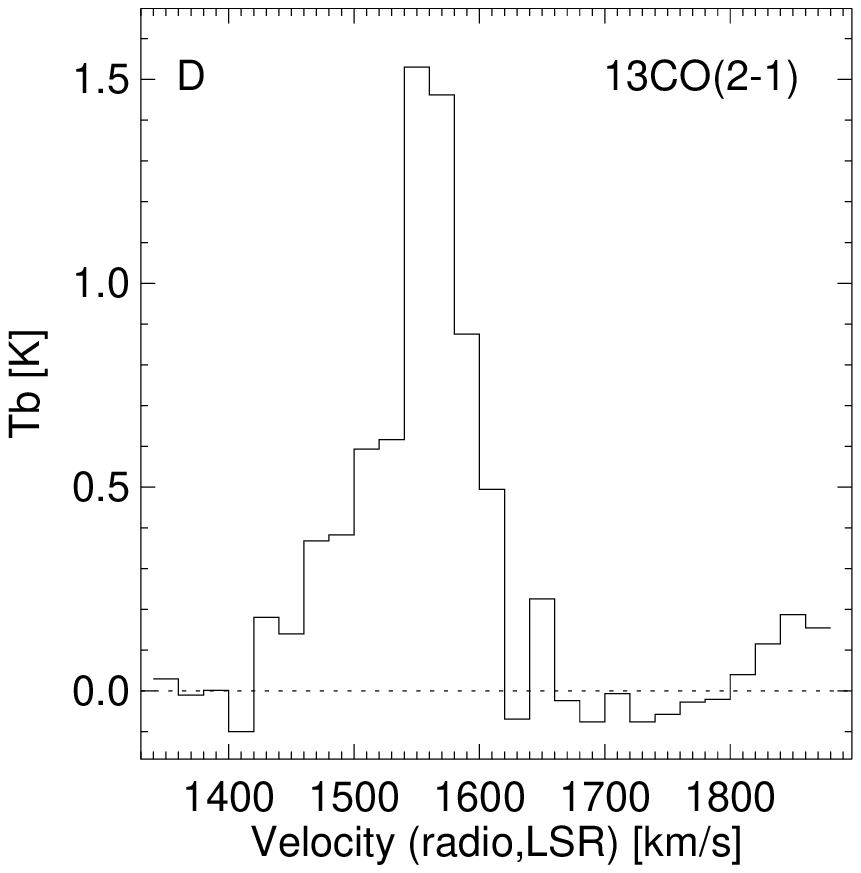} 
\plotone{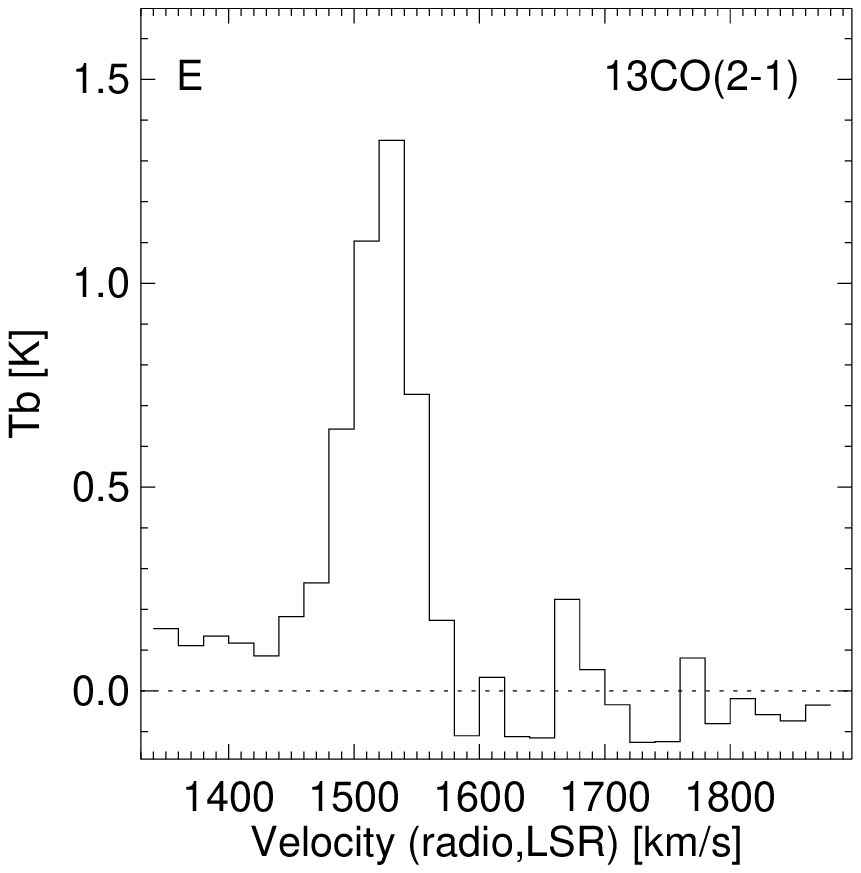} 
\plotone{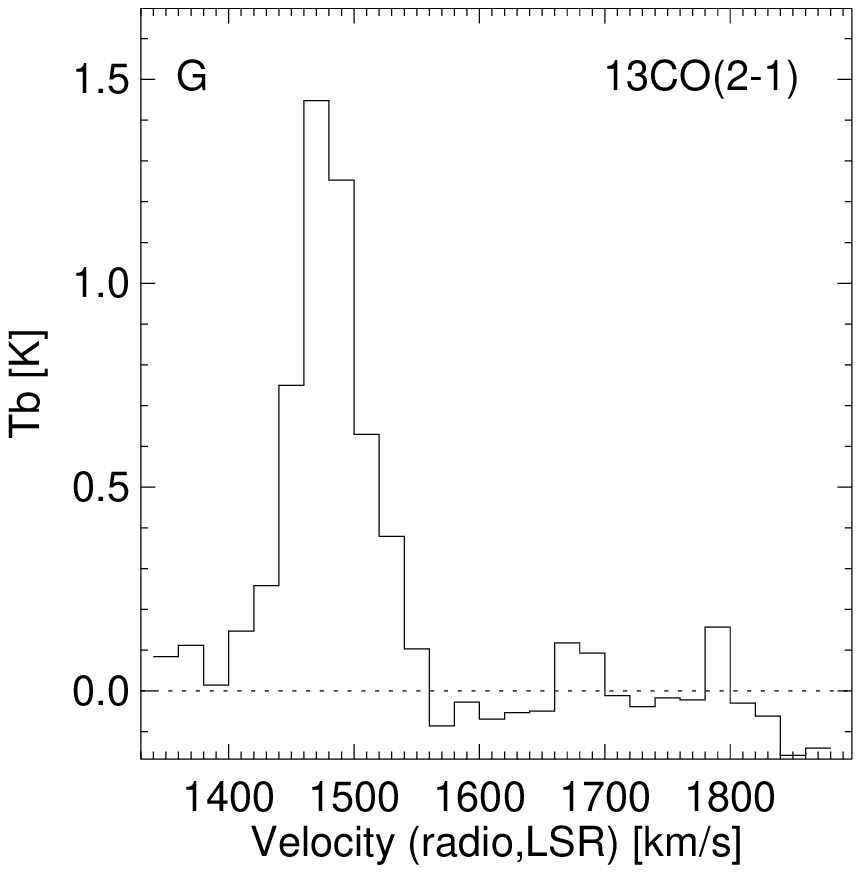}  
\plotone{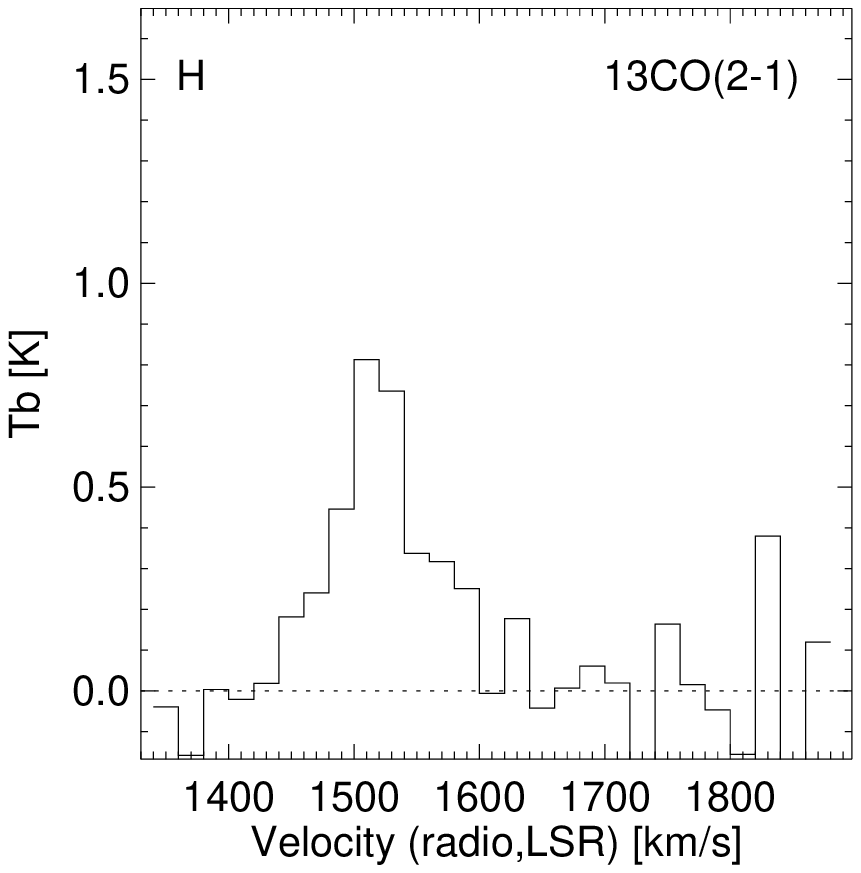} 
\plotone{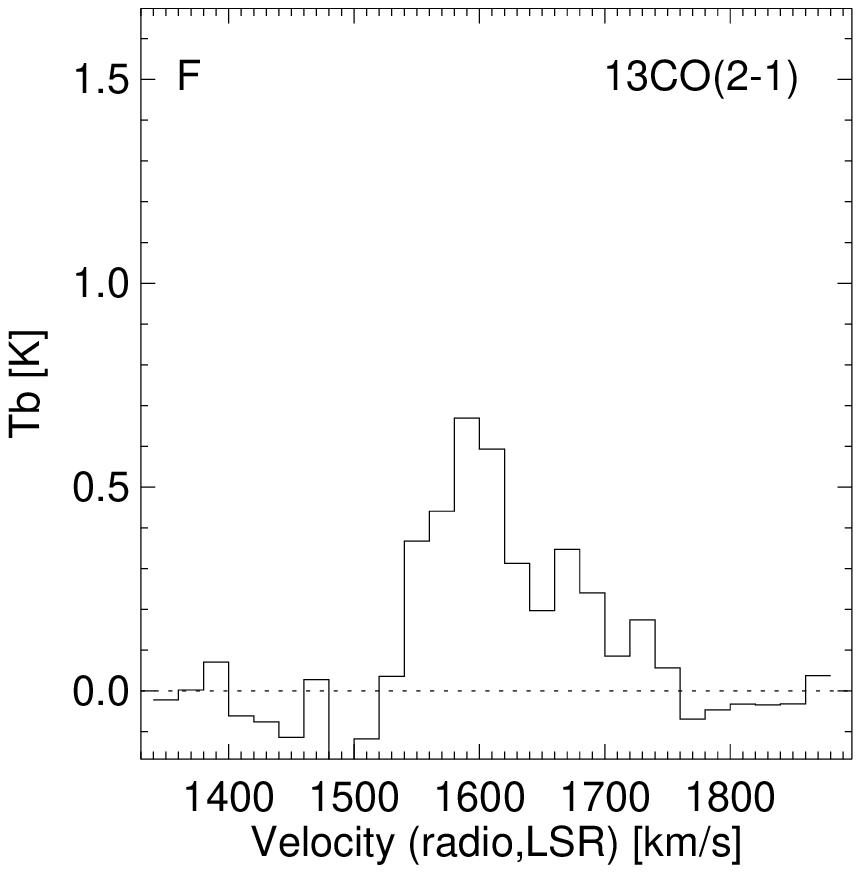}  
\\
\plotone{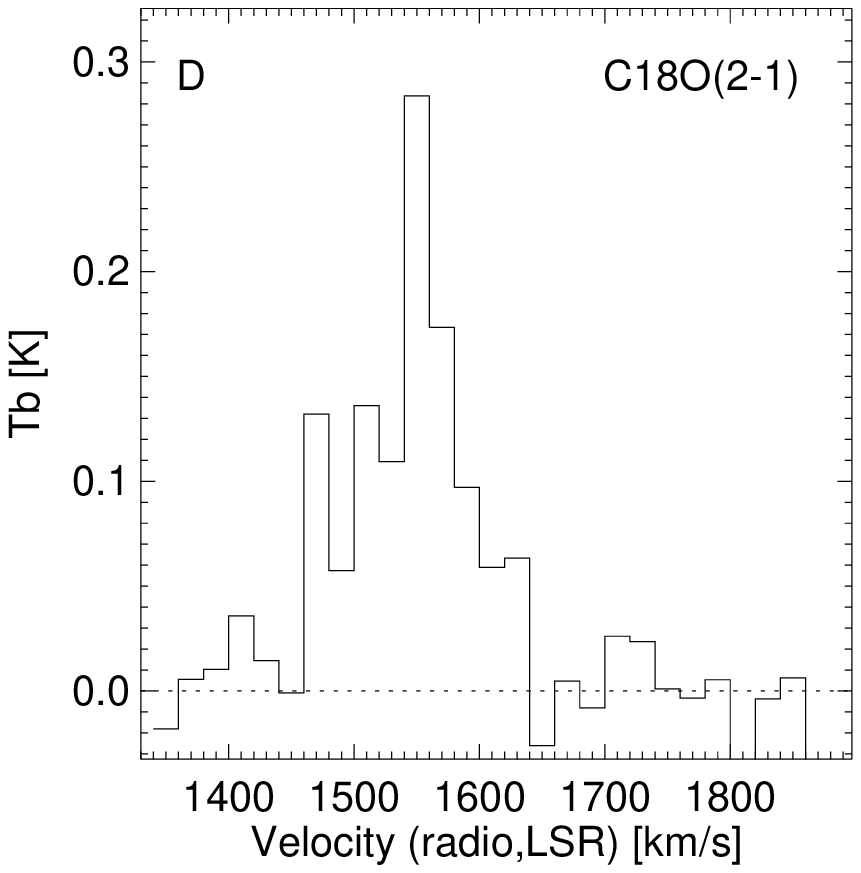} 
\plotone{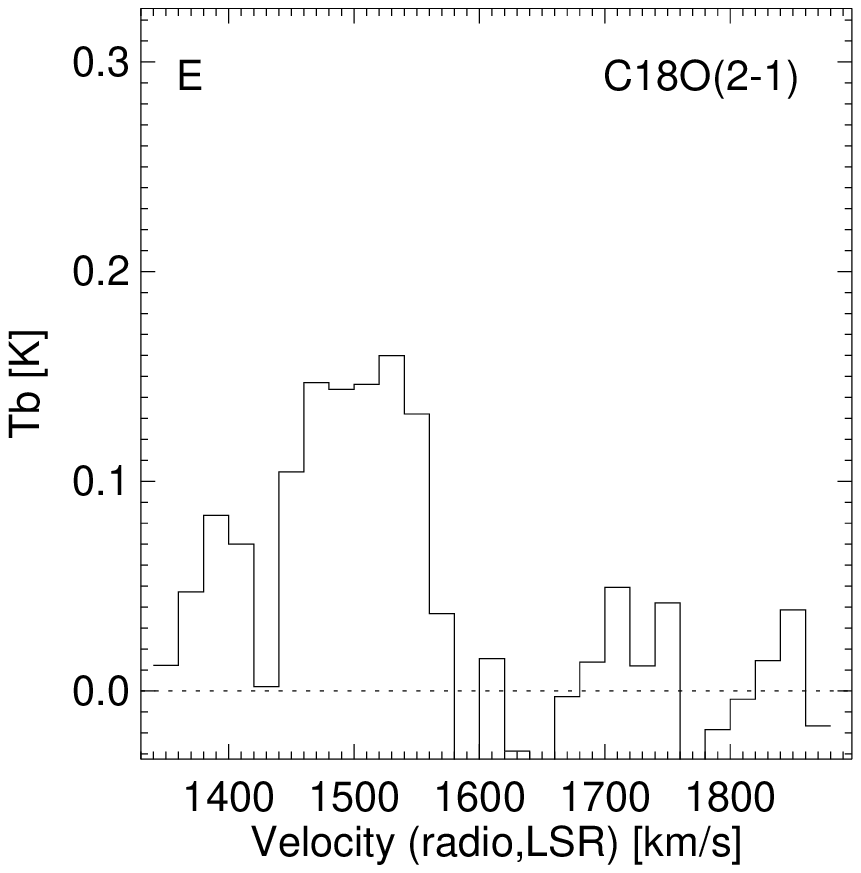} 
\plotone{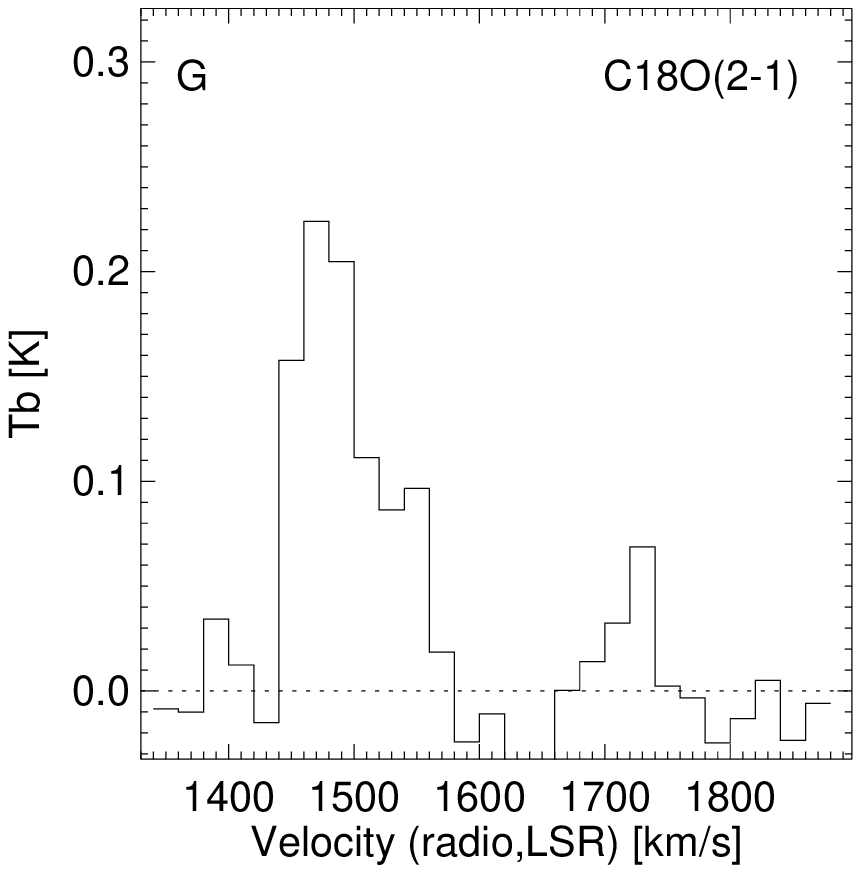} 
\plotone{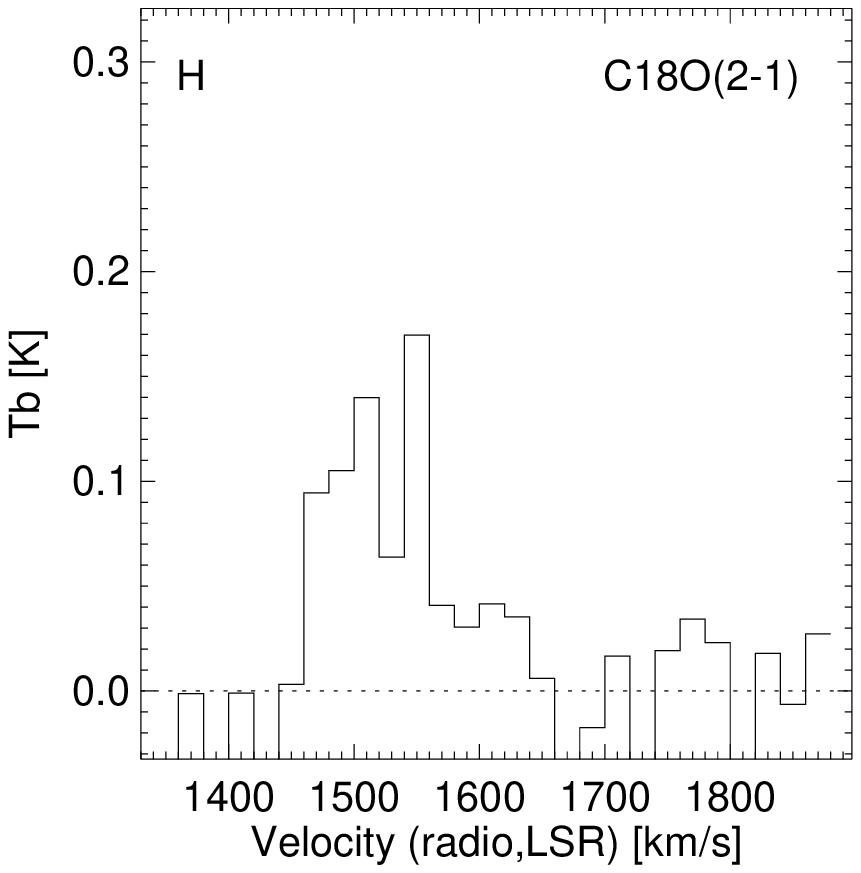} 
\plotone{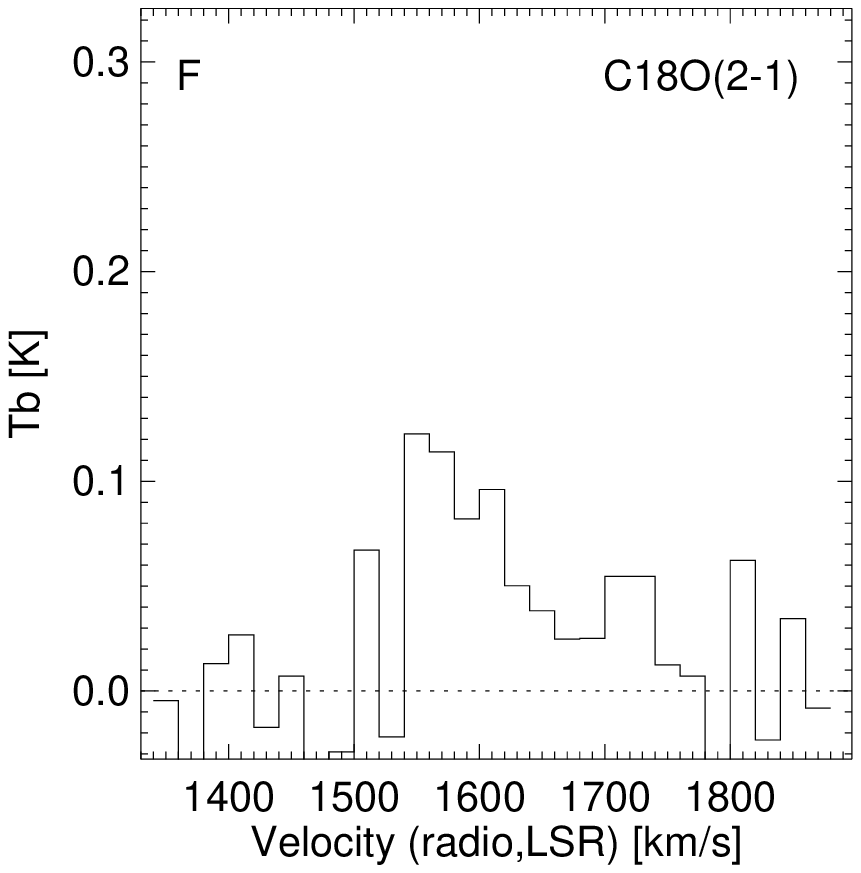} 
\epsscale{1.0}
\caption{Spectra of CO hotspots.
At the top of each panel are the radio name of the hotspot and the line name.
The spectra are sampled from the data cubes used for Fig. \ref{fig.comaps}
at the positions in Table \ref{t.clumps} after primary-beam correction.
Intensities are in brightness temperatures. 
The beam size, and hence the degree of beam dilution, are different for the three lines.
 \label{fig.peakspectra} }
\end{figure}

The CO spectra measured at the hotspots are predominantly single-peaked, 
and the peaks of different isotopic lines are at the same velocity for each hotspot,
as shown in Figure \ref{fig.peakspectra}.
This suggests that each hotspot is a single subregion in each gas complex.
The properties of CO emission measured at the CO hotspots are summarized in Table \ref{t.clumps}.

\subsection{Seyfert Nucleus}
The active nucleus is in the CO deficient area inside the molecular ring, and
no CO peak is at the nucleus.
The integrated intensity and peak brightness temperature of \twelveCO(2--1)
at the AGN position are 9.3 Jy \kms\ \persquarearcsec\ and 2.0 K, respectively, in
our 2\farcs6 $\times$ 1\farcs8 resolution data.
The surface density of molecular gas in the galaxy plane is
$2\times 10^{2}$ \Msol\ \persquarepc\ in the central 200 pc for the adopted \xco.
The mean total extinction in our line of sight
is $\sim$12 mag (i.e., $N_{\rm H_2} = 1.2\times 10^{22}$ \persquarecm). 
The column density  and extinction in front of the AGN would
be half of these if the nucleus were at the mid-plane of a uniform gas layer.
The mass of molecular gas within 100 pc of the nucleus 
is $6\times 10^{6}$ \Msol.
The supposed molecular torus confining the conical plasma outflow should be in this region, 
considering the opening angle of the flow \citep[$\sim$50\arcdeg ;][]{Hjelm96} 
and the typical scale height of the molecular gas layer ($\lesssim$ 100 pc) 
in the centers of disk galaxies.
Note that these mass-related quantities inherit the uncertainty of the conversion factor,
and that the factor quite possibly changes between the starburst ring and the 
vicinity of the AGN.

We detect no kinematical anomaly at the nucleus, such as 
high velocity gas suggestive of fast rotation around the AGN
or a kinematical sign of  a molecular outflow that may be entrained 
by the plasma outflow. 
Our spatial resolution, however, is probably insufficient to see gas motion
dominated by the central object.
The full-width at zero intensity of \twelveCO\ line  is 270 \kms\ at the nucleus.
This implies a dynamical mass of $M_{\rm dyn} (R_g \leq 0.1 \;\kpc) \sim 1\times 10^9 \Msol$
for gas in circular motion in the galaxy plane.
This is an order of magnitude larger than the mass of the central engine 
inferred from the $K$ magnitude of the galaxy by \citet{Risaliti05}. 
Judging from the PV diagram (Fig. \ref{fig.majpv}), 
a factor of $\gtrsim$3 higher resolution is likely needed before
one starts to see the gas motion influenced by the expected black hole.

The properties of molecular gas around the AGN can be constrained, though weakly, 
from the low brightness temperature of \twelveCO.
The physical temperature of molecular gas in the central 200 pc would be 6 K 
if the gas is uniformly distributed in the region and if \twelveCO\ is optically thick and in LTE. 
Judging from the low temperature, it is more likely that one or more of these conditions is not satisfied;
i.e., the gas has clumps, a \twelveCO(2--1) opacity $\lesssim 1$, 
subthermal excitation of CO, or a combination of these.
The clumpy distribution of the ISM could reduce the extinction toward the AGN, to
be more consistent with the optically visible nucleus.

\section{Bar-Driven Gas Dynamics } \label{s.bardynamics} 
The central concentration of molecular gas, with an extent of a few kpc and a mass of $\sim$$10^{9}$ \Msol, 
is almost an expected feature for NGC 1365 considering the long and prominent bar in the galaxy.
A stellar bar is known to drain the gas in the region it sweeps toward the galactic center 
\citep[e.g.,][]{Matsuda77, Sakamoto99b, Sheth05}.
The bar in NGC 1365 has apparently worked as a powerful gas funnel.

The overall distribution of molecular gas in the central concentration can be understood
in the framework of the bar-driven gas dynamics,
in which the dominant pattern of gas flow
changes between the so-called \xone-type and \xtwo-type in the circumnuclear region.
The \xone\ and \xtwo\ are families of stable, prograde, and periodic orbits for stars 
in the nonaxisymmetric potential of a galaxy bar \citep[see][]{Binney87}.
The former orbits are elongated along the bar and fill most of it, 
and the latter, which can exist closer to the galactic center under certain conditions, 
are elongated perpendicular to the bar.
Gas follows similar orbits or streamlines, but because gas is collisional unlike stars
the angle between \xone\ and \xtwo\ for gas can be oblique and less than 90\arcdeg\
measured in the direction of galaxy rotation \citep{Wada94}.
As has been modeled specifically for this galaxy by \citet{Teuben86} and \citet{Lindblad96} and for generic
barred galaxies by many authors, if there exist \xtwo\ orbits, gas can form a circumnuclear ring or spiral pattern
connected to a pair of straighter gas lanes in the bar.
What we see in NGC 1365 matches the model gas morphology and 
what has been seen in many other barred galaxies.
The leading-edge gas lanes on the bar are part of the \xone-like orbits 
and the gas ring, 
whose major axis makes an oblique angle less than 90\arcdeg\ with the bar, 
corresponds to the \xtwo-like orbits.

\begin{figure}[t]
\epsscale{0.5}
\plotone{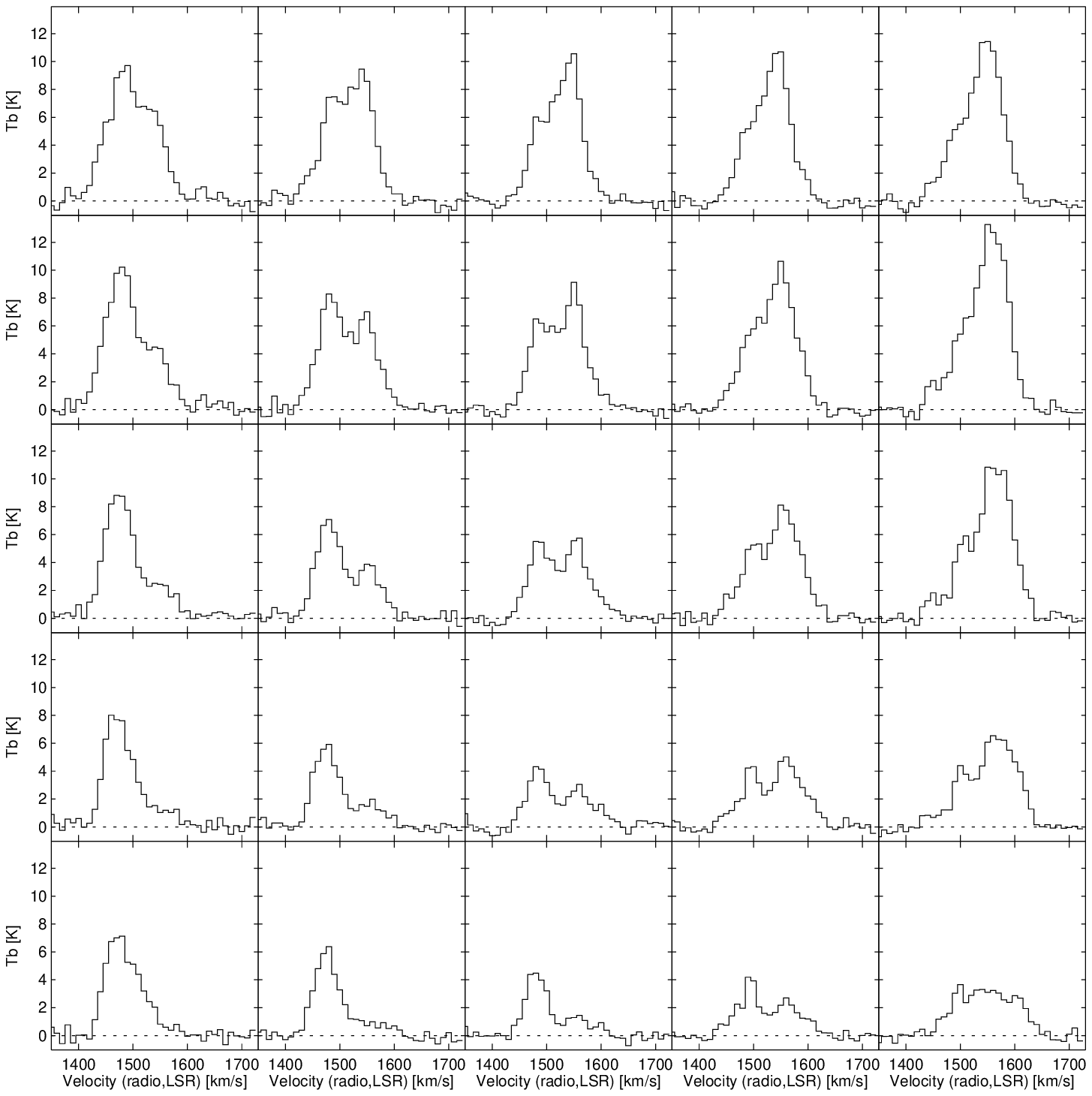} 
\\
\epsscale{0.6}
\plottwo{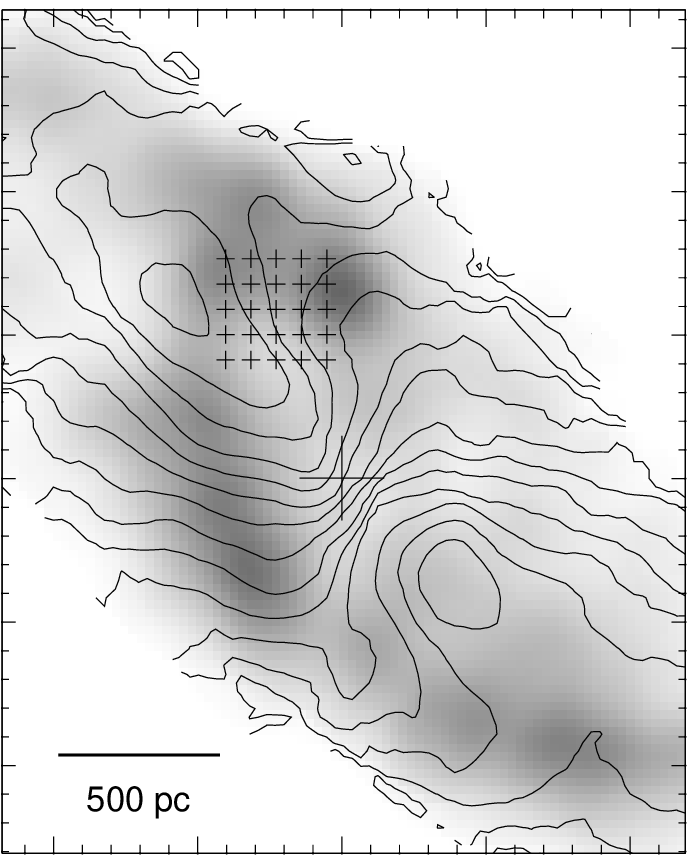}{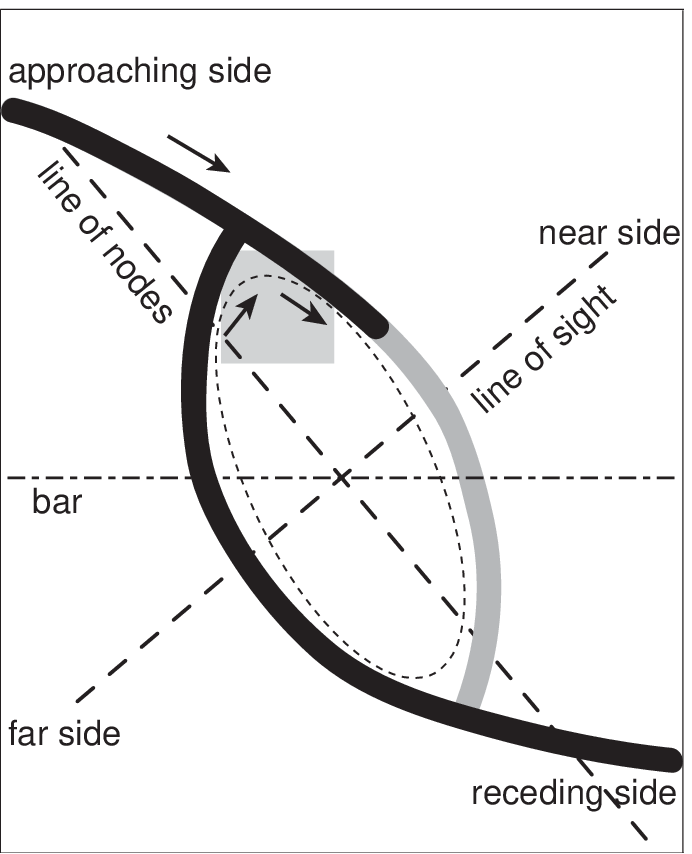}  
\caption{(a) Spectra of \twelveCO\  near an apocenter of the circumnuclear ring.
The double-peak profiles indicate the motion of gas roughly along the circumnuclear ring.
(b) Sampling positions of the line profiles, shown as a group of small plus signs. 
The background image is the \twelveCO\ integrated intensity, and the contours are 
the \twelveCO\ mean velocity in 20 \kms\ steps.
(c) Illustration of the gas motion. 
The double-peaked profiles in the sampling region (= the gray square) are due to
the distinctively different velocities in our line of sight 
between the gas approaching on the ring to the apocenter and the gas leaving it along the ring.
The gas streamline on the ring can be oval similar to the dotted ellipse, or it may have a cusp at a shock. 
In either case, 
its major axis makes an oblique angle  with the stellar bar ($\lesssim 90$\arcdeg\
measured in the direction of the galaxy rotation, or clockwise), 
indicating that the gas streamline is \xtwo-like.
Around the apocenter, 
the gas flow on the molecular ring and another flow roughly along 
a leading-edge gas lane on the bar (almost) converge, likely leading to frequent cloud collisions.
The major and minor axes of the galaxy are shown as the dashed lines 
and the stellar bar as the dot-dashed line.
\label{fig.ispec_x1x2} }
\end{figure}

Kinematically, there are two pieces of evidence that support the gas dynamical model mentioned above.
First, the {\sf S} shaped isovelocity contours around the systemic velocity (\S \ref{s.kinematics}) 
are explicable by the transition of dominant orbits between \xone-like outside the ring and \xtwo-like  around the ring 
as seen in the models with the \xone--\xtwo\ transition, or an inner Lindblad resonance (ILR),
in Fig. 4 of \citet{Teuben86} and Fig. 20b of \citet{Lindblad96}.
The noncircular motion can be caused purely by the transition \citep{Lindblad94}, 
but gas shocks may be also contributing.
Secondly, as shown in Fig. \ref{fig.ispec_x1x2}, 
the double-peaked line profiles near an apocenter of the circumnuclear ring are 
consistent with oval gas motion along the ring, whose major axis makes 
an oblique angle to the bar. 
This again suggests that \xtwo-like gas stream exists on or slightly inside the molecular ring.
We note that these kinematical and morphological clues for the presence of a \xone--\xtwo\ transition
are probably more robust than an analysis of the rotation curve by using the resonance condition
of $2(\Omega - \Omega_{\rm b})=\kappa$, where $\Omega$, $\Omega_{\rm b}$, and $\kappa$ are
angular frequency, the pattern speed of the bar, and epicyclic frequency, respectively.
The resonance condition is known to be a poor indicator for the presence of \xtwo\ orbits in a strong bar
\citep{Contopoulos80}.

\section{Gas Properties}  \label{s.gas_properties} 
\subsection{Average Properties in the Galactic Center \label{s.gas_average_properties}}
The observed \twelveCO-to-\thirteenCO\ intensity ratio, 
$R_{12/13}^{(2-1)}=8.7\pm2.8 $, is normal for the centers
of local spiral galaxies including the ones with moderate starburst 
($\log (L_{8-1000\;\mu {\rm m}}/\Lsol) \lesssim 10.5$).
For example, the ratio is about 10 at the centers of
our Galaxy, IC342, NGC 253, and M82 \citep{Sawada01,Meier00,Mao00,Sakamoto06a}.
The ratio separates NGC 1365 from the more infrared luminous mergers  
($\log (L_{8-1000\;\mu {\rm m}}/\Lsol) \gtrsim 11.5$), which are known to
show higher ratios of $R_{12/13}^{(2-1)} \gtrsim 20$ \citep{Casoli92b,Aalto95,Glenn01,Sakamoto06b}.

There are three basic circumstances under which the \twelveCO-to-\thirteenCO\ intensity ratio
is about 10, although some of their intermediate situations also produce the same ratio.
The three correspond to the three parts of the formal expression of the ratio:
\[
	R_{12/13}
	=
	\frac{(1-e^{-\tau_{12}})}{(1-e^{-\tau_{13}})}
	\frac{[J(T_{\rm ex,12}) - J(T_{\rm CMB})]}{[J(T_{\rm ex,13}) - J(T_{\rm CMB})]}
	\frac{f_{\rm b, 12}}{f_{\rm b, 13}},
\]
where $\tau$, $T_{\rm ex}$, and $f_{\rm b}$ are optical depth, excitation temperature, and beam
filling factor, respectively, and $J(T)$ is the effective radiation temperature
$(h\nu/k)/[\exp(h\nu/kT)-1]$ for the line frequency denoted as $\nu$.
First, if the CO is thermalized at least up to the J=2 level and 
if the two isotopic lines have almost the same beam-filling factors,
then the ratio is determined by the optical depths. 
For example, the ratio of 8.7 corresponds to $\tau_{^{13}{\rm CO}(2-1)} = 0.12$ and 
\thirteenCO\ column density of $5\times 10^{15}$ \persquarecm\ \perkms\ for a 50 K gas in LTE.
The molecular gas needs to be turbulent, with a characteristic velocity gradient of 
$dv/dr \sim 15$ \kms\ \perpc, to have a hydrogen density exceeding $n_{\rm H_2} \sim 10^4$ \percubiccm\
to maintain the excitation.
Secondly, the ratio of \tilde10 can also be observed when the \thirteenCO\ 
excitation is highly subthermal.
In the non-LTE analysis using the large velocity gradient model \citep{Goldreich74}, 
the mean ratio of 8.7 requires a gas density of
$n_{\rm H_2} \sim 3 \times 10^2$ \percubiccm\ for the abundances of  [\HH/\twelveCO] $= 10^{4}$
and [\twelveCO/\thirteenCO] = 40 and a velocity gradient in clouds of 1 \kms\ \perpc.
Gas temperatures in the range of 10 -- 300 K produce the same result.
In these conditions, the excitation to J=2 is subthermal with \thirteenCO\ having lower
excitation temperature than \twelveCO, and the \thirteenCO(2--1) opacity is 0.5--0.9.
Finally, different beam filling factors of the two lines can 
affect the ratio if molecular clouds in the galactic center have a core-envelope structure
as has been observed in the Galactic center and suggested in luminous mergers
\citep{Huttemeister95,Aalto95}. In this case, \twelveCO\ traces lower density envelopes
while \thirteenCO\ is  emitted from higher density cores.

Our mass estimate of molecular gas assumes the nearly thermalized excitation of the
first case rather than the very subthermal one in the second.
In the latter,  \CeighteenO(2--1)\ excitation temperature
is about a tenth of the thermal gas temperature for a 50 K gas.
Even though we already allowed for the contribution of thin gas, our mass estimates would need to be 
doubled if most of the \CeighteenO\ and \thirteenCO\ emission is from very poorly excited molecules.
Large turbulence in molecular gas, which separates the thermalized case from the subthermal one,
has been suggested to be a characteristic feature of the Galactic center \citep{Sawada01}.
The turbulence is measured with a velocity gradient in the escape probability approximation
and is larger than 10 \kms\ \perpc.
We expect the starburst nucleus of NGC 1365 has no less turbulence than the Galactic center,
and hence keep our mass estimates.
The very low excitation of \CeighteenO\ and \thirteenCO\ can not be ruled out, however,
without further observations.

\subsection{Spatial Variation}
The variation of $R_{12/13}^{(2-1)}$ within the circumnuclear region,
by about a factor of  3 at our resolution, allows multiple interpretations
corresponding to the three cases mentioned above.
In the thermalized case, a factor of 3 lower ratio can mean a
gas column density per velocity ($N_{\rm H_2}/d v$) increased by the same factor, 
or a gas temperature reduced by about a factor of 2.
In the subthermal case, a higher gas density reduces the ratio.
In the filling-factor model, molecular clouds in lower $R_{12/13}$ regions have smaller 
envelopes with respect to their cores.

Our data show the lack of an organized pattern in the $R_{12/13}^{(2-1)}$ ratio,
except for the slight difference between the near  and far sides.
This makes it difficult to assess the gas properties by combining the line ratio and
other information.
For example, it is possible that molecular gas is relatively dense in the
twin peaks owing to the convergence of gas flow \citep{Ott05}.
One of them, the gas complex containing radio sources D, E, and G, 
shows a lower ratio of \tilde8 than other bright \twelveCO\ complexes, 
which have ratios $\gtrsim$ 10 (Fig. \ref{fig.ratiomap}).
This is consistent with the higher density, which also makes $N_{\rm H_2}/d v$
higher for a constant turbulence ($dv/dr$).
However, there are regions having even lower ratios without obvious dynamical reasons.
Quite possibly, multiple parameters change at the same time for multiple reasons, 
making any large-scale ratio pattern less clear.

On smaller scales, our data also show no distinctive correlation 
between  $R_{12/13}^{(2-1)}$ and the radio, optical, and IR sources.
This likely means that the effect of these sources on the surrounding molecular gas is
either small in terms of the ratio or is heavily diluted in our 200 pc beam.
The former is possible, for example, if higher gas temperature around a cluster,
which could reduce the optical depths of the CO lines for a constant column density,
is offset by the gas concentration (i.e., larger column density) around the clusters.
The  ratio $R_{12/13}$, which traces the \thirteenCO\ optical depth in the thermalized case, 
would then be comparable toward the clusters and the surrounding area.
Finally, it is noteworthy that the high ratio of $\geq 20$ observed in luminous mergers
in kpc beams is not observed in any part of the starburst region.

\section{Circumnuclear Star Formation}  \label{s.starformation} 
\subsection{Star Formation Law}
The circumnuclear starburst in NGC 1365  follows the two empirical scaling relations 
of \citet[hereafter K98]{Kennicutt98} for starburst galaxies. 
Taking 1 kpc as the radius of the starburst region, 
and attributing half of the far-IR luminosity of the galaxy to the starburst (see \S\ref{s.n1365}), 
the star formation rate in the region is 9 \Msol\ \peryr\ 
and its surface density is \sigmasfr\ = 3 \Msol\ \peryr\ \persquarekpc\
for the star formation rate -to- far-IR luminosity relation of K98.
The mean surface density of \HH\ in the region is \sigmahtwo=$1\times10^3 (\xco/2.8)$ 
\Msol\ \persquarepc; K98 adopted  $\xco=2.8$.
Fig. \ref{fig.schmidt} shows that NGC 1365 with these surface densities follows the correlation,
or the Schmidt law, seen among circumnuclear starbursts.
The galaxy also follows the same correlation for a combined sample of starbursts and normal galaxy disks 
(see Fig. 6 in K98).
In addition, NGC 1365 follows another correlation between gas surface density normalized by the dynamical
time and the star formation surface density. 
The former is \sigmahtwo/\tdyn=$4\times 10^{1} (\xco/2.8)$ \Msol\ \peryr\ \persquarekpc\ 
for the starburst region having the \tdyn\ calculated in \S\ref{s.kinematics}.
This and the abovementioned \sigmasfr\ place NGC 1365 on the tight correlation between the two quantities
seen in Fig. 7 of K98, 
with the abscissa and ordinate  being
3.6 and 0.5, respectively, in the figure's units.

The fact that the circumnuclear starburst in NGC 1365 follows these empirical correlations is not
affected by the choice of the conversion relations used to derive \HH\ mass and star formation rate,
because we use the same conversion factors as in K98. 
Removing the conversion relations and stated in observed quantities, 
the circumnuclear starburst in the central 2 kpc of NGC 1365 follows the correlation, 
seen among starbursts and galactic disks, between 
far-IR luminosity averaged over the star forming region and CO (J=1--0 equivalent) luminosity averaged
in the same area. 
The second correlation that NGC 1365 follows can be also stated only using observed quantities.
The factor of \tilde6 difference in the CO to \HH\ conversion factor between this paper and K98 
only affects such quantities as gas consumption time and gas-to-dynamical mass ratio.
The time in which the current amount of gas would be consumed by the star formation at the current rate
is $1\times 10^{8}$ yr in the 2 kpc-diameter starburst region 
for the conversion factor we adopted,  and it is 6 times longer for the disk $X_{\rm CO}$  
used in K98. 
Note that \citet{Kennicutt98} used the disk conversion factor for his sample 
only for the sake of simplicity, being fully aware that the factor can change in starburst environment.

\subsection{Reasons for the Large Luminosity}
\begin{figure}[t]
\epsscale{0.7}
\plotone{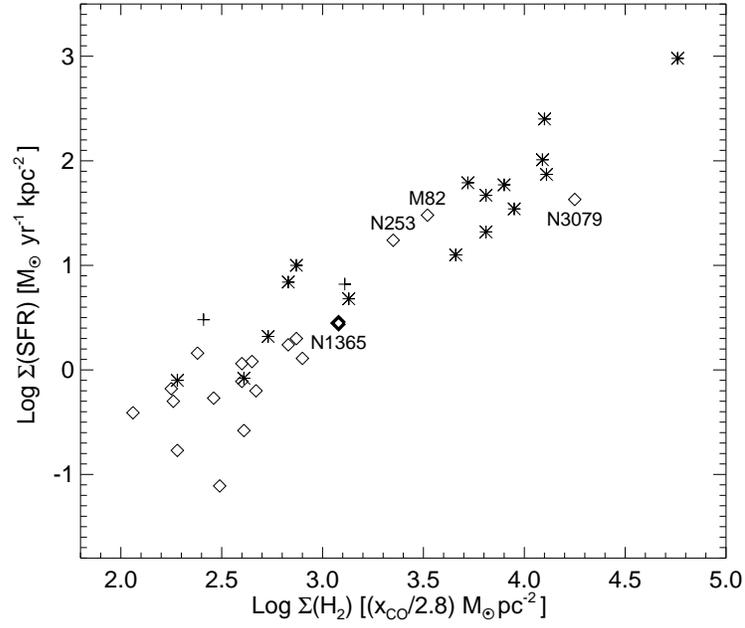} 
\caption{NGC 1365 in the Schmidt-law plot of \citet[Fig. 5]{Kennicutt98} for starburst nuclei. 
The gas surface density and the star formation rate are averaged over the circumnuclear
starburst region of each galaxy. The size of the starburst region is individually determined and is
different for each galaxy. It is 2 kpc in diameter for NGC 1365 and is taken from K98 for others.
The gas surface densities are normalized with \xco= 2.8 only to make a fair comparison.
Asterisks are mergers and diamonds are non-merging galaxies. 
The plus symbols are for two distant ($>100$ Mpc) galaxies that
are likely mergers judging from their large far-IR luminosities
 \citep[$\log (L_{8-1000 \mu{\rm m}}/\Lsol) \geq 11.80$;][]{Sanders03} 
but lack morphological evidence for merging.
\label{fig.schmidt} }
\end{figure}

The reasons for the starburst in NGC 1365 being one of the most luminous in non-merging galaxies
while following the normal star formation laws are 
the high gas surface density in the starburst and the
large size of the starburst region.
The mean gas surface density in the starburst is among the highest observed
in non-merging starburst galaxies, 
as seen in Fig. \ref{fig.schmidt} where the galaxy has the fourth highest \sigmahtwo, only 
after NGC 3079, M82, and NGC 253, among 17 non-merging starbursts.
The central gas-surface density of NGC 1365 also stands out when compared to
those in the NRO--OVRO survey of nearby galaxies \citep{Sakamoto99a,Sakamoto99b}.
The high-resolution survey gives gas surface densities in the central kiloparsec of 20 spirals,
including some starbursts but no merger.
The gas surface density in the central kpc of NGC 1365 is 
$\Sigma_{\rm mol} (R_g \leq 500\; {\rm pc}) = 4 \times 10^{2} (\xco /0.5)$ \Msol\ \persquarepc\
and is twice larger than the largest in the survey if the same \xco\ is used for all galaxies.
Meanwhile, 
the extent of the circumnuclear starburst in NGC 1365, \tilde2 kpc, is among the largest
compared to those in K98 and surveys by \citet[Fig. 5]{Jogee05} and \citet[Fig. 1]{Knapen06}.
Obviously, the high gas surface density in the large area needs a large amount of gas;
$M_{\rm mol}(r_g < 1\; \kpc) \sim 10^{9}$ \Msol\  (\S \ref{s.gas_distribution}).
For this accumulation, the large size of the galaxy 
\citep[$L_{\rm B}\approx 6\times 10^{10} \Lsol \approx 3 L_{\rm B}^{*}$, $D_{25}=59$ kpc;][]{rc3}
and the long bar ($l_{\rm bar}\sim 18$ kpc) were probably crucial through 
the bar-driven transport of disk gas.  
Thus the combination could be regarded as a deeper cause for the luminous starburst.

The luminous starburst in NGC 1365 contains subregions comparable 
to the entire starbursts in M82 and NGC 253.
Most notably, the northeastern edge of the circumnuclear ring 
has $\sigmahtwo \approx 3\times 10^{3} (\xco/2.8)$ \Msol\ \persquarepc, 
or the same gas surface density as in those two starbursts
for the same extent of \tilde400 pc and for the same \xco.
The radio and mid-IR sources in the region suggest enhanced \sigmasfr\ there.
With a few such subregions, it is reasonable that NGC 1365's starburst is a factor of 2--4
more luminous than those in M82 and NGC 253.
Although the empirical star formation laws in the previous section and the
scaling observation like this leave out the physics of star formation
on smaller scales, they are still useful to obtain an overview ---  star formation
in this luminous starburst is not discrete in nature from those in less luminous starbursts.

It remains to be found out why the luminous starburst  is seen {\it now} even though
it can last only for \tilde$10^{8}$ yr.
\citet{Roy97} suggested that the bar was young (age \tilde 1 Gyr) and started the 
gas transport in the recent past on the basis of a bend in the radial gradient of O/H abundance, 
but the existence of the bend has been disputed by \citet{Pilyugin03}.
The galaxy is a member of the Fornax cluster, but lacks an obvious interacting partner
for strong gravitational disturbance.
\ion{H}{1} imaging by \citet{Jorsater95} does not reveal a sign of recent galaxy interaction either,
though the outer \ion{H}{1} disk shows a hint of interaction with the intergalactic medium.
It may be that the starburst has been turning on and off without external triggers, 
perhaps owing to a threshold in the star formation mechanism \citep{Jogee05}. 
If so, the star formation law must switch between the one for the starburst state (e.g., Fig. \ref{fig.schmidt}) 
and the other for non-starburst.

\subsection{CO Hotspots and Super Clusters}
We saw in \S \ref{s.comparison.hotspot_ssc} that the molecular ring contains several CO hotspots
and that three of them are spatially associated with the radio/mid-IR sources D/M4, E/M5, and G/M6
that are suggested to be super clusters of \tilde$10^6$ \Msol, \tilde$10^9$ \Lsol\ 
and ages of  3--6 Myr by \citet{Galliano05}.
Another source, H/M8, also coincides with a CO hotspot, but it is fainter
and does not have age and mass estimates.
It is possible that each of the radio/mid-IR sources is a group of clusters rather than
a single cluster, since their sizes are only constrained to be $\leq$ 20--60 pc  \citep{Sandqvist95,Galliano05}
while super clusters have sizes of $\lesssim$ 10 pc.
We refer each source as a super cluster for simplicity, but
`a cluster' in the following can be `a cluster complex'.

The empirical star formation laws that we found applicable to the circumnuclear region as a whole
allow the amount of star formation needed for the clusters.
A star formation rate of $\Sigma_{\rm SFR} \approx 20 $ \Msol\ \peryr\ \persquarekpc\
is expected from the star formation laws in K98 for the typical CO integrated intensity on the
circumnuclear ring,  $I_{\rm CO2-1}\approx 150 $ Jy \kms\ \perbeam\ equivalent to
an inclination corrected surface density of
$\Sigma_{\rm H_2} \approx 3\times 10^3 (\xco /2.8) $ \Msol\ \persquarepc.
Thus, if the star formation laws apply to 100 pc scales, 
$10^{6}$ \Msol\ of stars will form every 5 Myr in every 100 pc by 100 pc
area in the molecular ring.
The empirical laws, however, do not tell whether and where a compact and massive cluster forms.
The spatial association of the putative \tilde$10^{6}$ \Msol\ clusters and the CO hotspots 
may be a valuable observational clue to address these points.

The significance of the hotspot-cluster association is that the CO hotspots were
very likely the sites of cluster formation in the recent past for two reasons.
First, the clusters' relative motion with respect to the ambient gas in their
\tilde5 Myr lives would be $\lesssim 50$ pc  if we assume 
the clusters' relative velocities against the ambient gas to be \tilde10 \kms, which is a typical
velocity dispersion of molecular clouds in galaxies. 
According to this estimate, the CO hotspots
seen in our \tilde200 pc resolution images are composed of gas that was in the immediate vicinity of 
the formation sites of the super clusters.
Second, given the non-trivial configuration of the three hotspot-cluster pairs, it is unlikely that
the three clusters formed elsewhere and have migrated to the CO hotspots, unless the clusters
themselves affected their ambient gas to create the hotspots. 
This argument reinforces the first one, in which the drift velocity might be larger than assumed
because the three clusters are in the region with two types of orbits
and because only gas is subject to hydrodynamical forces. 
Regarding the possibility that the clusters caused the hotspots
rather than the gas hotspots formed the clusters,
we favor the latter interpretation though our data do not rule out the former.
Our preference is because three hotspots are visible in \thirteenCO\ and one even in \CeighteenO.
Elevation of gas temperature, the most likely effect of a young cluster
on its ambient gas, does not increase the intensity of an optically thin CO line per unit mass of molecular gas
if the isotope CO is well excited, which we favorably discussed in \S\ref{s.gas_average_properties}.

In our more favored model, the CO hotspots are peaks of gas surface density.
The surface density is either that for the entire velocity or for a 10 or 20 \kms\ range
depending on whether the hotspot is in the integrated map or in the peak temperature map.
Gas temperature can also be higher in the hotspots seen in \twelveCO\ 
because the brightness temperature of the line
is the gas temperature multiplied by the beam filling factor. 
The gas surface densities in the hotspots are $\Sigma_{\rm mol} \sim 10^{3}$ \Msol\ \persquarepc\ when averaged 
over 200 pc scale (see Fig. \ref{fig.coihisto}).
The CO hotspot-cluster associations suggest that the super clusters formed in such 
a high surface-density environment. 
If the high surface density of molecular gas is one of the requirements for the SSC formation,
then SSC formation predominantly in starbursts can be a consequence of the high gas
surface densities in starburst regions.

In the alternative interpretation, 
the CO hotspots are largely due to higher gas temperature there, plausibly owing to the 
luminous super clusters.
For optically thick \twelveCO, this straightforwardly explains the hotspots.
For \thirteenCO\ and \CeighteenO, 
this interpretation is possible only when the gas excitation is subthermal and the isotope lines have
moderate (\tilde1) optical depths as in the second case in \S\ref{s.gas_average_properties}.
In such a situation, 
a higher gas temperature can increase the excitation temperature of the subthermally-excited CO 
without significantly reducing the line optical depth, 
and hence can increase the line brightness temperature.
The gas surface densities at the hotspots in this case would be higher than in the thermalized case 
(\S \ref{s.gas_average_properties}),
even though the hotspots may not be local peaks of gas surface density.

\subsubsection{Similar Sources in Other Galaxies}
There have been a small number of cases, in other galaxies, that are similar to 
the CO hotspot-cluster associations.
Table \ref{t.othergalaxies} compares them with those in NGC 1365.
These cases show that at least some of the most massive and youngest clusters in starbursts are
accompanied by CO hotspots detectable at \tilde100 pc resolutions.

One case is in the \object{Antennae} merger (\object{NGC 4038}/\object[NGC 4039]{4039}).
The brightest 15 \micron\ source in the system, 
located in the overlap region of the colliding galaxies,  is associated with a molecular complex called SGMC4--5
\citep{Wilson00, Mirabel98}.
Their parameters are within a factor of a few from those of the CO hotspot-cluster associations in NGC 1365 
considering the larger CO beam and hence more beam dilution in the Antennae.
The source has been suggested to be a young (\tilde4 Myr), massive (\tilde$10^{7}$ \Msol), 
and compact ($R_{\rm half\; light} \sim 30$ pc) cluster or cluster complex 
with thousands of O stars and a total luminosity of \tilde$10^{9}$ \Lsol\ 
\citep{Gilbert00,Neff00,Haas00}. 
Since the cluster is in the molecular complex but not at its CO peak, the CO hotspot is most likely
a peak of gas surface density.
Statistically, \citet{Zhang01} showed that young clusters (of ages \tilde5 Myr, including
the one mentioned above) in this galaxy are better associated with mid/far-IR, CO, and radio sources 
than older clusters.

Another case is in the barred starburst spiral \object{M83}.
The brightest CO peak in the galaxy's circumnuclear region, 
in both integrated intensity and brightness temperature, 
coincides with a radio and mid-IR peak \citep{Sakamoto04,Turner94,Telesco93}.
The radio and mid-IR emission is due to star formation, though there may be
contribution from a radio supernova to the radio peak \citep{Turner94,Telesco93}.
The CO hotspot is on a circumnuclear gas ring and is where \xone-like and \xtwo-like gas streamlines converge.
The hotspot location with respect to the bar and the circumnuclear gas ring is 
therefore identical to that in NGC 1365.

\subsection{Cluster and Hotspot Evolution}
It has been proposed that gravitational collapse in an ILR gas ring, 
where a bar accumulates disk gas,
forms gas clumps and subsequently causes intense star formation in them
\citep{Wada92,Elmegreen94}. 
The CO hotspots with embedded young super clusters may be such gas clumps.
If the molecular hotspots are high surface-density regions of molecular gas, possibly
caused by gravitational instability, then
the formation of super clusters from such giant gas concentrations of \tilde100 pc and $>$$10^{7}$ \Msol\
scales is in accordance with the supergiant-cloud model for the formation of super clusters
\citep[and references therein]{Wilson03}.
The CO hotspot-cluster association, however, is also
compatible with another model for SSC formation where a super cluster forms from a 
moderate-size  molecular cloud ($\lesssim$$10^6$\Msol) by converting most of the gas into stars
\citep[and references therein]{Keto05}. 
Such clouds quite possibly exist in the high surface-density environment of 
the CO hotspots and could have had such triggers as cloud collision and compression by 
nearby \ion{H}{2} regions or supernova remnants.
High surface-density clouds with a few 10 pc sizes and 
velocity widths of the order of 10 \kms\  could cause the CO hotspots in the peak temperature maps.

The SSCs in the hotspots are expected to disperse ambient gas
as they age while blowing stellar winds and having supernovae.
A super cluster may cause an expanding bubble or even an outflow of the ISM from the disk 
\citep[e.g.,][in our SMA survey]{Sakamoto06a,Sakamoto06b}.
Such features may quite possibly be hidden within the CO hotspots.
The optically-visible super clusters that are not associated with 
CO hotspots, most notably M2/SSC6 and M3/SSC3,
may be older super clusters that have already dispelled their dust cocoons in which they formed.
Alternatively, they could have formed from isolated molecular clouds and have
mostly consumed them  at the time of formation.

\section{Summary}  \label{s.summary}
We have observed 1.3 mm CO lines and continuum in the center of 
the barred spiral galaxy NGC 1365 at high spatial resolutions.
The circumnuclear starburst in the galaxy is among the most luminous and intense
in non-merging galaxies.
Our main findings regarding the properties of molecular gas and their relation to the starburst
and to the Seyfert nucleus are as follows.

\begin{enumerate}
\item 
There is an oval ring of molecular gas around the nucleus, 
having a \tilde2 kpc extent and twin-peak enhancements at its apocenters,  
and connected to gas lanes on the leading side of the stellar bar.
The gas morphology and kinematics in the region are consistent with the model that
the ring is at the transition region between \xone- and \xtwo-like gas stremlines in the bar.
The structure is typical of barred galaxies.

\item 
The molecular ring coincides with the starburst ring seen in radio emission, suggesting
that the gas distribution largely determines the extent and the shape of the starburst.
Moreover, the intensity of the starburst averaged over the circumnuclear region is what is
expected from the molecular gas surface density of the region 
and the star formation laws in \citet{Kennicutt98}. 
A plausible synopsis of the galaxy's starburst
is as follows:
The long (\tilde18 kpc) stellar bar 
funneled a large amount of molecular gas (\tilde$10^{9}$ \Msol)
toward the galactic center, and the resulting high gas surface density 
(\tilde$10^{3}$ \Msol\ \persquarepc) in the relatively large 
circumnuclear ring ($r\tilde1$ kpc) resulted in the starburst (\tilde3 \Msol\ \peryr\ \persquarekpc) 
of large total luminosity (\tilde$10^{11}$ \Lsol) following the star formation laws.
---
A remaining question is why the transient starburst (gas consumption timescale \tilde$10^{8}$ yr)
is seen now.

\item
There are compact ($\lesssim 200$ pc) peaks of CO integrated intensity and brightness temperature,
which we call CO hotspots, on the circumnuclear ring. 
Some of the hotspots coincide with the radio and mid-IR sources that have been
identified as dust-enshrouded super star clusters. 
The association of the CO hotspots and super clusters allows two interpretations.
The first one is that the most luminous and presumably massive super clusters were formed in regions 
with the highest gas surface densities (\tilde$10^{3}$ \Msol\ \persquarepc\ in 200 pc scale).
The second one is that the luminous embedded clusters heated their ambient gas to form the hotspot.
The former, which is more conventional in terms of the interpretation of CO intensities at $\gtrsim 100$ pc 
scales, 
implies a relation between the cluster formation and gas complexes 
already identifiable at a 200 pc resolution.

\item 
The active nucleus coincides with the dynamical center of the galaxy
within 30 pc, according to the CO velocity field.
Neither integrated intensity nor brightness temperature of CO peaks at the nucleus.
We found no anomaly in molecular gas that may be related to the AGN.

\end{enumerate}

\acknowledgments
We thank the SMA staff for their help prior to and during our observations,
Jun-Ichi Morino and Andrew Baker for stimulating discussions, 
Raffaella Morganti for kindly providing her radio data,
and the anonymous referee for his/her helpful comments.
This research made use of NED, ADS, and the HST archive, in addition to the SMA;  we are
grateful to their operating, funding, and contributing people and agencies.

{\it Facilities:} \facility{SMA}, \facility{HST (WFPC2)}


\clearpage

\begin{deluxetable}{lcc}
\tablewidth{0pt}
\tablecaption{NGC 1365 parameters  \label{t.galparm} }
\tablehead{ 
	\colhead{parameter}       &
	\colhead{value} &
	\colhead{ref.}
}
\startdata
distance [Mpc] & 17.95 &  1\\
scale & 1\arcsec=86 pc, $11\farcs5$=1 kpc & \\
inclination [deg.]   & 40 &  2\\
major axis P.A. [deg.] & 220 & 2\\
bar P.A. [deg] & 90 & 3,4,5\\
nucleus\tablenotemark{a} R.A. (J2000) & \phs03\hr33\mn36\fs38 & 5,6,7 \\
\multicolumn{1}{r}{Dec. (J2000)} & \minus36\arcdeg08\arcmin25\farcs7\phn  & 5,6,7 \\
$\log (L_{\rm 8-1000\; \mu m}/\Lsol)$  & 11.00 & 8\\
Hubble type & SB(s)b & 9
\enddata
\tablerefs{
1. \citet{Freedman01}; 
2. \citet{Jorsater95}; 
3. \citet{Lindblad96};
4. \citet{Regan97};
5. Measured in the data  of \citet{Jarrett03};
6. \citet{Sandqvist82};
7. \citet{Stevens99};
8. \citet{Sanders03}; 
9. \citet{rc3}
}
\tablenotetext{a}{The adopted mean position is within $0\farcs45$ of the positions in three wavelengths.}
\end{deluxetable}

\begin{deluxetable}{lcc}
\tablewidth{0pt}
\tablecaption{SMA observation parameters  \label{t.obsparm} }
\tablehead{ 
	\colhead{parameter}       &
	\multicolumn{2}{c}{value} 
}
\startdata
date (UT)  & 2003-12-23 & 2004-02-02 \\
array configuration       & compact & extended \\
No. of antennas & 6 & 7 \\
$\tau_{225}$ \tablenotemark{a} & 0.08 & 0.05 \\
$\langle T_{\rm sys}$(DSB)$\rangle$\tablenotemark{b} [K] & 210 & 160 \\
$S_\nu$(J0132\minus169)\tablenotemark{c} [Jy] & 1.38/1.35 & 1.28/1.20 \\
$S_\nu$(J0457\minus234)\tablenotemark{c} [Jy] & 1.28/1.20 & 0.77/0.75 \\
baseline length [m] & 6.8--64.9 & 11.4--175.0  \\
integration time\tablenotemark{d} [hr] & 3.8 & 2.7 \\
center frequency, L/U [GHz] 	& \multicolumn{2}{c}{218.600/228.600} \\
bandwidth [GHz] 			& \multicolumn{2}{c}{2.0} \\
spectral resolution [MHz] 		& \multicolumn{2}{c}{0.8125} \\
mosaic center (J2000) 		& \multicolumn{2}{c}{$\alpha$=\phs03\hr33\mn36\fs40} \\
				    		& \multicolumn{2}{c}{$\delta$=\minus36\arcdeg08\arcmin25\farcs7\phn}  \\
primary beam\tablenotemark{e} [arcsec] 		& \multicolumn{2}{c}{54/52} \\				    
mosaic field of view  [arcsec$^2$]    			& \multicolumn{2}{c}{$100\times50$} \\
rms noise\tablenotemark{f} [mJy \perbeam] 	& \multicolumn{2}{c}{67, 28, 37, 4}  
\enddata
\tablenotetext{a}{Zenith opacity at 225 GHz measured at the Caltech Submillimeter Observatory
next to the SMA.}
\tablenotetext{b}{Median system temperature toward NGC 1365.}
\tablenotetext{c}{Flux density of the calibrator. LSB/USB.}
\tablenotetext{d}{Integration time on the galaxy.}
\tablenotetext{e}{The full width at half maximum of the primary beam approximated to be a circular Gaussian. LSB/USB.}
\tablenotetext{f}{Noise in the \twelveCO, \thirteenCO, \CeighteenO\ cubes and in the continuum data, respectively,
used for Fig. \ref{fig.comaps}. Those for line data are the ones in each channel. The velocity resolution is
10 \kms\ for \twelveCO\ and 20 \kms\ for other  lines.}
\end{deluxetable}

\begin{deluxetable}{lll}
\tablewidth{0pt}
\tablecaption{Measured and derived parameters  \label{t.parameters} }
\tablehead{ 
	\colhead{parameter}       &
	\colhead{value }   &
	\colhead{unit} 
}
\startdata
$S_{\rm ^{12}CO(2-1)}  (R_g \leq 1\; \kpc)$            & $4.3\times 10^{3}$                              & Jy \kms \\
$S_{\rm ^{13}CO(2-1)} (R_g \leq 1\; \kpc)$    & $3.8\times10^2$                                  & Jy \kms \\
$S_{\rm C^{18}O(2-1)} (R_g \leq 1\; \kpc)$    & $1.1\times10^2$                                  & Jy \kms \\
$\max I_{\rm ^{12}CO(2-1)}$                            & 42                                                        & Jy \kms\ \persquarearcsec \\
$\max T_{\rm b, ^{12}CO(2-1)}$                      & 13                                                        & K \\
$M_{\rm mol}(R_g \leq 1\; \kpc)$            &  $9 \times 10^{8}$                            & \Msol \\
$\max \Sigma_{\rm mol}$                         & $9 \times 10^{2}$                             & \Msol\ \persquarepc \\
$V_{\rm sys}$ (radio, LSR)                      & 1618                                                   & \kms   \\
R. A. dynamical center (J2000)               & \phs$03$  $33$  $36.35$                      & h, m, s  \\
Dec.  dynamical center (J2000)              & \minus36 $08$  $25.8$                        & \arcdeg, \arcmin, \arcsec \\
$T_{\rm dyn}(R_g=1\; \kpc)$        &  $3\times 10^{7} $                            & yr  \\
$M_{\rm dyn}(R_g\leq 1\; \kpc) $ & $1 \times 10^{10} $                         & \Msol \\
$M_{\rm mol}/M_{\rm dyn} $  $(R_g\leq 1 \; \kpc)$ & 0.1                         & \\
mean $R_{12/13}^{(2-1)}$                       & 8.7                                                      & \\
r.m.s.  $R_{12/13}^{(2-1)}$                       & 2.8                                                      & \\
$\Sigma_{\rm mol}/\Sigma_{\rm SFR}\, (R_g\leq 1\; \kpc)$ & $1\times10^{8}$  & yr
\enddata
\tablecomments{Parameters related to gas mass are derived from \twelveCO(2--1) data using
$N(\HH)/I_{\rm CO(2-1)} = 0.5 \times 10^{20}$ \unitofx\ and a factor of 1.36 for helium.
See text for other assumptions and methods used to derive the parameters, and for their uncertainties. }
\end{deluxetable}

\begin{deluxetable}{cllccccccccccc}
\tabletypesize{\small}
\tablewidth{0pt}
\tablecaption{Association of CO Peaks with Known Sources  \label{t.co_others} }
\tablehead{ 
	\multicolumn{3}{c}{sources} &
	\colhead{} &
	\multicolumn{6}{c}{association with a CO peak} &
	\colhead{} &
	\multicolumn{3}{c}{source parameters} 
	\\
	\cline{1-3}
	\cline{5-10}
	\cline{12-14}
	\\
	\colhead{radio}       &
	\colhead{mid-IR}      &
	\colhead{optical}       &
	\colhead{} &
	\colhead{$I_{12}$} 	     &
	\colhead{$T_{\rm b, 12}$} &
	\colhead{$I_{13}$}  &
	\colhead{$T_{\rm b, 13}$}  &
	\colhead{$I_{18}$}  &
	\colhead{$T_{\rm b, 18}$}	&
	\colhead{} &
	\colhead{$S_{\rm 3.5\; cm}$}   &
	\colhead{$S_{\rm 11.9\; \mu m}$} &
	\colhead{$m_{\rm B}$}	
	\\
	\colhead{(1)}       &
	\colhead{(2)}      &
	\colhead{(3)}       &
	\colhead{} &
	\colhead{(4)} 	     &
	\colhead{(5)} &
	\colhead{(6)}  &
	\colhead{(7)}  &
	\colhead{(8)}  &
	\colhead{(9)}	&
	\colhead{} &
	\colhead{(10)}   &
	\colhead{(11)} &
	\colhead{(12)}			   	
}
\startdata 
Nuc. & M1   &  Nucleus  &   &  N    &  N   &  N  &  N  &  N  &  N & &  0.24   & $510\pm100$ & $17.00$ \\  
A      & \nd   &  SSC 10    &    &  N    &  N   &  N  &  N  &  N  &  N & &  0.97   & \nd                     & $21.42$ \\  
\nd   & M2   &  SSC 6      &    &  N    &  N   &  N   & N  &  N  &  N & &  \nd      & $< 20$              & $17.21$ \\
\nd   & M3   &  SSC 3      &    &  N    &  N   &  N   & N  &  N  &  N & &  \nd      & $30\pm 6$       & $17.20$ \\
D      & M4   &  \nd\tm{a} &   &  Y    &  Y   &  Y    & Y  &  N  &  Y & &  1.47    & $40 \pm 8$      & \nd         \\    
E      & M5   &  \nd\tm{b}  &    &  Y    &  Y   &  Y? &  Y  &  N  &  N & &  0.80   & $120 \pm 20$  & \nd         \\   
F      & \nd   &  \nd            &    &  Y? &  Y? &  N   &  N  &  N  &  N & &  0.63   & \nd                     & \nd         \\   
G      & M6   &  \nd            &    &  Y    &  Y   &  Y   &  Y  &  N   &  Y? & &  1.56   & $140\pm 20$  & \nd         \\    
\nd   & M7   &  \nd            &    &  N    &  N   &  N   &  N  &  N  &  N & &  \nd     &  \nd                    & \nd         \\
H      & M8? &  \nd            &    &  Y    &  Y   &  N   &  Y  &  N  &  N & &  0.58   &  \nd                    & \nd         \\   
J       & \nd   &  \nd            &    &  N    &  N   &  N   &  N &  N  &  N & &  0.37   &  \nd                    & \nd 
\enddata
\tablecomments{
Col. (1) Radio sources \citep{Sandqvist95, Forbes98}. 
Col. (2) Mid-IR sources \citep{Galliano05}. M1 is the nucleus.
Col. (3) Optical sources \citep{Kristen97}.
Cols. (4)--(9) Association with a peak in integrated intensity ($I$) and 
with a peak in brightness temperature ($T_{\rm b}$). 
The subscripts 12, 13, and 18 indicate \twelveCO(2--1), \thirteenCO(2--1), and \CeighteenO(2--1),
respectively.
Col. (10) Flux density in mJy at 3.5 cm \citep{Stevens99}. The numbers here differ from those
in Table \ref{t.othergalaxies} because of the different methods used to estimate and subtract the background.
Col. (11) Flux density in mJy in the `11.9 \micron\ filter' of \citet{Galliano05}. 
Col. (12) B magnitude \citep{Kristen97}.
}
\tablenotetext{a}{\citet{Galliano05} pointed out a conical feature near M4 in a HST image. See Fig. \ref{fig.cohst}.  }
\tablenotetext{b}{\citet{Galliano05} pointed out a double source near M5 in a HST image. See Fig. \ref{fig.cohst}. }
\end{deluxetable}

\begin{deluxetable}{ccccccrrl}
\tablewidth{0pt}
\tablecaption{Properties of CO Hotspots with a Radio Counterpart \label{t.clumps} }
\tablehead{ 
	\colhead{R.A.}   &
	\colhead{Dec.} &
	\colhead{velocity}   &
	\colhead{$\max T_{\rm b, 12}$} &
	\colhead{$I_{12}$} &
	\colhead{$\Delta V_{12}$}  &
	\colhead{$T_{\rm b,12}/T_{\rm b,13}$} &		
	\colhead{$I_{12}/I_{13}$} &			
	\colhead{association}
\\
	\colhead{(1)}   &
	\colhead{(2)}   &
	\colhead{(3)} &
	\colhead{(4)} &
	\colhead{(5)}  &
	\colhead{(6)} &		
	\colhead{(7)} &			
	\colhead{(8)} &
	\colhead{(9)}	
}
\startdata 
36\fs39  &  18\farcs7  &  1560 &  12.7 & 42  &  70  &   $7.3\pm0.6$  & $7.7\pm0.5$   & D, M4 \\
36\fs61  &  16\farcs0  &  1540 &  13.3 & 34  &  60  &   $7.8\pm0.7$  & $9.3\pm0.8$   & E, M5 \\
36\fs77  &  18\farcs5  &  1480 &  10.2 & 33  &  80  &   $5.5\pm0.4$  & $8.1\pm0.7$   & G, M6 \\
36\fs76  &  23\farcs6  &  1510 &  10.5 & 32  &  70  &   $9.9\pm1.4$  & $11.2\pm1.3$ & H, M8? \\
36\fs64  &  28\farcs0  & 1590  &    9.8 & 37  &  90  & $12.9\pm2.2$  & $14.3\pm1.7$ & F
\enddata
\tablecomments{
Cols. (1)--(2) Position. $\alpha$=0\hr33\mn, $\delta= -36\arcdeg08\arcmin$ (J2000.0).
Peak positions measured in the peak temperature maps of \twelveCO\ and \thirteenCO\ 
are averaged for the CO hotspots corresponding to D,E,G, and H.
The radio position is used for F.
Col. (3) Velocity at which the peak brightness temperature of \twelveCO(2--1) emission is observed. 
The unit is \kms. 
Columns from (3) to (6) are measured in the data cubes of $2\farcs6 \times 1\farcs8 \times 10\; \kms$ resolution.
Col. (4) Peak brightness temperature of \twelveCO(2--1) emission in K.
Col. (5) Integrated \twelveCO(2--1) intensity in the units of Jy \kms\ \persquarearcsec.
Col. (6) Full width at half maximum of the \twelveCO(2--1) line. The unit is \kms.
Col. (7) Ratio of \twelveCO(2--1) to \thirteenCO(2--1) brightness temperature at the velocity 
where \twelveCO(2--1) has the maximum intensity. 
The ratios in this and the next columns are measured in $3\farcs3 \times 2\farcs0 \times 20\; \kms$ 
resolution cubes made using visibilities in the same \uv\ range. The error is 1 $\sigma$.
Col. (8) Ratio of \twelveCO(2--1) to \thirteenCO(2--1) integrated brightness temperatures 
and its 1$\sigma$ error.
Col. (9) Spatial association with sources in other wavelengths. 
D through H are radio sources \citep{Sandqvist95, Forbes98} and M\# are mid-IR sources \citep{Galliano05}.
}
\end{deluxetable}

\begin{deluxetable}{lcccl}
\tablewidth{0pt}
\tablecaption{CO Hotspot -- Cluster Associations  \label{t.othergalaxies} }
\tablehead{ 
	\colhead{}   &
	\colhead{NGC 1365}   &
	\colhead{Antennae}   &		
	\colhead{M83} &  
	\colhead{(unit)}
}
\startdata 
distance\tablenotemark{a} & 17.9 & 21.5 & 4.5 & Mpc \\
$\log L_{\rm 8-1000\; \mu m}$\tablenotemark{a} & 11.00  & 10.84 &  10.29 &  \Lsol \\
name(s) & D/M4,E/M5,G/M6 & WS95-80/SGMC4-5\tablenotemark{b} & M83-2\tablenotemark{c} &  \\
$S_{\rm 3.5\,cm}$ & 1.8--4.2\tablenotemark{d} & 4.7\tablenotemark{e} & 4.8\tablenotemark{f} & mJy \\
$P_{\rm 3.5\,cm}$ & 0.7--1.5 & 2.6 & 0.11 & $10^{20}$ W \perhertz \\
$S_{\rm 11.9\,\mu m}$ & 40--140 & 70\tablenotemark{g} & \tilde1000\tablenotemark{h} & mJy \\
$P_{\rm 11.9\,\mu m}$ & 15--54 & 39 & \tilde24 & $10^{20}$ W \perhertz \\
$\max T_{\rm b}({\rm CO})$\tablenotemark{i} & 10--13 & 4 & 11 & K \\
$\max[\int\! T_{\rm b}({\rm CO}) dv]$\tablenotemark{j} & 9--11 & 3 & 7 & $10^2$ K\kms \\
$\Sigma_{\rm mol}$\tablenotemark{k} & 1.9--2.4 & 0.7 & 1.5 & $10^3 x_{\rm CO}$ \Msol\ \persquarepc \\
$\theta_{\rm CO\;beam}$\tablenotemark{l} & 1.8 & 4.1 & 0.7 & $10^2$ pc
\enddata
\tablenotetext{a}{The galaxy distances and luminosities are from \citet{Sanders03}, except that M83's 
distance is from \citet{Thim03} and its luminosity is scaled for the distance.}
\tablenotetext{b}{The names are from \citet{Whitmore95} and \citet{Wilson00}. The latter
lists other names of the source. }
\tablenotetext{c}{In Table 2 of \citet{Turner94}. }
\tablenotetext{d}{These flux densities are measured from the 3.5 cm map of \citet{Morganti99}. 
The spatial resolution of the data is 1\farcs6$\times$0\farcs8.
High-pass filtering is made for background subtraction
using the AIPS task {\sf MWFLT} with a 4\farcs5$\times$4\farcs5 filtering box,
to be consistent with the Antennae data. }
\tablenotetext{e}{The flux density of the source reported in \citet{Neff00}. 
A 1\farcs4$\times$1\farcs0 resolution image was used after high-pass filtering.}
\tablenotetext{f}{Interporated from the 2 cm and 6 cm data  in \citet{Turner94}.  
The data resolution is 2\farcs5 $\times$1\farcs6. No background subtraction is made and
the source is assumed to be unresolved. }
\tablenotetext{g}{Flux in a 6\arcsec\ pixel on the source \citep{Mirabel98, Wilson00}.}
\tablenotetext{h}{Flux in a 4\arcsec\ pixel on the source \citep{Telesco93}.}
\tablenotetext{i}{Peak excess brightness temperature of \twelveCO\ emission. 
The transition of CO is J=2--1 for NGC 1365 and M83 and J=1--0 for the Antennae.
The Antenna data are from \citet{Wilson00}  and M83 data from \citet{Sakamoto04}.}
\tablenotetext{j}{Integrated \twelveCO\ intensity.}
\tablenotetext{k}{Peak surface density of molecular gas at the CO hotspot. It is estimated from the
peak CO integrated intensity. Correction for inclination has not been made.}
\tablenotetext{l}{Resolution of the CO data, i.e., the FWHM of the observing beam at the distance of the galaxy.}
\end{deluxetable}

\clearpage

\clearpage
\begin{thebibliography}{}	
\bibitem[Aalto et al.(1995)]{Aalto95} 
	Aalto, S., Booth, R. S., Black, J. H., and Johansson, L. E. B.
	1995, \aap, 300, 369
\bibitem[Barth et al.(1995)]{Barth95} 
	Barth, A. J., Ho, L. C., Filippenko, A. V., and Sargent, W. L.
	1995, \aj, 110, 1009	
\bibitem[Binney \& Tremaine(1987)]{Binney87}
	Binney, J. and Tremaine, S.
	1987, Galactic Dynamics, (Princeton: Princeton Univ. Press)
\bibitem[Bridle \& Greisen(1994)]{Bridle94}
	Bridle, A. H., and Greisen, E. W. 
	1994, NRAO AIPS Memo 87
\bibitem[Bryant \& Scoville(1996)]{Bryant96}
	Bryant, P. M., and Scoville, N. Z.
	1996, \apj, 457, 678		
\bibitem[Casoli et al.(1992)]{Casoli92b}  
	Casoli, F., Dupraz, C., and Combes, F.
	1992, \aap, 264, 55  				
\bibitem[Contopoulos \& Papayannopoulos(1980)]{Contopoulos80}
	Contopoulos, G., Papayannopoulos, T.
	1980, \aap, 90, 33
\bibitem[Curran et al.(2001)]{Curran01}  
	Curran, S. J., Polatidis, A. G., Aalto, S., and Booth, R. S.
	2001, \aap, 368, 824	
\bibitem[Dahmen et al.(1998)]{Dahmen98}
	Dahmen, G., H\"{u}ttemeister, S., Wilson, T. L., and Mauersberger, R.
	1998, \aap, 331, 959	
\bibitem[Dame et al.(2001)]{Dame01}
	Dame, T. M., Hartmann, D., and Thaddeus, P.
	2001, \apj, 547, 792	
\bibitem[de Grijp et al.(1985)]{deGrijp85}
	de Grijp, M. H. K., Miley, G. K., Lub, J., and de Jong, T.
	1985, \nat, 314, 240	
\bibitem[de Vaucouleurs et al.(1991)]{rc3}  
	de Vaucouleurs, G., de Vaucouleurs, A., Corwin, H. G., Buta, R. J., 
	Paturel, G., \& Fouque, P. 1991, 
	"Third Reference Catalogue of Bright Galaxies", (Springer Verlag)	
\bibitem[Elmegreen(1994)]{Elmegreen94} 
	Elmegreen, B. G.
	1994, \apjl, 425, L73	
\bibitem[Forbes \& Norris(1998)]{Forbes98}
	Forbes, D. A., Norris, R. P.
	1998, \mnras, 300, 757			
\bibitem[Freedman et al.(2001)]{Freedman01}
	Freedman, W. L., et al. 2001, \apj, 553, 47
\bibitem[Galliano et al.(2005)]{Galliano05}
	Galliano, E., Alloin, D., Pantin, E., Lagage, P. O., and Marco, O.
	2005, \aap, 438, 803		
\bibitem[Gilbert et al.(2000)]{Gilbert00}
	Gilbert, A. M. et al.
	2000, \apj, 533, L57	
\bibitem[Glenn \& Hunter(2001)]{Glenn01}
	Glenn, J., and Hunter, T. R.
	2001, \apjs, 135, 177	
\bibitem[Goldreich \& Kwan(1974)]{Goldreich74}
	Goldreich, P. \& Kwan, J.
	1974, \apj, 189, 441
\bibitem[Haas et al.(2000)]{Haas00}
	Haas, M., Klaas, U., Coulson, I., Thommes, E., and Xu, C.
	2000, \aap, 356, L83	
\bibitem[Henkel \& Mauersberger(1993)]{Henkel93}
	Henkel, C., and Mauersberger, R.
	1993, \aap, 274, 730	
\bibitem[Hjelm \& Lindblad(1996)]{Hjelm96}
	Hjelm, M., and Lindblad, P. O.
	1996, \aap, 305, 727		
\bibitem[Ho et al.(2004)]{Ho04}
	Ho, P. T. P., Moran, J. M., Lo, K. Y.
	2004, \apjl, 616, L1	
\bibitem[Hunter et al.(1997)]{Hunter97}
	Hunter, S. D., et al. 
	1997, \apj, 481, 205
\bibitem[H\"{u}ttemeister et al.(1995)]{Huttemeister95} 
	H\"{u}ttemeister, S., Wilson, T. L., Mauersberger, R., Lemme, C., Dahmen, G., and Henkel, C.
	1995, \aap, 294, 667		
\bibitem[Iyomoto et al.(1997)]{Iyomoto97}
	Iyomoto, N., Makishima, K., Fukazawa, Y., Tashiro, M., and Ishisaki, Y.
	1997, \pasj, 49, 425	
\bibitem[Jarrett et al.(2003)]{Jarrett03}
	Jarrett, T. H., Chester, T., Cutri, R., Scheneider, S. E., and Huchra, J. P. 
	2003, \aj, 125, 525	
\bibitem[Jogee et al.(2005)]{Jogee05}
	Jogee, S., Scoville, N., and Kenney, J. D. P.
	2005, \apj, 630, 837
\bibitem[J\"{o}rs\"{a}ter \& van Moorsel(1995)]{Jorsater95}
	J\"{o}rs\"{a}ter, S. and  van Moorsel, G. A.
	1995, \aj, 110, 2037
\bibitem[Kenney et al.(1992)]{Kenney92} 
	Kenney, J. D. P., Wilson, C. D., Scoville, N. Z., Devereux, N. A., and Young, J. S.
	1992, \apjl, 395, L79	
\bibitem[Kennicutt(1998)]{Kennicutt98} 
	Kennicutt, R. C. 
	1998, \apj, 498, 541 (K98)	
\bibitem[Keto et al.(2005)]{Keto05}
	Keto, E., Ho, L. C., and Lo, K. -Y.
	2005, \apj, 635, 1063	
\bibitem[Knapen et al.(2006)]{Knapen06}
	Knapen, J. H., Mazzuca, L. M., B\"{o}ker, T., Shlosman, I., Colina, L., Combes, F., and Axon, D. J.
	2006, \aap, 448, 489		
\bibitem[Komossa \& Schulz(1998)]{Komossa98}
	Komossa, S. and Schulz, H.
	1998, \aap, 339, 345	
\bibitem[Kristen et al.(1997)]{Kristen97}
	Kristen, H., Jorsater, S., Lindblad, P. O., and Boksenberg, A.
	1997, \aap, 328, 483	
\bibitem[Kronberg et al.(1985)]{Kronberg85}
	Kronberg, P. P., Biermann, P., and Schwab, F. R.
	1985, \apj, 291, 693	
\bibitem[Lindblad et al.(1996)]{Lindblad96}
	Lindblad, P. A. B., Lindblad, P. O., and Athanassoula, E.
	1996, \aap, 313, 65		
\bibitem[Lindblad \& Lindblad(1994)]{Lindblad94} 
	Lindblad, P.~O., and Lindblad, P.~A.~B. 
	1994, ASP Conf.~Ser.~ 66: Physics of the Gaseous and Stellar Disks of the Galaxy, 66, 29 	
\bibitem[Lindblad(1999)]{Lindblad99}
	Lindblad, P. O.,
	1999, \aaps, 9, 221	
\bibitem[Maloney \& Black(1988)]{Maloney88}
	Maloney, P., and Black J. H.
	1988, \apj, 325, 389	
\bibitem[Mao et al.(2000)]{Mao00}
	Mao, R. Q. et al.
	2000, \aap, 358, 433	
\bibitem[Matsuda \& Nelson(1977)]{Matsuda77} 
	Matsuda, T., and Nelson, A.~H. 
	1977, \nat, 266, 607 
\bibitem[Matsushita et al.(2006)]{Matsushita06}
	Matsushita, S., Saito, M., Sakamoto, K., Hunter, T.R.,
         Patel, N.A., Sridharan, T.K., and Wilson, R.W.  
	2006, Proc. SPIE, 6275, 62751W
\bibitem[Meier et al.(2000)]{Meier00} 
	Meier, D. S., Turner, J. L., and Hurt, J. L.
	2000, \apj, 531, 200		
\bibitem[Meier \& Turner(2001)]{Meier01}
	Meier, D. S., and Turner, J. L.
	2001, \apj, 551, 687		
\bibitem[Meier \& Turner(2004)]{Meier04}
	Meier, D. S., and Turner, J. L.
	2004, \aj, 127, 2069	
\bibitem[Miley et al.(1985)]{Miley85}
	Miley, G. K., Neugebauer, G., and Soifer, B. T.
	1985, \apjl, 293, L11	
\bibitem[Mirabel et al.(1998)]{Mirabel98}
	Mirabel, I. F. et al.
	1998, \aap, 333, L1		
\bibitem[Morgan(1958)]{Morgan58} 
	Morgan, W. W.
	1958, \pasp, 70, 364	
\bibitem[Morganti et al.(1999)]{Morganti99}
	Morganti, R., Tsvetanov, Z. I., Gallimore, J., and Allen, M. G.
	1999, \aaps, 137, 457	
\bibitem[Neff \& Ulvestad(2000)]{Neff00}
	Neff, S. G., and Ulvestad, J. S.
	2000, \aj, 120, 670	
\bibitem[Oka et al.(1998)]{Oka98}
	Oka, T., Hasegawa, T., Hayashi, M., Handa, T., and Sakamoto, S.
	1998, \apj, 493, 730		
\bibitem[Ott et al.(2005)]{Ott05}
	Ott, J., Wei\ss, A., Henkel, C., and Walter, F.
	2005, in AIP Conf. Proc. 783, The Evolution of Starbursts, ed. S. H\"uttemeister et al., 141
\bibitem[Papadopoulos \& Seaquist(1998)]{Papadopoulos98}
	Papadopoulos, P. P., and Seaquist, E. R.
	1998, \apj, 492, 521	
\bibitem[Phillips et al.(1983)]{Phillips83}
	Phillips, M. M., Turtle, A. J., Edmunds, M. G., and Pagel, B. E. J.
	1983, \mnras, 203, 759	
\bibitem[Pilyugin(2003)]{Pilyugin03}
	Pilyugin, L. S.
	2003, \aap, 397, 109	
\bibitem[Regan \& Elmegreen(1997)]{Regan97}
	Regan, M., and Elmegreen, D. M.
	1997, \aj, 114, 965
\bibitem[Risaliti et al.(2005)]{Risaliti05}
	Risaliti, G., Elvis, M., Fabbiano, G., Baldi, A., and Zezas, A.
	2005, \apj, 623, 93		
\bibitem[Roussel et al.(2001)]{Roussel01}
	Roussel, H., et al.
	2001, \aap, 372, 406	
\bibitem[Roy \& Walsh(1997)]{Roy97}
	Roy, J. -R., and Walsh, J. R.	
\bibitem[Saikia et al.(1994)]{Saikia94}
	Saikia, D. J., Pedler, A.,  Unger, S. W., and Axon, D. J.
	1994, \mnras, 270, 46
\bibitem[Sakamoto et al.(1999a)]{Sakamoto99a}  
	Sakamoto, K., Okumura, S.~K., Ishizuki, S., and Scoville, N.~Z. 
	1999a, \apjs, 124, 403 	
\bibitem[Sakamoto et al.(1999b)]{Sakamoto99b}  
	Sakamoto, K., Okumura, S.~K., Ishizuki, S., and Scoville, N.~Z. 
	1999b, \apj, 525, 691 
\bibitem[Sakamoto et al.(2004)]{Sakamoto04}
	Sakamoto, K., Matsushita, S., Peck, A. B., Wiedner, M. C., and Iono, D.
	2004, \apjl, 616, L59		
\bibitem[Sakamoto et al.(2006a)]{Sakamoto06a} 
	Sakamoto, K. et al. 2006a, \apj, 636, 685	
\bibitem[Sakamoto et al.(2006b)]{Sakamoto06b} 
	Sakamoto, K., Ho, P. T. P., and Peck, A. B.
	2006b, \apj, 644, 862
\bibitem[Sanders et al.(1984)]{Sanders84}
	Sanders, D. B., Solomon, P. M., and Scoville, N. Z.
	1984, \apj, 276, 182			
\bibitem[Sanders et al.(2003)]{Sanders03}  
	Sanders, D. B., Mazzarella, J. M., Kim D. -C., Surace, J. A., and Soifer, B. T.
	2003, \aj, 126, 1670		
\bibitem[Sandqvist et al.(1982)]{Sandqvist82}
	Sandqvist, A., J\"{o}rs\"{a}ter, S., and Lindblad, P. O.
	1982, \aap, 110, 336	
\bibitem[Sandqvist et al.(1988)]{Sandqvist88}
	Sandqvist, A., Elfhang, T., and J\"{o}rs\"{a}ter, S.
	1988, \aap, 201, 223		
\bibitem[Sandqvist et al.(1995)]{Sandqvist95}
	Sandqvist, A., J\"{o}rs\"{a}ter, S., and Lindblad, P. O.
	1995, \aap, 295, 585
\bibitem[Sandqvist et al.(1999)]{Sandqvist99}
	Sandqvist, A.
	1999, \aap, 343, 367	
\bibitem[Sault et al.(1995)]{Sault95}
	Sault, R. J., Teuben P. J., and Wright M. C. H. 1995, 
	in ASP Conf. Ser. 77, Astronomical Data Analysis Software and Systems IV, 
	ed. R. Shaw et al. (San Francisco: ASP), 433	
\bibitem[Sawada et al.(2001)]{Sawada01}
	Sawada, T. et al. 
	2001, \apjs, 136, 189		
\bibitem[Scoville et al.(1993)]{Scoville93}
	Scoville, N. Z., Carlstrom, J. E., Chandler, C. J., Phillips, J. A., Scott, S. L., Tilanus, R. P. J., and Wang, Z.
	1993, \pasp, 105, 1482
\bibitem[Sersic \& Pastoriza(1967)]{Sersic67}
	S\'{e}rsic, J. L., and Pastoriza, M.
	1967, \pasp, 79, 152	
\bibitem[Sheth et al.(2005)]{Sheth05}
	Sheth, K., Vogel, S. N., Regan, M. W., Thornley, M. D., and Teuben, P. J.
	2005, \apj, 632, 217 
\bibitem[Sodroski et al.(1995)]{Sodroski95}
	Sodroski, T. J. et al. 
	1995, \apj, 452, 262
\bibitem[Stevens et al.(1999)]{Stevens99}
	Stevens, I. R., Forbes, D. A., and Norris, R. P.
	1999, \mnras, 306, 479
\bibitem[Strong et al.(2004)]{Strong04}
	Strong, A. W., Moskalenko, I. V., Reimer, O., Digel, S., and Diehl, R.
	2004, \aap, 422, L47		
\bibitem[Stutzki \& G\"{u}sten(1990)]{Stutzki90}
	Stutzki, J., and G\"{u}sten, R. 
	1990, \apj, 356, 513		
\bibitem[Telesco et al.(1993)]{Telesco93} 
	Telesco, C. M., Dressel, L. L., and Wolstencroft, R. D.
	1993, \apj, 414, 120
\bibitem[Teuben et al.(1986)]{Teuben86}
	Teuben, P. J., Sanders, R. H., Atherton, P. D., and van Albada, G. D.
	1986, \mnras, 221, 1	
\bibitem[Thean et al.(2000)]{Thean00}
	Thean, A., Pedlar, A., Kukula, M. J., Baum, S. A., O'Dea, C. P.
	2000, \mnras, 314, 573	
\bibitem[Thim et al.(2003)]{Thim03}
	Thim, F., Tammann, G. A., Saha, A., Dolphin, A., Sandage, A., Tolstoy, E., and Lbhardt, L.
	2003, \apj, 590, 256		
\bibitem[Turner \& Ho(1994)]{Turner94} 
	Turner, J. L., and Ho, P. T. P.
	1994, \apj, 421, 122
\bibitem[Ulvestad \& Antonucci(1997)]{Ulvestad97}	
	Ulvestad, J. S., and Antonucci, R. R. J.
	1997, \apj, 488, 621	
\bibitem[Veilleux et al.(2003)]{Veilleux03}
	Veilleux, S., Shopbell, P. L., Rupke, D. S., Bland-Hawthorn, J., and Cecil, G.
	2003, \aj, 126, 2185	
\bibitem[Wada(1994)]{Wada94}   
	Wada, K.
	1994, \pasj, 46, 165 
\bibitem[Wada \& Habe(1992)]{Wada92}  
	Wada, K., \& Habe, A. 
	1992, \mnras, 258, 82 	
\bibitem[Weiler et al.(2002)]{Weiler02}
	Weiler, K. W., Panagia, N., Montes, M. J., and Sramek, R. A.
	2002, \araa, 40, 387	
\bibitem[Whitmore \& Schweizer(1995)]{Whitmore95} 
	Whitmore, B. C., and Schweizer, F.
	1995, \aj, 109, 960	
\bibitem[Wild et al.(1992)]{Wild92}
	Wild, W., et al.
	1992, \aap, 265, 447	
\bibitem[Wilson et al.(2000)]{Wilson00}
	Wilson, C. D., Scoville, N. Z., Madden, S. C., and Charmandaris, V.
	2000, \apj, 542, 120	
\bibitem[Wilson et al.(2003)]{Wilson03}
	Wilson, C. D., Scoville, N. Z., Madden, S. C., and Charmandaris, V.
	2003, \apj, 599, 1049
\bibitem[Wilson \& Rood(1994)]{Wilson94}
	Wilson, T. L., and Rood, R.
	1994, \araa, 32, 191
\bibitem[Zhang et al.(2001)]{Zhang01}
	Zhang, Q., Fall, S. M., and Whitmore, B. C.
	2001, \apj, 561, 727			
\end{thebibliography}
\end{document}